\documentclass[twocolumn]{aastex631}

\newcommand{\pkg}[1]{\texttt{#1}}

\usepackage{amssymb}
\usepackage{amsmath}
\usepackage{float}
\usepackage{hyperref}
\usepackage{rotating}
\usepackage{xcolor}
\hypersetup{
    colorlinks,
    linkcolor={blue!50!black},
    citecolor={blue!50!black},
    urlcolor={blue!80!black}
}

\usepackage{graphicx}
\usepackage{verbatim}
\usepackage{placeins}
\usepackage{multirow}
\usepackage{tabularx}
\usepackage{footmisc}

\begin{document}

\title{Interpolation techniques for reconstructing Galactic Faraday rotation}

\author[0009-0001-2196-8251]{Affan Khadir}
\affiliation{David A. Dunlap Department of Astronomy and Astrophysics, University of Toronto, 50 St. George Street, Toronto, M5S 3H4, ON, Canada}
\affiliation{Dunlap Institute for Astronomy and Astrophysics, University of Toronto, 50 St. George Street, Toronto, M5S 3H4, ON, Canada}
\affiliation{Department of Physics, University of Toronto, Toronto, ON M5R 2M8, Canada}
\correspondingauthor{Affan Khadir}
\email{affan.khadir@mail.utoronto.ca}

\author[0000-0002-8897-1973]{Ayush Pandhi}
\affiliation{David A. Dunlap Department of Astronomy and Astrophysics, University of Toronto, 50 St. George Street, Toronto, M5S 3H4, ON, Canada}
\affiliation{Dunlap Institute for Astronomy and Astrophysics, University of Toronto, 50 St. George Street, Toronto, M5S 3H4, ON, Canada}

\author[0000-0002-6952-9688]{Sebastian Hutschenreuter}
\affiliation{University of Vienna, Department of Astrophysics, T\"urkenschanzstrasse 17, 1180 Vienna, Austria}

\author[0000-0002-3382-9558]{B. M. Gaensler}
\affiliation{Dunlap Institute for Astronomy and Astrophysics, University of Toronto, 50 St. George Street, Toronto, M5S 3H4, ON, Canada}
\affiliation{David A. Dunlap Department of Astronomy and Astrophysics, University of Toronto, 50 St. George Street, Toronto, M5S 3H4, ON, Canada}
\affiliation{Division of Physical and Biological Sciences, University of California Santa Cruz, Santa Cruz, CA 95064, USA}

\author[0009-0004-7773-1618]{Shannon Vanderwoude}
\affiliation{David A. Dunlap Department of Astronomy and Astrophysics, University of Toronto, 50 St. George Street, Toronto, M5S 3H4, ON, Canada}
\affiliation{Dunlap Institute for Astronomy and Astrophysics, University of Toronto, 50 St. George Street, Toronto, M5S 3H4, ON, Canada}

\author[0000-0001-7722-8458]{Jennifer L. West}
\affiliation{National Research Council Canada, Herzberg Research Centre for Astronomy and Astrophysics, Dominion Radio Astrophysical Observatory, PO Box 248, Penticton, BC V2A 6J9, Canada}

\author[0000-0002-3968-3051]{Shane P. O'Sullivan}
\affiliation{Departamento de Física de la Tierra y Astrofísica \& IPARCOS-UCM, Universidad Complutense de Madrid, 28040 Madrid, Spain
}



\begin{abstract}

The line-of-sight structure of the Galactic magnetic field (GMF) can be studied using Faraday rotation measure (RM) grids. We analyze how the choice of interpolation kernel can affect the accuracy and reliability of reconstructed RM maps. We test the following kernels: inverse distance weighting (IDW), natural neighbour interpolation (NNI), inverse multiquadric interpolation (IM), thin-plate spline interpolation (TPS), and a Bayesian rotation measure sky (BRMS); all techniques were tested on two simulated Galactic foreground RMs (one assuming the GMF has patchy structures and the other assuming it has filamentary structures) using magnetohydrodynamic simulations. Both foregrounds were sampled to form RM grids with densities of $\sim$40 sources deg$^{-2}$ and area $\sim$144 deg$^2$. The techniques were tested on data sets with different noise levels and Gaussian random extragalactic RM contributions. The data set that most closely emulates expected data from current surveys, such as the POlarization Sky Survey of the Universe's Magnetism (POSSUM), had extragalactic contributions and a noise standard deviation of $\sim 1.5$ rad m$^{-2}$. For this data set, the accuracy of the techniques for the patchy structures from best to worst was: BRMS, NNI, TPS, IDW and IM; while in the filamentary simulate foreground it was: BRMS, NNI, TPS, and IDW. IDW is the most computationally expensive technique, while TPS and IM are the least expensive. BRMS and NNI have the same, intermediate computational cost. This analysis lays the groundwork for Galactic RM studies with large radio polarization sky surveys, such as POSSUM.

\end{abstract}

\keywords{Astrostatistics (1882); Radio astronomy (1338); Milky Way magnetic fields (1057)}


\section{Introduction} \label{sec:intro}
Understanding the Galactic magnetic field (GMF) is an important doorway to understanding other phenomena, such as the behaviour of ultra-high energy cosmic rays \citep{KACHELRIES_2007}, the diffusion of galactic cosmic rays \citep{Shukurov_2017}, the diffusive the properties of the interstellar medium (ISM) \citep{RevModPhys.73.1031}, and star formation \citep[e.g.,][]{2019FrASS...6....7K, pandhi_2023}. 

One method of probing the line-of-sight (LOS) GMF is by using the Faraday rotation observed in background linearly polarized radio sources. 
When linearly polarized light propagates through a magnetoionic plasma, the angle $\psi$ of the plane of polarization is rotated due to Faraday rotation. The change in $\psi$ due to this rotation is given by: 
\begin{eqnarray}
    \label{eq:rm}
     \psi - \psi_0 =  \frac{e^3\lambda^2}{2\pi m_e^2 c^4} \int_{0 }^d n_e(l)B_\parallel(l)dl, 
\end{eqnarray}

where $\psi_0$ is the initial polarization angle in radians, $l$ is the position along the LOS in cm, $n_e(l)$ is the thermal electron density in cm$^{-3}$, $B_\parallel(l)$ is the parallel component of the magnetic field strength at position $l$ in gauss, and $d$ is the distance from the radio source in cm \citep{2021MNRAS.507.4968F}. 

Rotation measure (RM) is a variable that quantifies the extent of this Faraday rotation as a function of the observing wavelength:
\begin{eqnarray}
     \psi = \psi_0 + \mathrm{RM}\lambda^2.
\end{eqnarray}
First, the polarization angle is measured: 
\begin{eqnarray}
    \psi = \frac{1}{2} \arctan\left(\frac{U}{Q}\right),
\end{eqnarray}
where $Q, U$ are the two linear Stokes parameters \citep{Brentjens_2005}. The predominant algorithms for computing RMs of linearly polarized sources are RM synthesis \citep{Brentjens_2005} and QU fitting \citep{Farnsworth}. Alternatively, the RM can also be found by fitting the slope for a $\psi$ vs. $\lambda^2$ plot. 

Some early compilations of RM sources include those by \citet{1981ApJS...45...97S}, \citet{1988Ap&SS.141..303B}, and \citet{Han_1999}. 
Currently, the most widely used single RM survey is the NRAO VLA Sky Surveys (NVSS) RM catalogue. The survey covered the sky north of declination $-40$ deg, with a density of $\sim$ 1 source deg$^{-2}$ \citep{Condon_1998, taylor2009rotation}. In contrast, the upcoming Australian Square Kilometre Array Pathfinder (ASKAP) Polarization Sky Survey of the Universe's Magnetism (POSSUM), focuses on the southern sky, and aims to achieve a density of $\sim$ 40 sources deg$^{-2}$ \citep{2010AAS...21547013G, 2024AJ....167..226V}. 
Using these RM data, it is possible to probe the LOS GMF at much smaller spatial scales than is currently possible. Reconstructing the true foreground RM sky from these discrete data points is done using interpolation. Having more accurate foreground RM maps will help us constrain and refine models of the GMF \citep[e.g.,][]{1980ApJ...242...74S, 1989ApJ...343..760R, Jansson_2012, 2019A&A...623A.113S,Unger}.

In an early work, \citet{Frick_2001} analyzed large-scale magnetic structures using RM data from \citet{1981ApJS...45...97S} and \citet{1988Ap&SS.141..303B}, using interpolation with wavelet analysis on the Orion arm, the Sagittarius arm, and the radio Loop I. Another work focusing on the Orion arm and the radio Loop I carried out by \citet{Dineen} applied an interpolation technique that relies on calculating spherical basis functions from the data set, and using these to calculate coefficients of the spherical harmonics of the foreground RM. A simple interpolation technique was implemented by \citet{taylor2009rotation} on the NVSS RM catalogue. They determined the RM at an interpolation point by taking the median RM of all data points in a 4 deg radius of the interpolation point.

In recent years, many interpolation attempts of the RM sky have used a Bayesian statistical approach. An early attempt at such a reconstruction was carried out by \citet{10.1214/07-BA226} who implemented Bayesian convolution using Markov Chain Monte Carlo, with the assumption of a sparse Gaussian process model for the RM sky, to obtain a reconstructed RM map.

Another Bayesian interpolation technique that is based on Bayesian maximum likelihood analysis to determine the magnetic-field power spectra, similar to Cosmic Microwave Background (CMB) analysis, was developed by \citet{Vogt_2005}. A similar Bayesian inference algorithm was used to reconstruct the Galactic RM sky by \citet{Oppermann_2012} and has since been improved with updated methodology and larger data sets \citep{refId0, Hutschenreuter_2022}; \citet{Hutschenreuter_2022} have used these results to show that the RM structure of the Galactic disk is far thinner and more prominent than found in past works. This inference technique can also fold in dispersion measure (DM) observations of pulsars and fast radio bursts to produce maps of the mean LOS magnetic field, $\langle B_\parallel \rangle$ \citep{pandhi2022method, hutschenreuter2023disentangling}.

In this paper, we will be discussing how the choice of an interpolation prior affects the reconstructed result. We will be analyzing the following interpolation techniques: inverse distance weighting (IDW), natural neighbour interpolation (NNI), inverse multiquadric interpolation (IM), thin plate spline interpolation (TPS), and Bayesian RM sky (BRMS). We have adapted the BRMS technique to work on a patch of the sky (that can be of any size) from \citet{Hutschenreuter_2022}, where it was applied over the full sky. The reconstruction results from BRMS (over the full sky) have been proven to be accurate; however, the purpose of this paper is to test different priors, which we will explore through testing different inference algorithms. In addition, we will be testing the reconstruction on small scales ($\lesssim 10^{1}$ deg). This is because previous reconstructions of the Galactic RM have focused on the larger scale features, having a spacing of $\lesssim$ 0.1145 deg between pixels
\citep{Hutschenreuter_2022}, while being constrained by an RM grid with density of only about 1 RM per square degree.
Future RM surveys \citep{2010AAS...21547013G} will come much closer to the nominal resolution of \citet{Hutschenreuter_2022}. We also note that these interpolation techniques are not specific to reconstruction of foreground RM, and that the conclusions from this paper can be applied to any 2D interpolation of astronomical data that are signed scalars.  

This paper is structured as follows: Section \ref{sec:methods} describes the interpolation techniques, Section \ref{sec:data} gives a description of the simulated data used in the testing, Section \ref{sec:results} presents the results that were obtained from the reconstructions, and Section \ref{sec:discussion} provides a discussion of the results. 

\section{Methods}\label{sec:methods}

In all interpolation techniques we assume that the foreground Galactic RM, hereafter denoted as $\mathrm{RM_{gal}}$, is some function of the on-sky coordinates $x = (\alpha, \delta)$, where $\alpha$ represents the right ascension and $\delta$ represents the declination. A summary of the interpolation techniques used is provided in Table \ref{tab:summary}. For further discussion on interpolation techniques refer to Appendix \ref{app:discussion}.

\subsection{Inverse Distance Weighting (IDW)}
IDW relies on the idea that for a relatively smooth function, the function value can be approximated as a weighted sum of the function values at the positions observed, with a weight, $w_i$, that decreases with increasing distance from the interpolation point \citep{10.1145/800186.810616},  
\begin{eqnarray}
    w_i = \frac{1}{||x_i - x||^p},
\end{eqnarray}
where $x_i$ is the position of the data point, $x$ is the position of the interpolation point, and $p > 0$ is the power parameter. $p$ determines how sharp the peaks of the function at the data points are; higher values of $p$ have sharper peaks and as $p \to \infty$, the radius of the peaks goes to zero. We will be testing with $p=1, 2$ in order to test a global and local interpolation kernel, respectively. A power parameter of $p = 2$ is most commonly used for two-dimensional interpolation \citep{10.1145/800186.810616}. Our motivation behind choosing $p = 1$ is to demonstrate a more global interpolation kernel, while also allowing the RM map to be differentiable everywhere, as power parameters less than this result in maps that are no longer differentiable at the data points \citep{10.1145/800186.810616}. Hereafter, these two interpolation kernels will be referred to as IDW1 and IDW2. The function value at the interpolation point $x$ is given by: 
\begin{eqnarray}
    \mathrm{RM_{gal}}(x) = \begin{cases}\frac{\sum_{i=1}^N w_i \mathrm{RM_{gal}}_i}{\sum_{i=1}^N w_i} \ &: ||x_i -x|| \neq 0\\
     \mathrm{RM_{gal}}_i \ &: ||x_i-x|| = 0 ,
    \end{cases}
\end{eqnarray}
\label{eq4}
where $\mathrm{RM_{gal}}_i$ are the values of the function at $x_i$. We implemented this algorithm using \pkg{NumPy V1.21.5} \citep{harris2020array} . 

\subsection{Natural Neighbour Interpolation (NNI)}
\begin{figure*}
    \centering
    \gridline{\fig{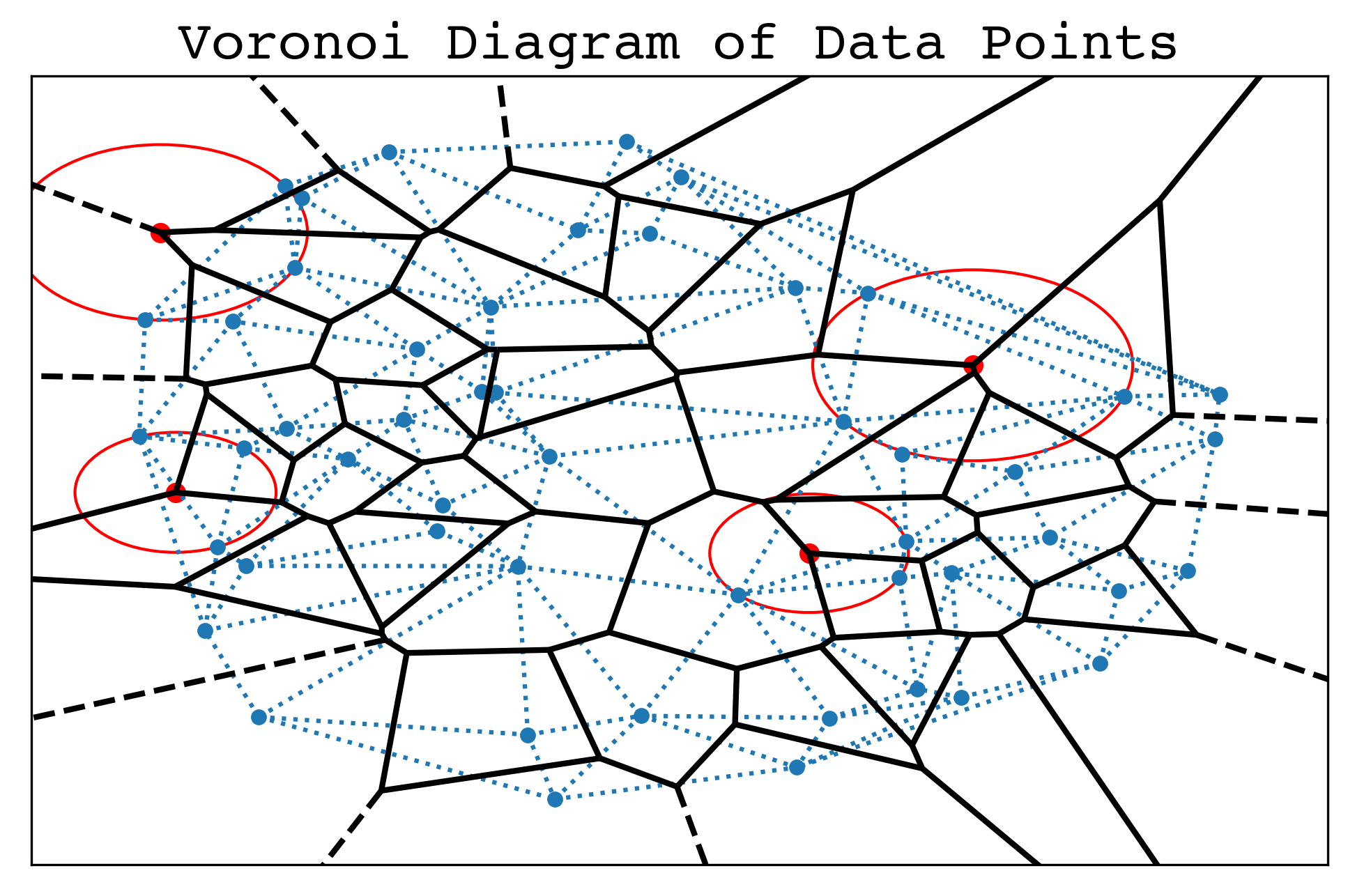}{0.48\textwidth}{} \fig{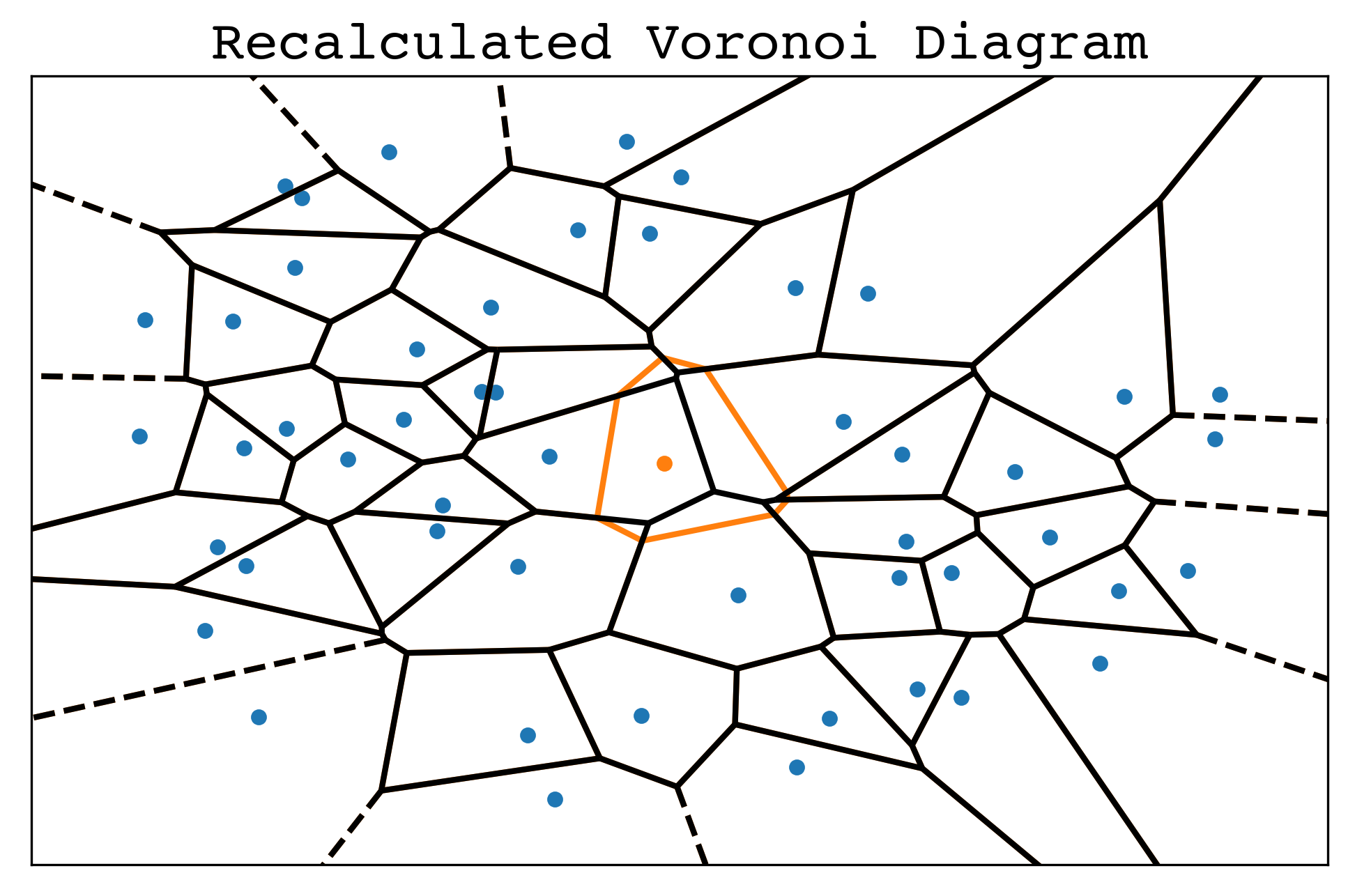}{0.48\textwidth}{}}
    \caption{The plot on the left shows the data points (in blue). The Delaunay triangulation is shown in dotted blue lines, with a few selected circumcircles portrayed in red. The black lines show the edges of the Voronoi polygons. The plot on the right shows the recalculated Voronoi diagram when an interpolation point (in orange) is inserted into the figure. The orange lines show the edges of the Voronoi polygon of the interpolation point.} 
    \label{fig:1}
\end{figure*}

NNI \citep{sibson} is similar to inverse distance weighting in that it aims to calculate the function value at an interpolation point using a weighted average. However, the method for calculating the weights relies on areas, instead of distances. \\
For every data point, we can consider a polygon that contains points that are closer to this data point than they are to any other data point. Creating such polygons for every data point, we can arrive at a \textit{Voronoi grid}, as shown in the left panel of Figure \ref{fig:1}. To calculate the function value at the interpolation point, we simply insert the interpolation point into the grid and re-calculate the Voronoi diagram, as shown in the right panel of Figure \ref{fig:1}. Here the weights are given by: 
\begin{eqnarray}
    w_i = A_i,
    \label{eq:6}
\end{eqnarray}
where $A_i$ is the area that the new polygon (corresponding to the interpolation point) captures from the $i$-th Voronoi polygon around the $i$-th data point. The function value is then given by: 
\begin{eqnarray}
    \mathrm{RM_{gal}}(x) =  \frac{\sum_{i=1}^N  w_i \mathrm{RM_{gal}}_i}{\sum_{i = 1}^N w_i},
\end{eqnarray}
where $\mathrm{RM_{gal}}_i$ is the function value at the data point $x_i$.

Producing a Voronoi diagram directly from a grid of data points is difficult because the distances from each interpolation point to every data point must be calculated. Once these distances have been obtained, the data point which the interpolation point is closest to must then be found--the interpolation point lies in the Voronoi polygon of this data point. This process must then be repeated for each of the interpolation points to form the Voronoi diagram, so the computation time increases considerably with each additional data point or interpolation point.
An effective and far quicker method to produce the Voronoi diagram is to first find the \textit{Delaunay triangulation} \citep{Liang2010ASA}. To explore the Delaunay triangulation, it is first necessary to introduce some notation. 

Following \citet{Liang2010ASA}, we define the \textit{circumcircle} to be the unique circle that passes through a set of three vertices. Now consider any triangulation of the grid of data points. The Delaunay criterion requires that the circumcircle around each triplet of vertices does not contain any other vertices aside from these three. The unique triangulation that satisfies the Delaunay criterion is the Delaunay triangulation of a grid. 

For each circumcircle, we can now define its center to be the \textit{circumcenter}. Then, the grid of circumcenters for a Delaunay triangulation forms the vertices of the Voronoi diagram. 

To implement the NNI algorithm, we have used \pkg{MetPy V1.4.0} \citep{metpy} (which uses the implementation of \citet{Liang2010ASA}), in particular, the \pkg{metpy.interpolate} module. \\
\subsection{Radial Basis Function (RBF) Interpolation }
We consider the true RM to be a function lying in a function space, and that it can be approximated as some linear combination of the basis functions of this space. Here, we assume that radially symmetric  functions from the data points, $\phi(||x - x_i||)$ form a reasonably accurate basis for this space \citep{buhmann_2000}. Then, the function value can be computed to be: 
\begin{eqnarray}
    \mathrm{RM_{gal}}(x)  = \sum_{i=1}^N w_i \phi(||x-x_i||),
\end{eqnarray}
where the weights $w_i$ are found by solving the following linear system: 
\begin{eqnarray}
    \begin{bmatrix} 
    \phi(||x_1 - x_1||) & \cdots & \phi(||x_1 -x_n ||)\\
    \vdots  & \vdots & \vdots \\
    \phi(||x_n- x_1||) & \cdots  & \phi(||x_n -x_n||)
    \end{bmatrix} \begin{bmatrix}
        w_1 \\
        \vdots \\
        w_n
    \end{bmatrix} = \begin{bmatrix}
        \mathrm{RM_{gal}}_1\\
        \vdots \\
        \mathrm{RM_{gal}}_n
    \end{bmatrix}.
\end{eqnarray}
We consider two kinds of RBFs: the first, where the RBFs are inverse multiquadric (IM), and the other where they are based on thin plate splines (TPS). In both cases, the algorithm was implemented using \pkg{SciPy V1.21.5} \citep{2020SciPy-NMeth}, particularly the \pkg{scipy.interpolate} module. 
\subsubsection{Inverse Multiquadric (IM)}
Here we consider RBFs that are given by: 
\begin{eqnarray}
    \phi(r_i) = \frac{1}{\sqrt{1+ (\epsilon r_i)^2}},
\end{eqnarray}
where $\epsilon \in \mathbb{R} $ is the shape parameter (in deg$^{-1}$) that determines the shape of the interpolation function, and $r_i$ (in deg) is the angular distance between the interpolation point and the data point $x_i$. Lower values of $\epsilon$ make the function flatter, and higher values give it a greater peak, with the function becoming constant as $\epsilon \to 0 $; as $\epsilon \to \infty$ the RM map approaches a map that is zero everywhere except at the data points $x_i$, where it takes on the RM value of the data point RM$_{\rm{gal}_i}$ \citep{mongillo2011choosing}. Qualitatively, lower values of $\epsilon$ resulted in smoothing of small scale ($\sim 1$ deg) structures, while higher values led to sharp small scale peaks in the reconstruction. After some testing, we decided to use $\epsilon = 0.1$, as lower or higher values cause us to lose small scale and large scale ($\sim 10^{-1}$ deg) structures, respectively, leading to large discrepancies from the simulated RM map.

\subsubsection{Thin-Plate Splines (TPS)}
TPS Interpolation \citep{eberly_1996} uses piecewise polynomials to fit the data. The piecewise function is chosen such that the function value at the position of the data points corresponds to the $z$ coordinate of the data points. For determining the optimal interpolation function, we minimize the bending energy,
\begin{eqnarray}
    E(\mathrm{RM_{gal}})  = \int_{\mathbb{R}^3} |D^2 \mathrm{RM_{gal}} |^2 dV. 
    \label{eq:11}
\end{eqnarray} 
Here, $D$ is the derivative operator in $\mathbb{R}^2$ and $dV$ is the infinitesimal volume element in  $\mathbb{R}^3$. With a little computation, one can express $\mathrm{RM_{gal}}$ as: 
\begin{eqnarray}
    \mathrm{RM_{gal}}(x) = \sum_{i=1}^N a_i G(x, x_i) + b_0 + \sum_{j=1}^2 b_j y_j, 
\end{eqnarray}
where $a_i, b_j$ are coefficients to be estimated,  \\$G(x, x_i) = - \frac{1}{8\pi^2} ||x-x_i||^2 \ln(||x-x_i||)$, and $y_j$ are the components of the interpolation point $x$. \\
To find the coefficients, we can solve the following linear system: 
\begin{eqnarray}
    \left(\begin{bmatrix}
        G(x_1, x_1)& \cdots& G(x_1,x_N)\\
        \vdots & \vdots & \vdots\\ 
        G(x_1, x_N) & \cdots & G(x_N, x_N)
    \end{bmatrix} + \psi I \right) \begin{bmatrix}
        a_1 \\
        \vdots \\
        a_N
    \end{bmatrix}\nonumber \\+ \begin{bmatrix}
        1 & x_{11} & x_{12}\\
        \vdots & \vdots & \vdots\\
        1 & x_{N1}  & x_{N2}
    \end{bmatrix} \begin{bmatrix}
        b_0 \\
        b_1 \\
        b_2
    \end{bmatrix} =
    \begin{bmatrix}
        f_1 \\
        \vdots \\
        f_N
    \end{bmatrix},
\end{eqnarray}
where $\psi $ is a smoothing parameter that is decided upon beforehand, $I_\mathrm{res}$ is the $N \times N$ identity matrix, and $x_{ij}$ is the $j$-th component of the $i$-th data point. To completely determine all of the coefficients, we need the additional orthogonality condition: 
\begin{eqnarray}
    \begin{bmatrix}
        1 & \cdots & 1 \\
        x_{11} & \cdots & x_{N1}\\
        x_{12} & \cdots & x_{N2}
    \end{bmatrix} \begin{bmatrix}
        a_1 \\
        \vdots \\
        a_N
    \end{bmatrix} = 0.
\end{eqnarray}

\subsection{Bayesian RM Sky (BRMS)}
 \citet{Hutschenreuter_2022} have developed a Bayesian algorithm to infer the all sky RM map from a catalog of extragalactic RM sources \citep{2023VanEck}.
We refer to this algorithm as the Bayesian RM Sky (BRMS) method in this work.
This method is an upgrade to previous versions \citep{Oppermann_2012, refId0, 2019Hutschenreuter} and is based on Information Field Theory (IFT). 
IFT is a general Bayesian statistical framework for signal processing and image reconstruction, often used to design inference algorithms in high dimensional parameter spaces and constrained by noisy data sets \citep{2019Ensslin}. The fundamental difference between BRMS and the other algorithms described is that BRMS fits the parameters of the interpolation kernel, while those for the others are set manually. 

The RM sky model specifically is motivated by the functional form of Equation \ref{eq:rm} and the fact that the RM can be interpreted as the product of the DM and the electron weighted LOS component of the magnetic field.
The RM sky is hence modelled as the product of two sky maps 
\begin{eqnarray}
    \mathrm{RM_{gal}} = e^\rho \chi, 
\end{eqnarray}\\
where both $ \rho$ and $\chi$ are Gaussian processes. 
The log-normal field $e^\rho$ covers possible amplitude variations of the RM sky over several magnitudes (mostly, but not completely caused by the DM), while $\chi$ mainly models the sign variations caused by the averaged magnetic field.   
The correlation structure of these Gaussian processes is unknown and inferred as well, using the model of \citet{2022Arras}.
The BRMS model of \citet{Hutschenreuter_2022} has an additional component to effectively down-weight data points that are inconsistent with the rest of the sky, thereby implicitly reducing the effect of non-Galactic RM components. 
We have turned off this feature in this work, and instead rely on the explicit outlier removal outlined in Section \ref{sec:data} in order to make the various interpolation methods more comparable. In addition, BRMS takes the observational noise of each data point into account when computing the reconstructed map. This feature is not relevant for our work specifically, as we have assumed a constant standard deviation for the RM error across our simulated data. However, it may be an additional benefit of BRMS when applied to real data that can have, for example, errors that depend on the region of the sky being probed or the signal-to-noise ratios of the data points.

The posterior distribution resulting from the setup outlined above is very high dimensional and non-Gaussian. 
This setup is hence evaluated using variational inference schemes \citep{2019Knollmuller, 2021Frank}, which approximate this posterior with simpler, tractable probability distributions. Through this process, BRMS produces statistical uncertainties for the reconstructed RM map, which are not output by the other algorithms being tested.  
The BRMS algorithm is implemented with the python package \pkg{Nifty V8.4} \citep{2019Nifty}.

\begin{table*}[!htb]
\centering
\textbf{Interpolation Techniques Used}
    \begin{tabular}{|c|c|l|}
    \hline
        \textbf{Name} & \textbf{Acronym} & \multicolumn{1}{c|}{\textbf{Description}}\\
    \hline
    \hline
     Inverse & & Weights the effect that the RM of the data points have on the interpolation\\
      Distance  & IDW & point by the inverse of the distance between the points. The power parameter\\
     Weighting& & $p$ controls how strongly the weight falls of as a function of distance. \\
     \hline 
     Natural& & Weights the effect that the RM of the data points have on the interpolation\\
      Neighbour  & NNI  & point by the area that the Voronoi polygon of the interpolation  \\
     Interpolation & & captures from the Voronoi polygon of the data points.\\
     \hline
     & & Uses radial basis functions to reconstruct the foreground RM. The shape\\
    Inverse Multiquadric & IM & parameter $\epsilon$ determines how sharply peaked the RM around the data points\\ && is in the reconstructed map. \\
    \hline
    Thin Plate Splines & TPS & Uses radial basis functions that minimize the bending energy (Eq \ref{eq:11}).\\
    \hline
    Bayesian Rotation& \multirow{2}*{BRMS}& Uses Information Field Theory and fits the data by modelling the amplitude\\ Measure Sky && and the sign variations of the RM as Gaussian processes.\\
    \hline
    \end{tabular}
\caption{The interpolation techniques that will be tested in this work.}
    \label{tab:summary}
\end{table*}

\section{Simulated Data}
\label{sec:data}
To test the accuracy of the above methods, we apply the techniques to simulated data. We have generated the data using simulated cubes from \citet{2021PhRvF...6j3701S} (which attempt to simulate the ISM in a volume), where they have simulated data for the supersonic turbulent plasma of the ISM on a grid of $512^3$ data points. The turbulent flow is driven solenoidally (divergence-free). In observations, we expect a combination of compresssive and solenoidal flow, with most of the large scale structure being driven by compressive flow (due to shocks from supervnovae that are pure divergence flows); however, the smaller scale structure can demonstrate solenoidal driven properties \citep{2010A&A...512A..81F}. The reason for choosing solenoidal driving is because it is more efficient in driving a dynamo \citep{2011PhRvL.107k4504F}. \citet{2021PhRvF...6j3701S} define the Mach number of the simulated ISM as follows: 
\begin{align}
    \mathrm{Mach} = \frac{u_{\rm{rms}}}{c_s},
\end{align}
where $u_{\rm{rms}}$ is the root mean squared turbulent velocity in the plasma, and $c_s$ is the sound speed. In their simulation, high Mach numbers result in filamentary ISM structures and low Mach numbers correspond to patchy RM structures.

The warm ISM typically has low Mach numbers of $\sim 1 $ \citep{Ferrière_2020}. However, it must be noted that in the ISM, a low Mach number does not necessarily imply RM structures to be non-filamentary; filaments can be formed by supernovae \citep{1975MNRAS.171..263G}, stellar winds \citep{1996ApJ...470L..49R}, and other processes external to the plasma of the ISM. 

To test the techniques in different foreground RM structures, we have simulated two data sets, one of which contains patchy structures and the other of which contains filamentary structures. The purpose of this was to test the interpolation techniques on distinct underlying priors on the small-scale RM structures. These two data sets demonstrate two extreme cases of spatial correlation structures, and challenge the interpolation algorithms. Filaments are present in the ISM on all scales \citep{hacar}; we hence view them as a good geometric structure to challenge the interpolation techniques. 


The simulated foreground RM is produced by `stacking' simulated cubes. By `stacking' the simulated cubes we attempt to simulate column depth and generate more structure in the foreground. It must be noted that by stacking we produce an ISM column, while in reality when observing we would see an ISM cone (therefore the RM is dominated by foreground sources), but this distinction is not important for testing interpolation kernels. The simulated thermal electron density follows the same density profile as the gas density (because the simulations are isothermal) and has a lognormal power spectrum. In addition to this, the magnetic field is assumed to be a Gaussian random field. Then, the RM is calculated using Equation \ref{eq:rm}, assuming that there is a uniform distribution of polarized background radio sources. The path length $d$, is taken to be the length of all the stacked cubes.

\begin{figure*}
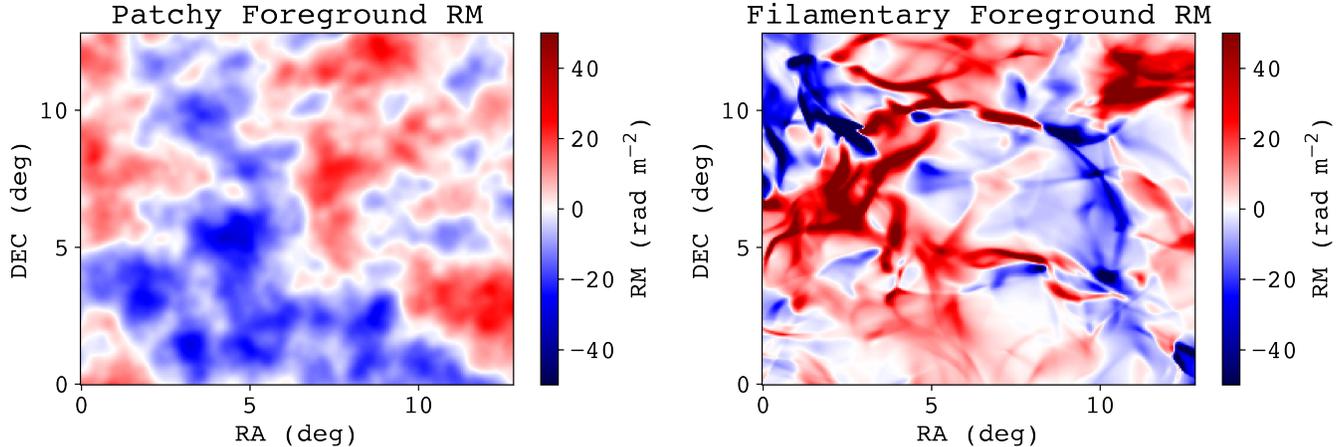

    \gridline{\fig{Patchy_sky.png}{0.49\textwidth}{} \fig{Fil_sky.png}{0.49\textwidth}{}}
    \caption{Simulated foreground RM maps for patchy (left) and filamentary (right) RM structures. The color bars represent RM in units of rad m$^{-2}$ and are saturated from $-50$ to $+50$ rad m$^{-2}$.}
    \label{fig:3}
\end{figure*}

Simulated cubes with a low Mach number have patchy RM structures while, higher Mach numbers mean the cubes will be more turbulent, giving rise to filamentary RM structures. In this work, we stack two simulated cubes and use a Mach number of 0.1 for patchy RM structures (this corresponds to $\mathcal{P}(n_e) \propto k^{-1.71}$ and $\mathcal{P}(B) \propto k^{-0.81}$) and 10 for filamentary RM structures (this corresponds to $\mathcal{P}(n_e) \propto k^{-0.92}$ and $\mathcal{P}(B) \propto k^{-0.77}$). This results in RM maps with power spectra $\mathcal{P(\rm{RM}}) \propto k^{-1.68}$ and $\mathcal{P}(\mathrm{RM}) \propto k^{-1.58}$, where we have estimated the power indices of the RM maps using Equation 15 of \citet{10.1093/mnras/stac2972}. We restrict ourselves to a patch of the sky that has an area $\sim$144 deg$^2$. The two simulated RM maps that will be considered are shown in Figure \ref{fig:3}.

The simulated cubes that are derived by \citet{2021PhRvF...6j3701S} do not include extragalactic RM. Because of this, we added extragalactic RM to the simulated RM foregrounds. The extragalactic RM was modelled to be Gaussian with a mean of 0 rad m$^{-2}$ and standard deviation of 7 rad m$^{-2}$ \citep[following][]{refId0}. In addition to this, simulated observational noise was added with a mean of 0 rad m$^{-2}$ and a standard deviation of 1.5 rad m$^{-2}$, which is the expected noise from POSSUM low-band data \citep{2024AJ....167..226V}. In addition to this, we've tested varying the standard deviation in the noise and the results obtained are described in Appendix \ref{app:c}. We note here that the assumption of homogeneous Gaussian noise in our model is a simplification and observed RM data of background radio galaxies likely have more complex noise statistics.

Once this was done, data points were randomly sampled throughout the interpolation region (which is $\sim 144$ deg$^{2}$) for both patches with a total of 6459 and 6556 data points for the  patchy and filamentary RM skies, respectively. This yields a density of $\sim 45$ points deg$^{-2}$, which is similar to POSSUM Pilot 1 data \citep{2024AJ....167..226V}. 

\begin{figure*}
    \gridline{\includegraphics[width = 0.49\textwidth]{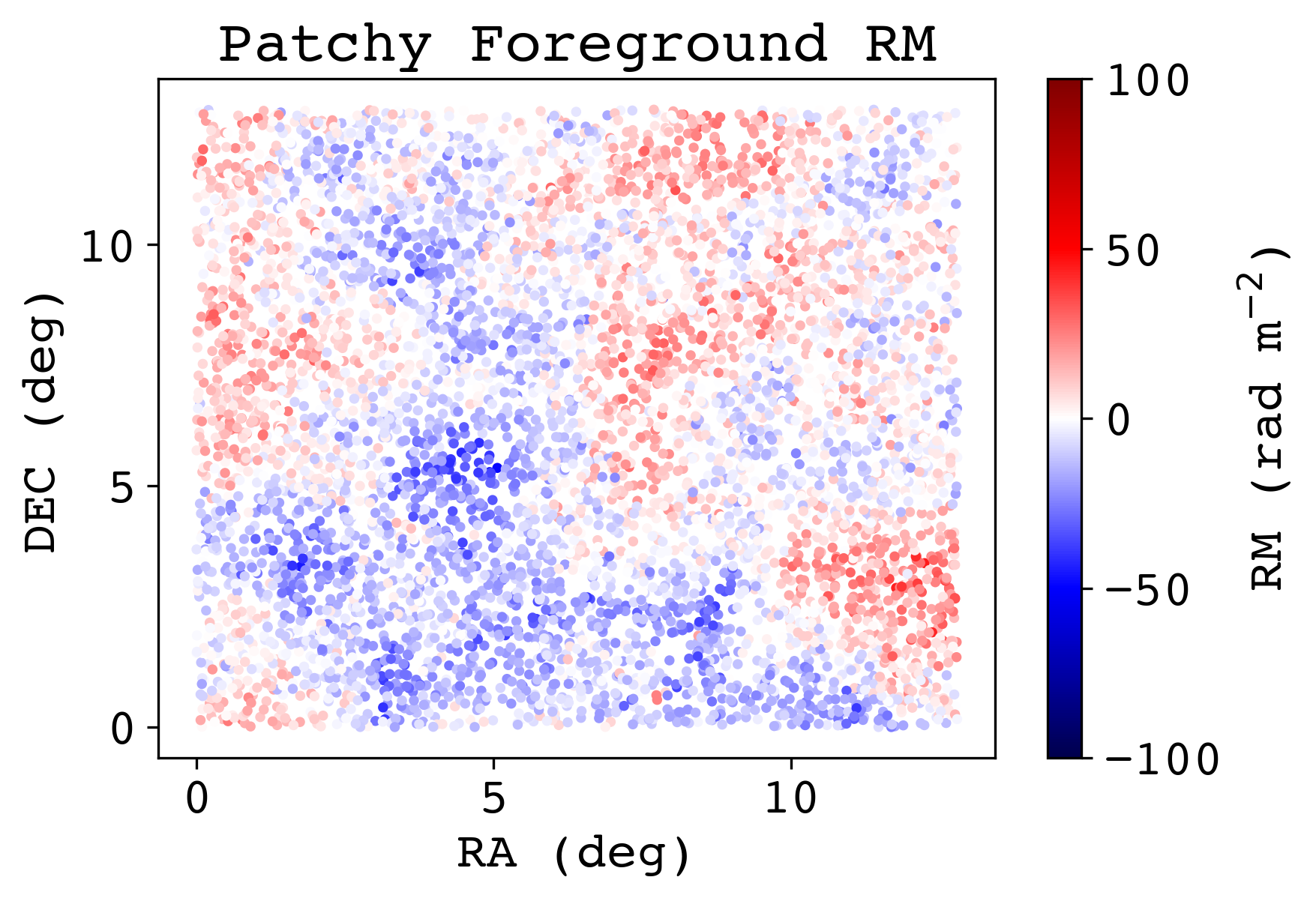} \includegraphics[width = 0.49 \textwidth]{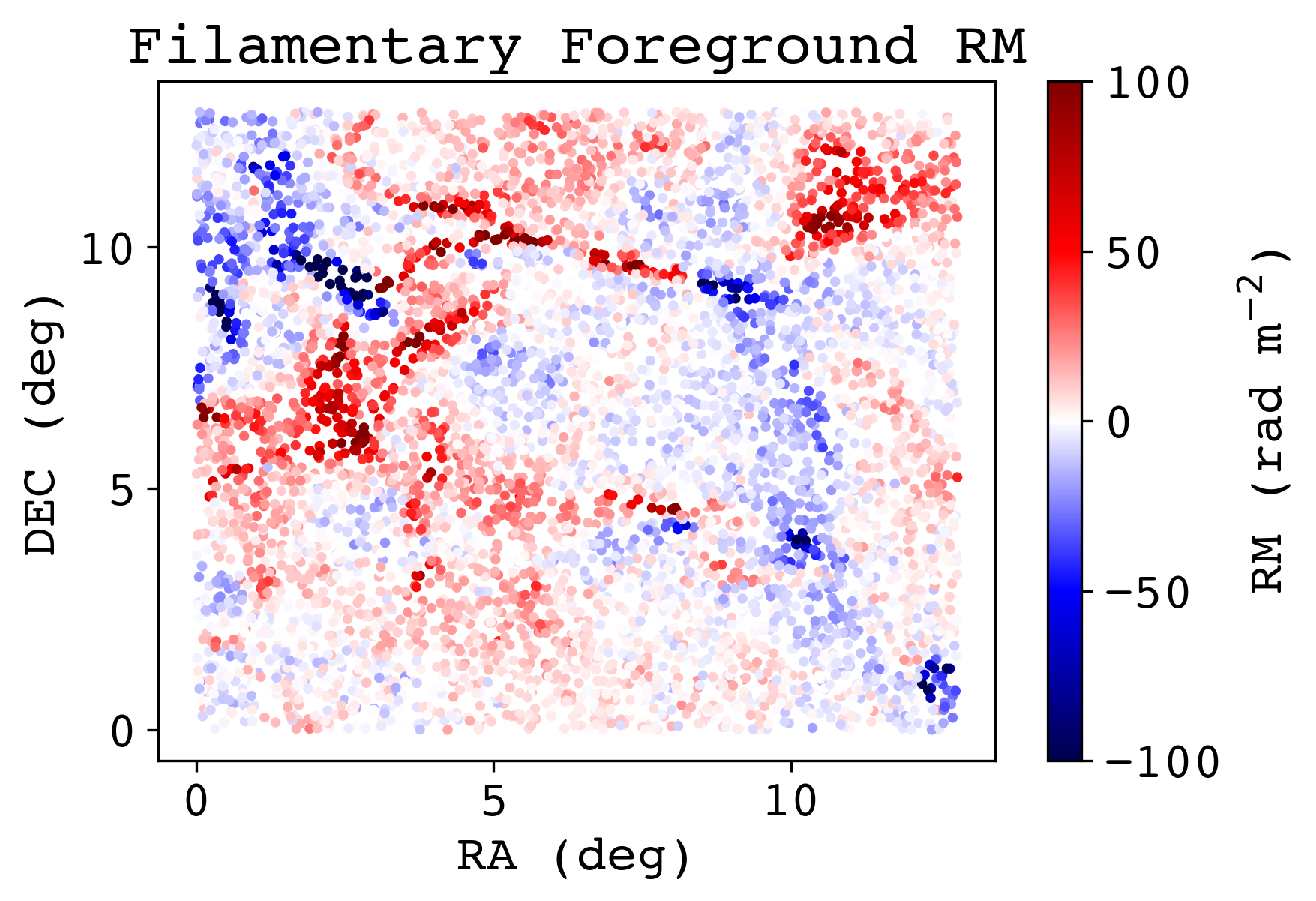}}
    \caption{The sampling of the RM data points from the patchy (left) and filamentary (right) foreground RM, saturated from $-100$ to $+100$ rad m$^{-2}$ after the $3\sigma$ cut has been made.}
    \label{fig:sample}
\end{figure*}

To accurately generate a smooth foreground RM, this extragalactic RM must first be removed from the data points. To correct for this, we place the following requirement on each data point: its RM value must be within $3\sigma $ of the mean RM of its ten nearest data points. While this threshold can be changed, the main motivation behind this choice was that a smaller coefficient for $\sigma$ removed the small scale variability in RM, and a higher value led to too many sources dominated by extragalactic RM being included in our analysis. Figure \ref{fig:sample} indicates the sampling of the data points in the grid, after the $3\sigma$ cut has been made. Initially, the patchy simulated foreground and filamentary simulated foreground had 6509 and 6586 points, respectively. After the $3\sigma$ cut, the number of data points were reduced to 6310 and 6436 in the patchy and filamentary foregrounds, respectively. It appears that the algorithm has removed far more data points in the patchy case than in the filamentary case. 

To explain this, we have to remind ourselves that when applying interpolation algorithms to real data, we do not know the underlying RM structures that are present. Because of this, we cannot objectively decide the best interpolation parameters (such as the one above for removing extragalactic RM). In our tests, we treat all parameters in the interpolation algorithms as priors that we enforce on the RM structures, and check how these different priors affect the reconstructions in different underlying simulated RM structures (filamentary and patchy RM structures). 

We also computed the second-order RM structure function (SF), using a 2 point stencil (for details see \citet{10.1093/mnras/stac2972}), defined as: 
\begin{align}
    \mathrm{SF}_{\mathrm{RM_{\rm{obs}}}} (\Delta \theta)= \langle |\mathrm{RM}(\theta) - \mathrm{RM} (\theta + \Delta \theta)|^2 \rangle,  
\end{align}
where $\Delta \theta$ is the angular separation, $\rm{RM}_{\rm{obs}}$ is the observed RM and the mean is taken over all positions. Following the procedure outlined in Appendix A of \citet{2004ApJ...609..776H}, we can correct for the noise in the RM to obtain: 
\begin{align}
    SF_{\rm{RM}}(\Delta\theta) = SF_{\rm{RM}_{obs}}(\Delta \theta) - SF_{\delta\mathrm{RM}}(\Delta \theta) ,
\end{align}
where $\delta{RM}$ is the error in the measured RMs.  

\begin{figure*}
    \centering
    \gridline{\includegraphics[width = 0.49\textwidth]{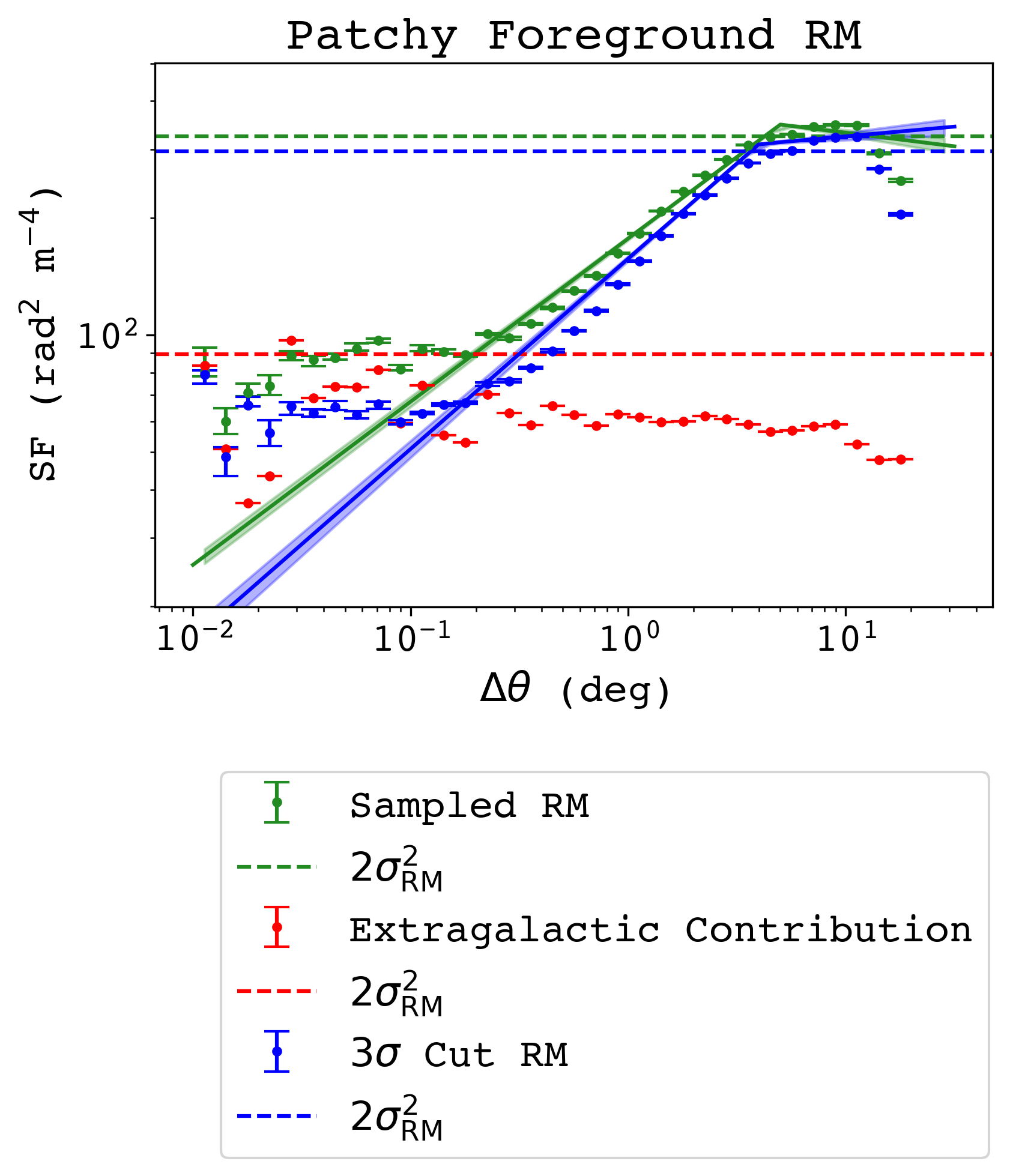} \includegraphics[width = 0.49\textwidth]{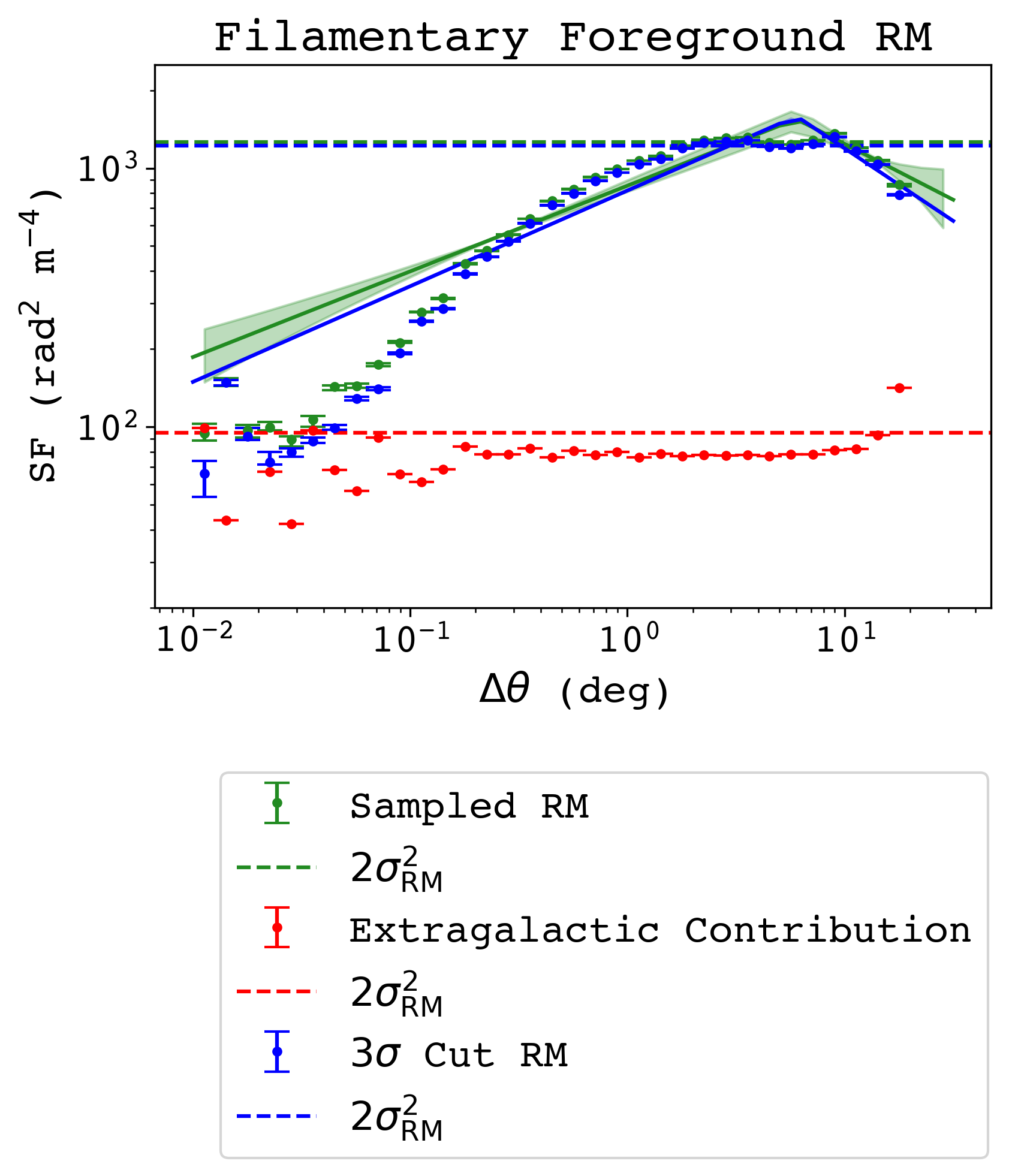}}
    \caption{The second order, 2 point stencil structure functions for the patchy (left) and filamentary (right) foreground RM. The green points indicate the  SF for the initial sampled data (with extragalactic contribution), the red points represent the SF from the Gaussian extragalactic contribution that was added, and the blue points represent the SF after the $3\sigma$ cut was made on the data. The dashed lines represent the $2\sigma_{\rm{RM}}^2$ asymptotes for each of the RMs considered. In the patchy foreground, the sampled RM has best fits of $A = 348.15^{+1.51}_{-2.60}$ rad$^2$ m$^{-4}$, $\Delta \theta_{\rm{break}} = 5.00^{+0.07}_{-0.07}$ deg, $\alpha_1 = 0.42^{+0.00}_{-0.00}, \alpha_2  = -0.07^{+0.01}_{-0.01}$ and $A = 295.07^{+1.02}_{-4.63}$ rad$^2$ m$^{-4}$, $\Delta \theta_{\rm{break}} = 4.00^{+0.05}_{-0.04}$ deg, $\alpha_1 = 0.49^{+0.00}_{-0.01}, \alpha_2 = 0.05^{+0.01}_{-0.01}$ for the $3\sigma$ cut RM. With the filamentary foreground, the sampled RM had  best fits of $A = 1549.08^{+1.50}_{-5.68}$ rad$^2$ m$^{-4}$, $\Delta \theta_{\rm{break}} = 6.00^{+0.05}_{-0.00}$ deg, $\alpha_1 = 0.33^{+0.00}_{-0.01}, \alpha_2 = -0.43^{+0.01}_{-0.00}$ for the sampled RM and $A = 1589.34^{+1.92}_{-2.59}$ rad$^2$ m$^{-4}$, $\Delta \theta_{\rm{break}} = 6.00^{+0.00}_{-0.02}$ deg, $0.37^{+0.00}_{-0.00}, \alpha_2 = -0.56^{+0.01}_{-0.01}$ for the $3\sigma$ cut RM.  }
    \label{fig:structure}
\end{figure*}

The SFs were computed using the python package \pkg{structurefunction V1.1.3} \citep{structure}, in bins of 0.1 deg. Once the SFs were computed, we fitted the ones for the sampled data and the $3\sigma$ cut data with broken power laws given by Equation \ref{eq:broken}. The results are displayed in Figure \ref{fig:structure}and the $2\sigma_{\rm{RM}}^2$ are reported in Table \ref{tab:2sigma}. Something to be noted is that the SF does not saturate at the $2\sigma_{\rm{RM}}^2$ value for the extragalactic RMs in either case which is because the extragalactic RMs were modelled as a multivariate Gaussian field.

\begin{equation}
    \mathrm{SF}(\Delta \theta) = \begin{cases}
        A (\Delta \theta)^{\alpha_1}: \Delta \theta < \Delta \theta_{\rm{break}}\\
         A (\Delta \theta)^{\alpha_2}: \Delta \theta < \Delta \theta_{\rm{break}}
    \end{cases}
    \label{eq:broken}
\end{equation}

\begin{table}[!htb]
    \centering
    \begin{tabular}{|c|c|c|}
    \hline
        \textbf{RM} & \textbf{Patchy} & \textbf{Filamentary} \\
        \hline
        \hline
        \textbf{Sampled} & 324.86 & 1263.84\\
        \textbf{Extragalactic} &89.32 & 94.75\\
        \textbf{3}$\mathbf{\sigma}$ \textbf{cut} & 296.94 & 1224.78 \\
        \hline
    \end{tabular}
    \caption{The 2$\sigma_{\rm{RM}}^2$ in rad$^2$ m$^{-4}$ for the patchy and filamentary simulated  foregrounds; in particular, for the sampled, extragalactic and $3\sigma$ cut RMs.}
    \label{tab:2sigma}
\end{table}

The structure function for the sampled data portrays how sharp the change in RM is. In the patchy foreground, we see that the structure function is convex, indicating that the RM increases gradually over spatial scales and has a break point at around $4.00^{+0.05}_{-0.04}$ deg. And for the filamentary foreground, the structure function appears to be concave, and asymptotes much quicker than that for the patchy foreground (at $\sim 1$ deg compared to the  deg for the patchy foreground). This indicates that over angular scales of $\sim 1$ deg there is a lot of variation in the RM for the filamentary foreground, which we see from Figure \ref{fig:3}. However, what must be noted is that the fitted model for the filamentary foreground prefers a break point at $6.00^{+0.00}_{-0.02}$ deg; this is likely because the fit is skewed by the increase in structure beyond the asymptote.

Additionally, there seems to be a rather large difference between the structure functions of the data and the $3\sigma$ cut data in the patchy case when compared to the filamentary case. This indicates that the extragalactic removal algorithm is better at identifying extragalactic contribution in the patchy case; a possible reason for this is that the sharp changes in the RM in the filamentary foreground make it difficult for the algorithm to differentiate between extragalactic RM and the galactic foreground RM. For this reason, we will also consider the case where there is no extragalactic RM contribution and check how this changes our results in Section \ref{sec:noegal}. 

\section{Results}
\label{sec:results}

In this section, we present the reconstructed RM maps that were obtained by implementing the techniques described in Section \ref{sec:methods} on the simulated data described in Section \ref{sec:data}. 

All of the interpolation algorithms were performed on a 2021 M1 MacBook Pro with 16 GB of unified memory, 8 core CPU, and 16 core GPU. All the reconstructed RM maps have an angular resolution of $20$ px/deg and a pixel resolution of $256\times 256$. The code used in this work is available on GitHub\footnote{ \rule[12pt]{0.4\textwidth}{0.2pt}\newline
\url{https://github.com/AffanKhadir/Interpol_Schemes}}.

\subsection{Patchy Foreground}

\begin{figure*}
\gridline{\fig{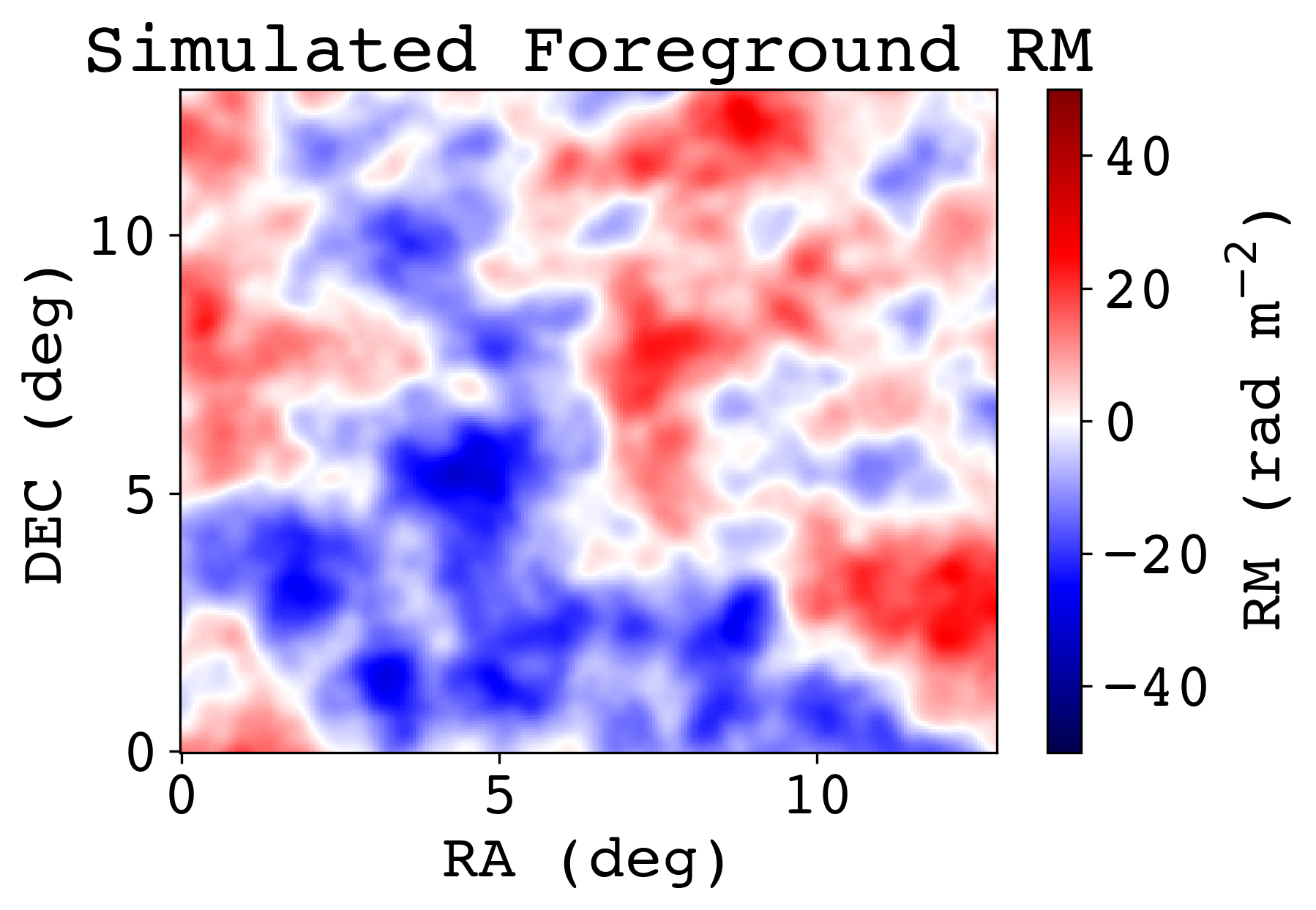}{0.33\textwidth}{}}
\gridline{\fig{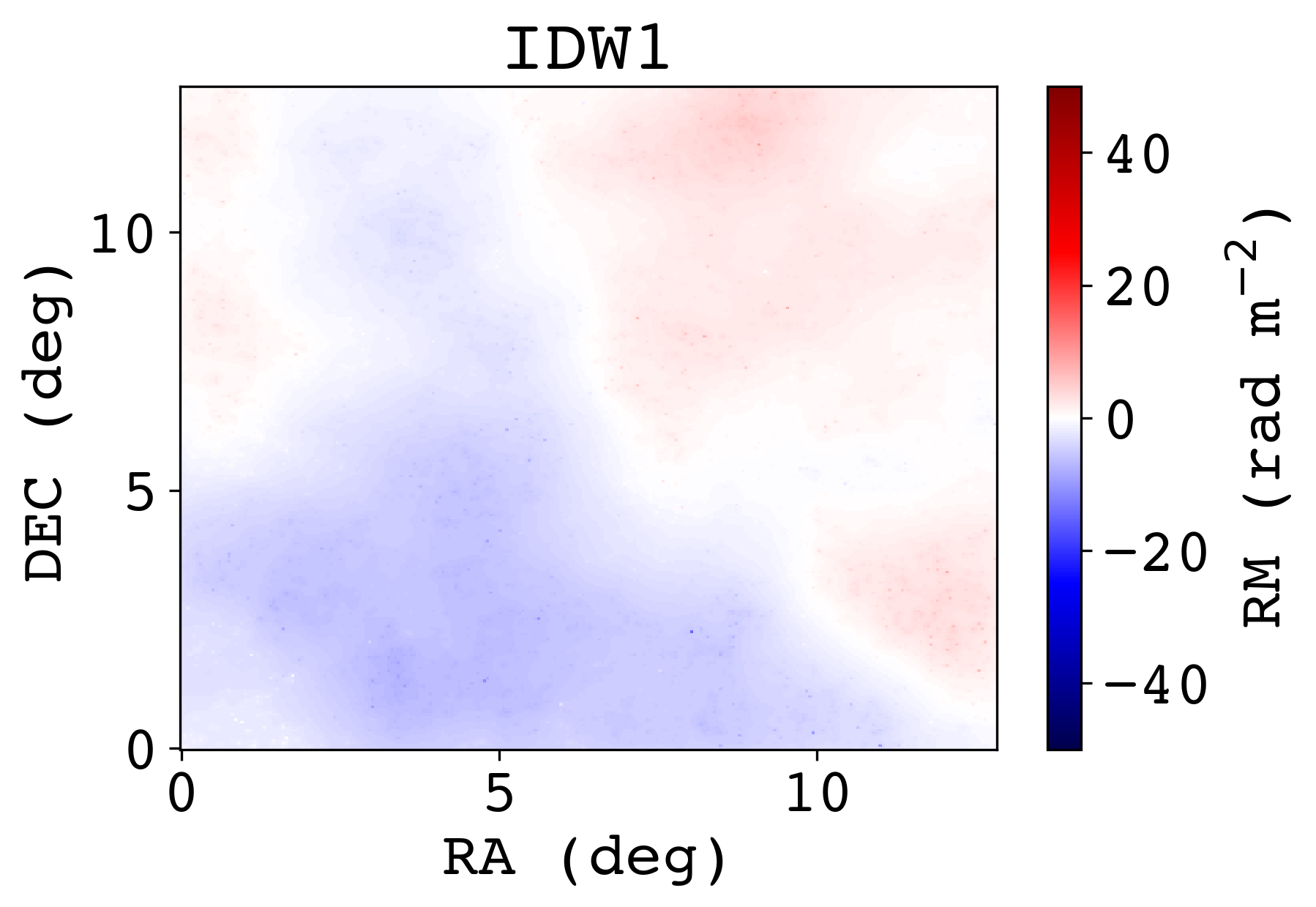}{0.33\textwidth}{}\fig{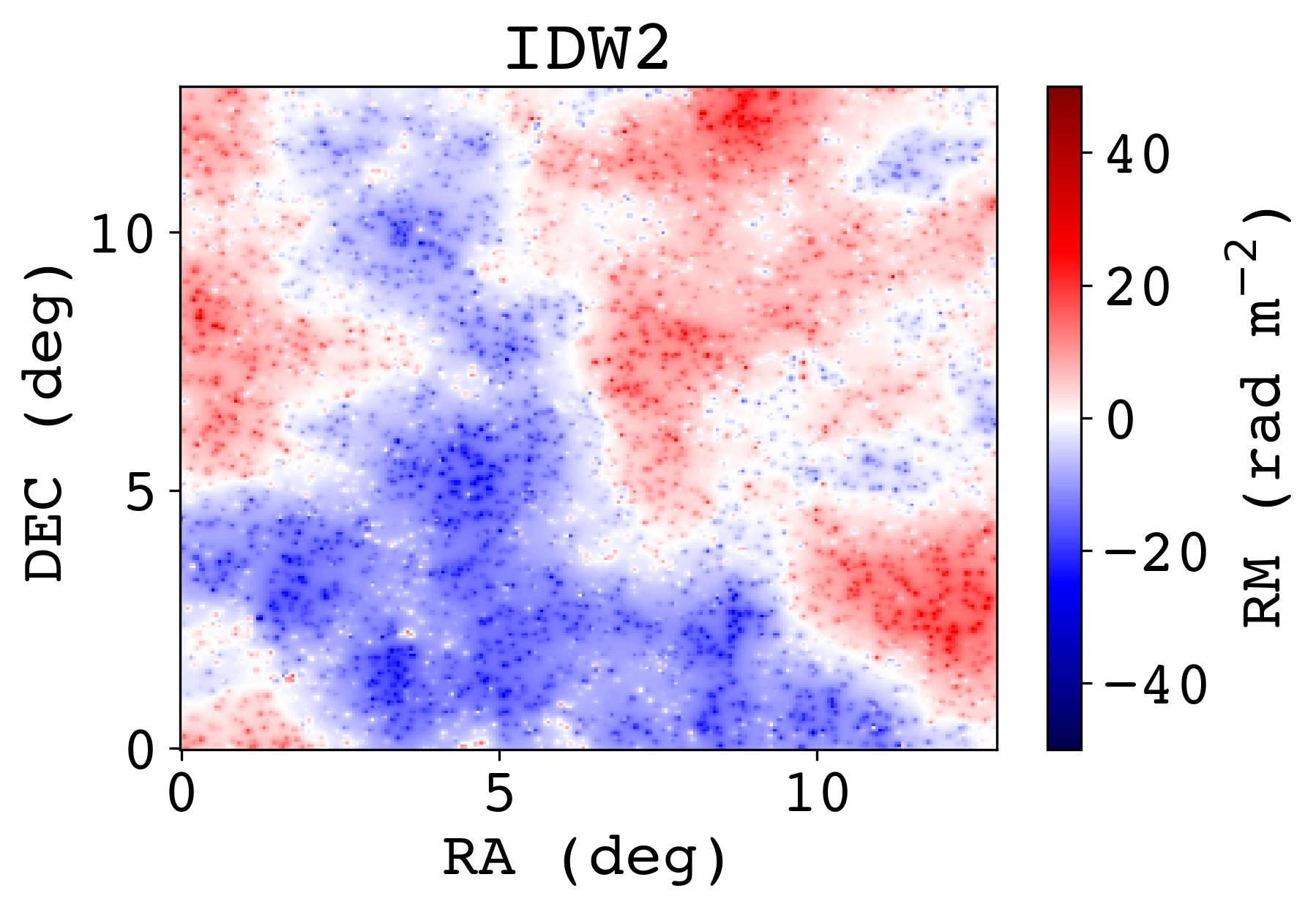}{0.33\textwidth}{}\fig{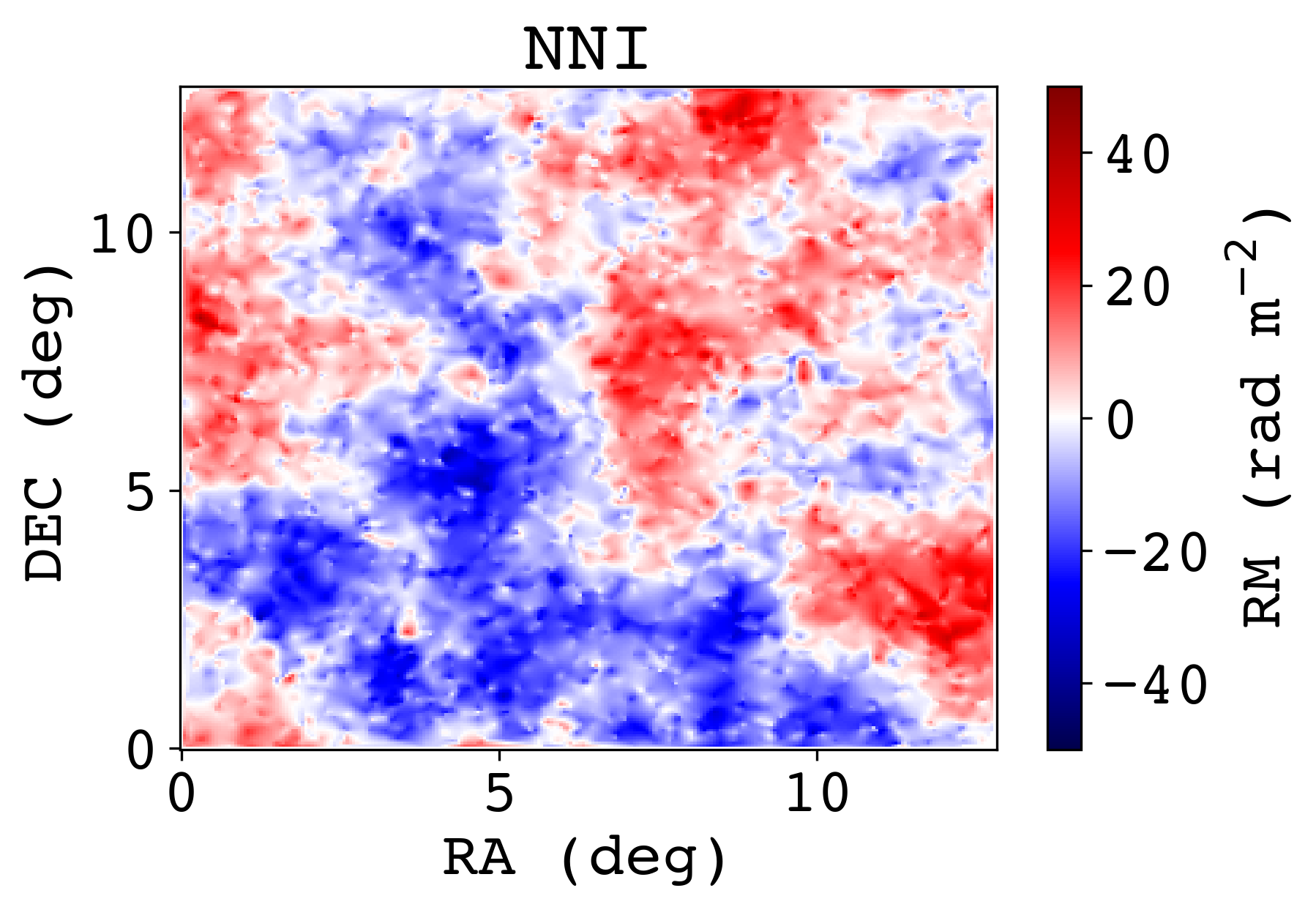}{0.33\textwidth}{}}
\gridline{\fig{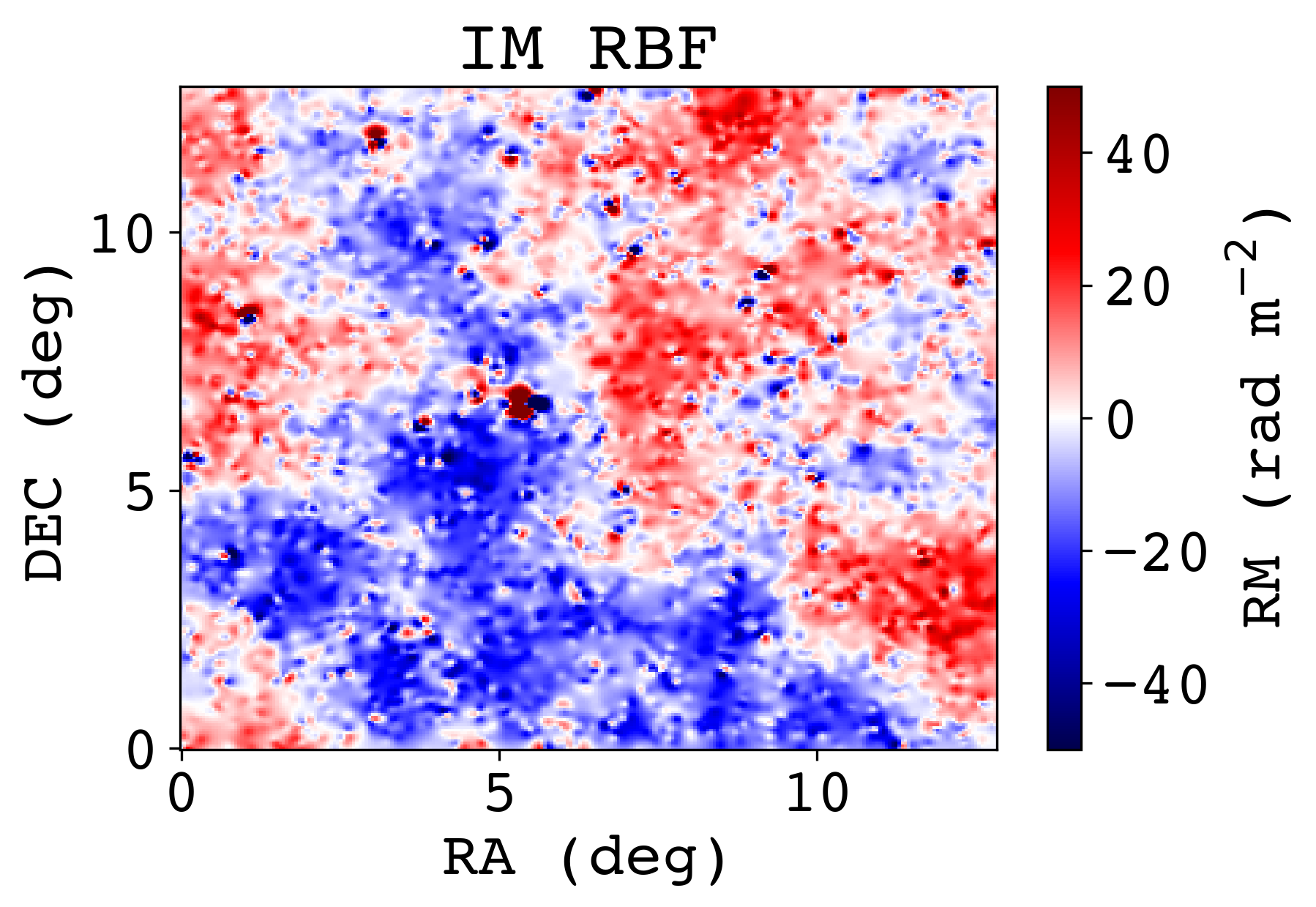}{0.33\textwidth}{}\fig{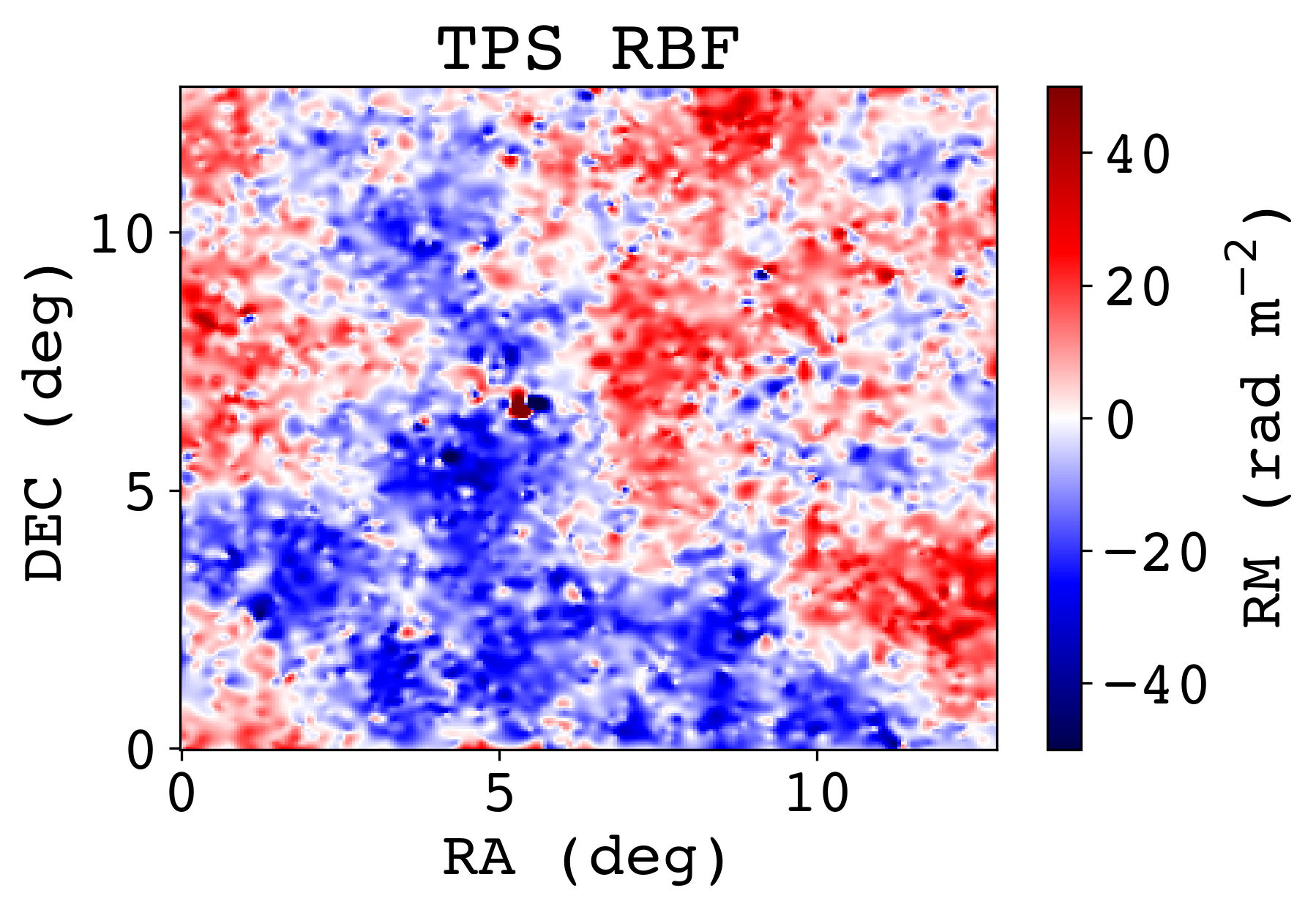}{0.33\textwidth}{}\fig{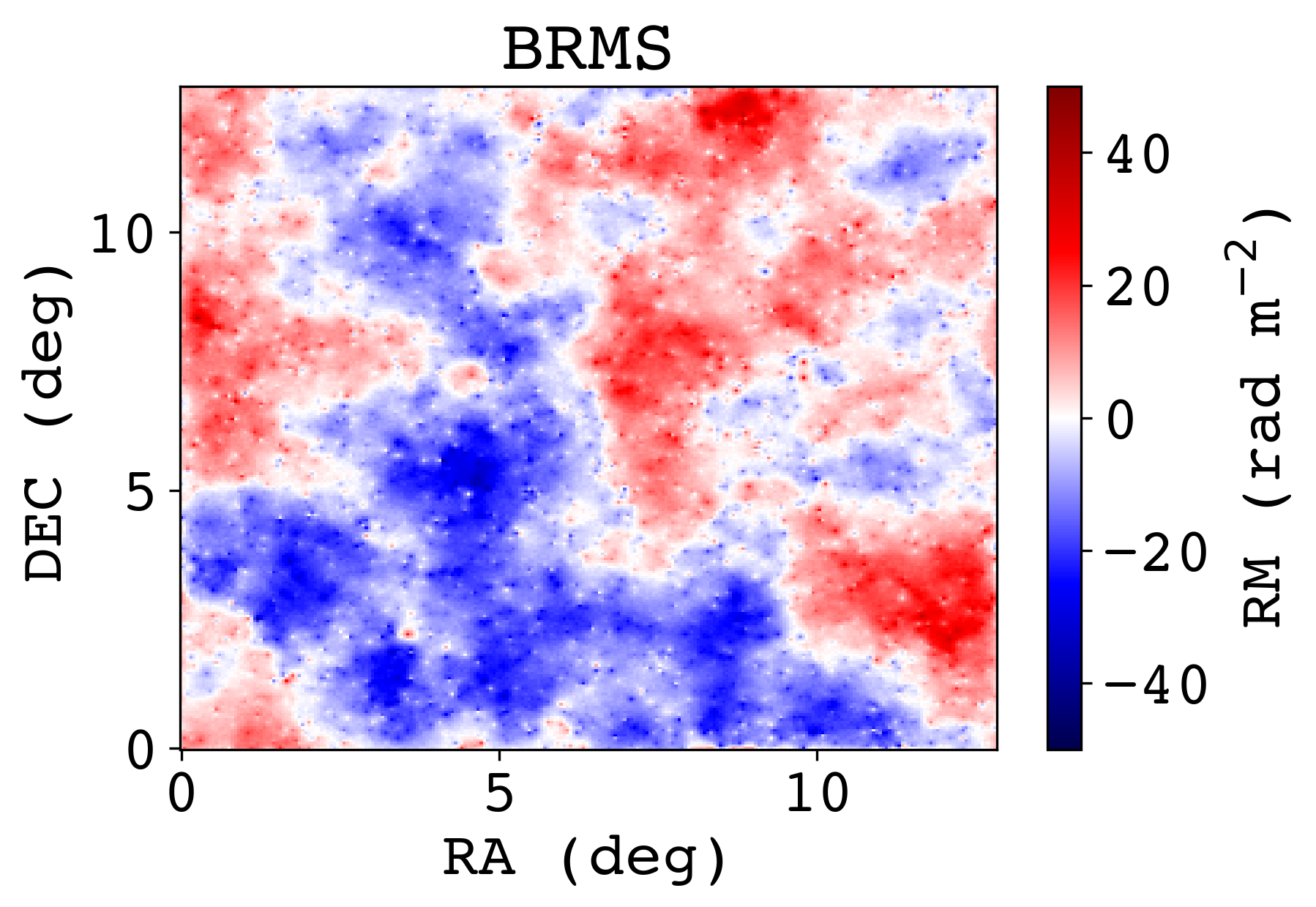}{0.33\textwidth}{}}
\caption{Reconstructed RM maps for each of the interpolation techniques for the patchy structures, saturated from $-50$ to $+50$ rad m$^{-2}$. The color map represents RM, with red being positive and blue being negative. At the top-centre, we present the simulated foreground RM. From top-left to bottom-right, the reconstructions are as follows: IDW1, IDW2, NNI, IM RBF, TPS RBF, BRMS.}
\label{fig:4}
\end{figure*}

In this subsection, we analyze the reconstruction of foreground RM for the patchy RM sky, which was produced by using the low Mach simulated cubes. To better visualize both high and low RM amplitude variations, we present our reconstructed results with a color bar saturated at $\pm 100$ rad m$^{-2}$ and at $\pm 20$ rad m$^{-2}$, respectively. Reconstructed maps using all of the aforementioned interpolation techniques, as well as the simulated foreground RM maps, are presented in Figure \ref{fig:4} (saturated at $\pm 50$ rad m$^{-2}$).

All of the interpolation techniques have the correct sign across most of the RM structures. The biggest outlier is IDW1 which fails to capture some of the positive RM at $\alpha <5$ deg. Similarly, all the techniques except IDW1 mostly capture the correct amplitude for RM structures. 

IDW1 reconstructs a map with a far lower $|\rm{RM}|$ at all points compared to the simulated foreground map and has the most smoothing of large scale structure of any of the techniques. IDW2 introduce less smoothing, however, it shows more sudden `jumps' in RM, corresponding to the positions of the input data points, compared to IDW1. 

NNI captures most of the large RM structure in the interior of the simulated foreground map, however, there are some fluctuations in the RM from the simulated foreground map at regions with lower $|\rm{RM}|$ values. In particular this occurs in transition regions where the RM sign changes. Notably, NNI (incorrectly) returns \pkg{nan} values at the edges of the interpolation region, which will be further discussed in Section \ref{sec:discussion}.

Both RBF interpolation techniques (IM and TPS) appear to reconstruct most of the large scale RM structures accurately. However, both techniques portray peaks in $|\rm{RM}|$ at some tightly concentrated regions, with IM RBF presenting more such fluctuations than TPS RBF. Similar to NNI, both RBF techniques also portray fluctuations in the reconstructed RM from the simulated foreground RM at regions of low $|\rm{RM}|$. 

BRMS appears to provide an accurate reconstruction of both large ($\sim 10^{0.8}$ deg) and small ($\sim 10^{0}$ deg) scale RM structures.

\subsection{Filamentary Foreground}
\begin{figure*}
\gridline{\fig{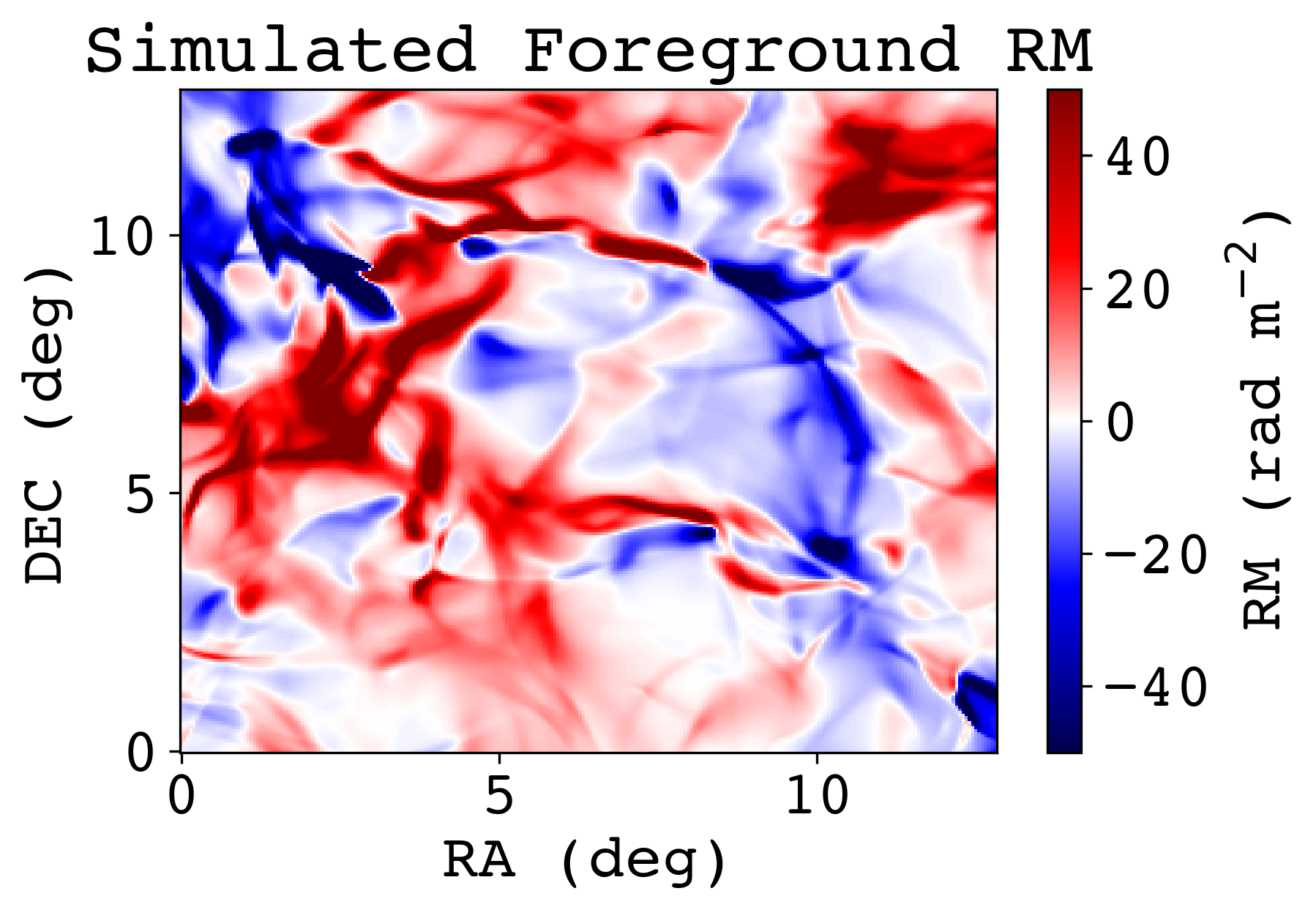}{0.33\textwidth}{}}
\gridline{\fig{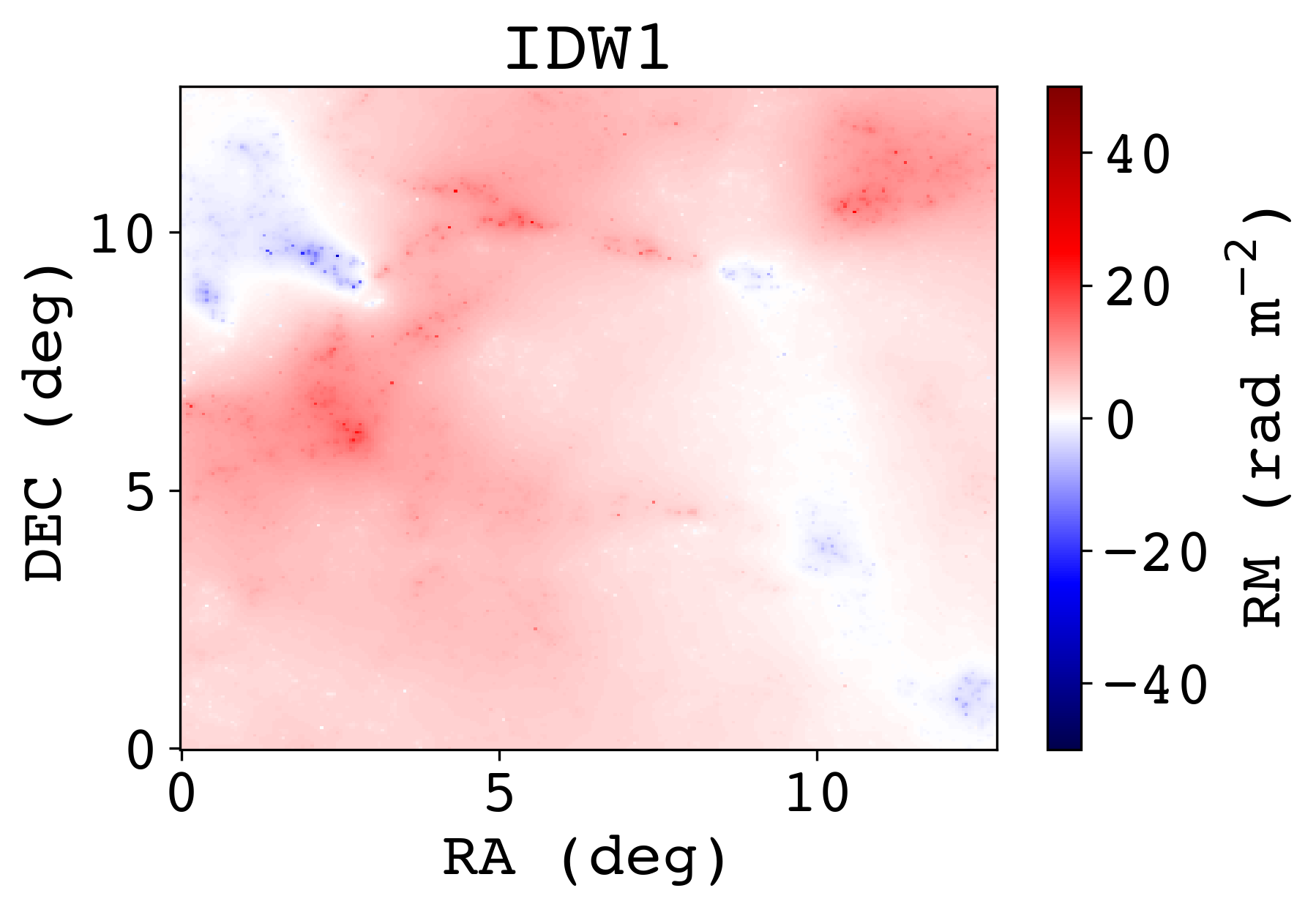}{0.33\textwidth}{}\fig{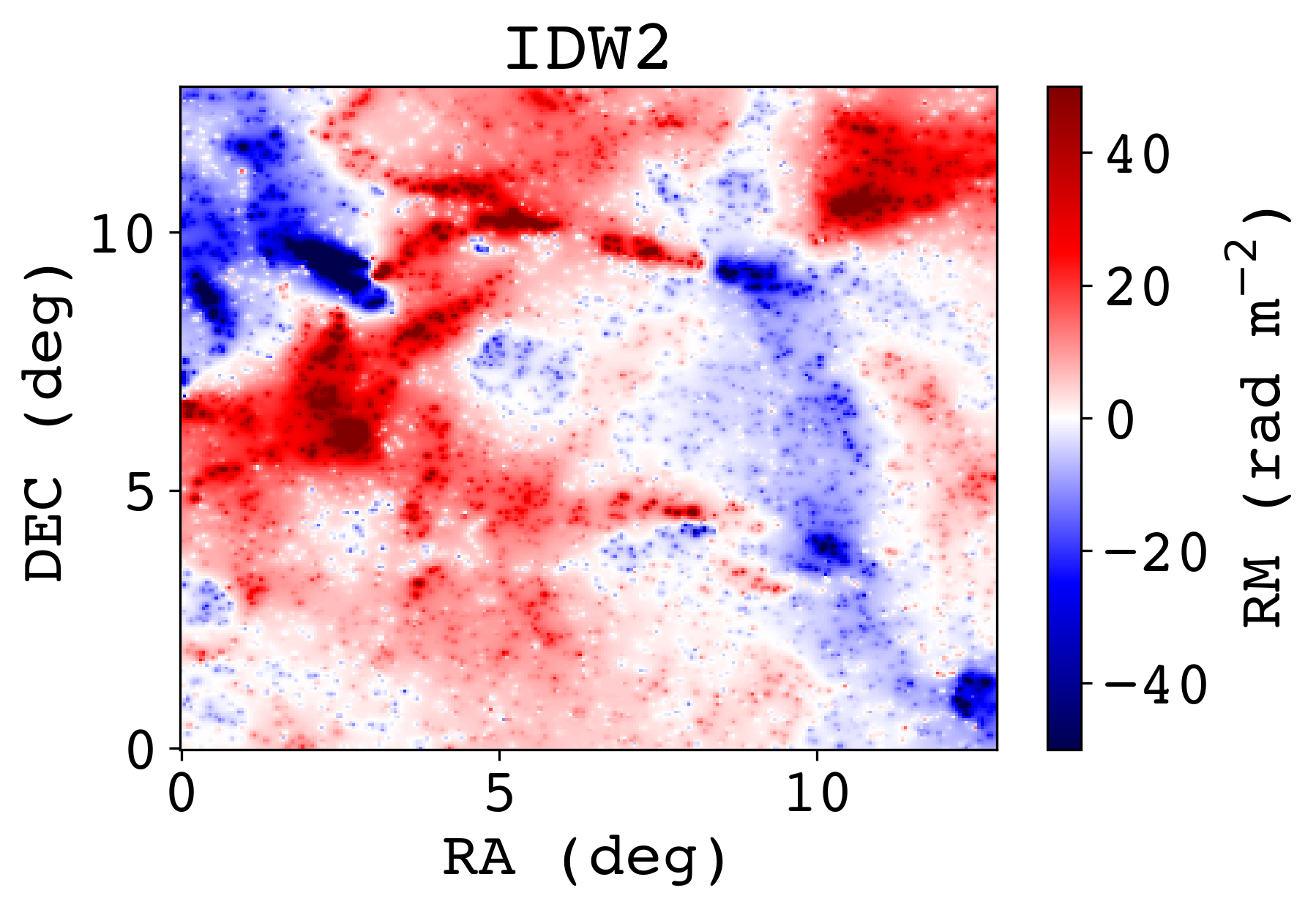}{0.33\textwidth}{}\fig{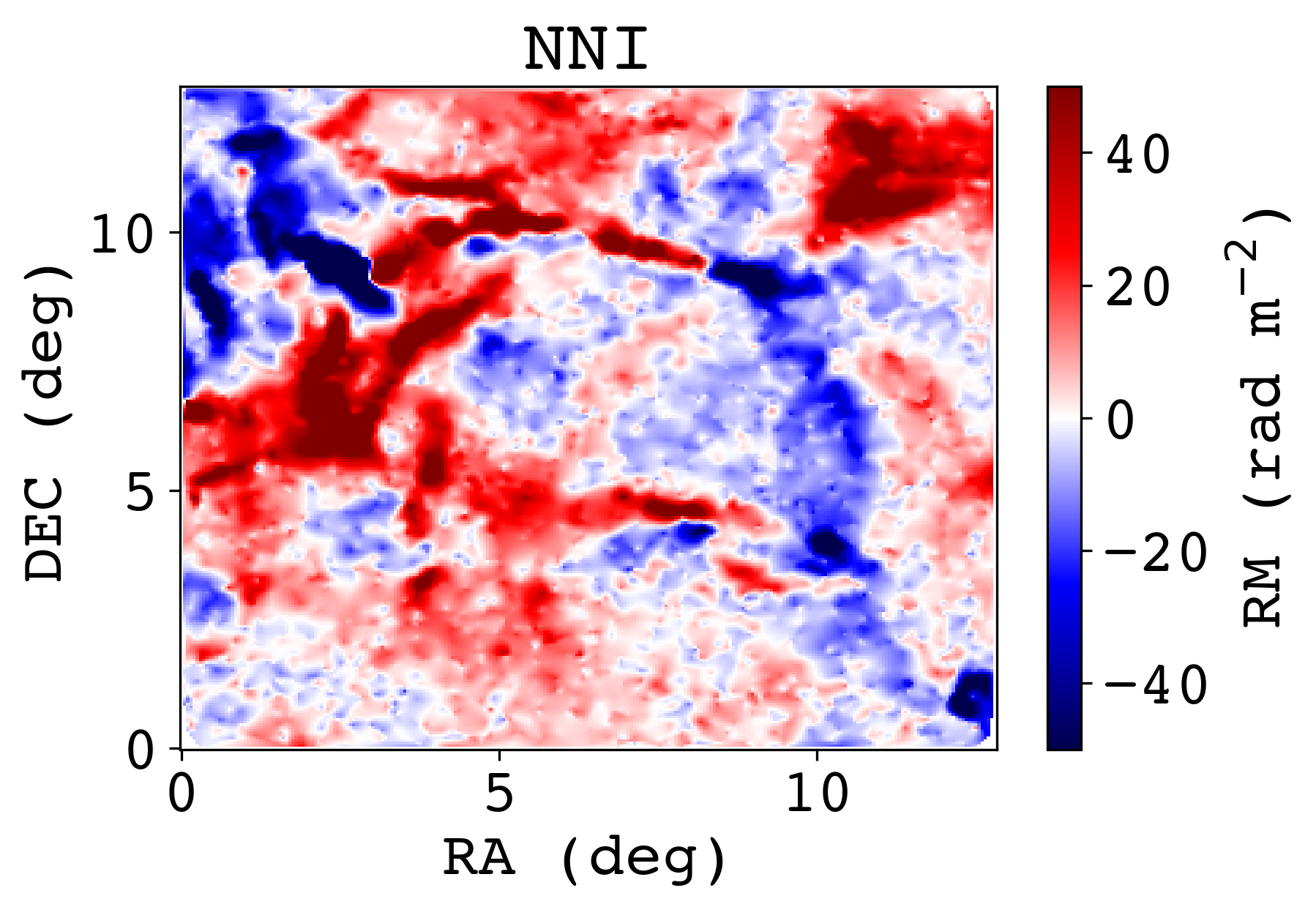}{0.33\textwidth}{}}
\gridline{\fig{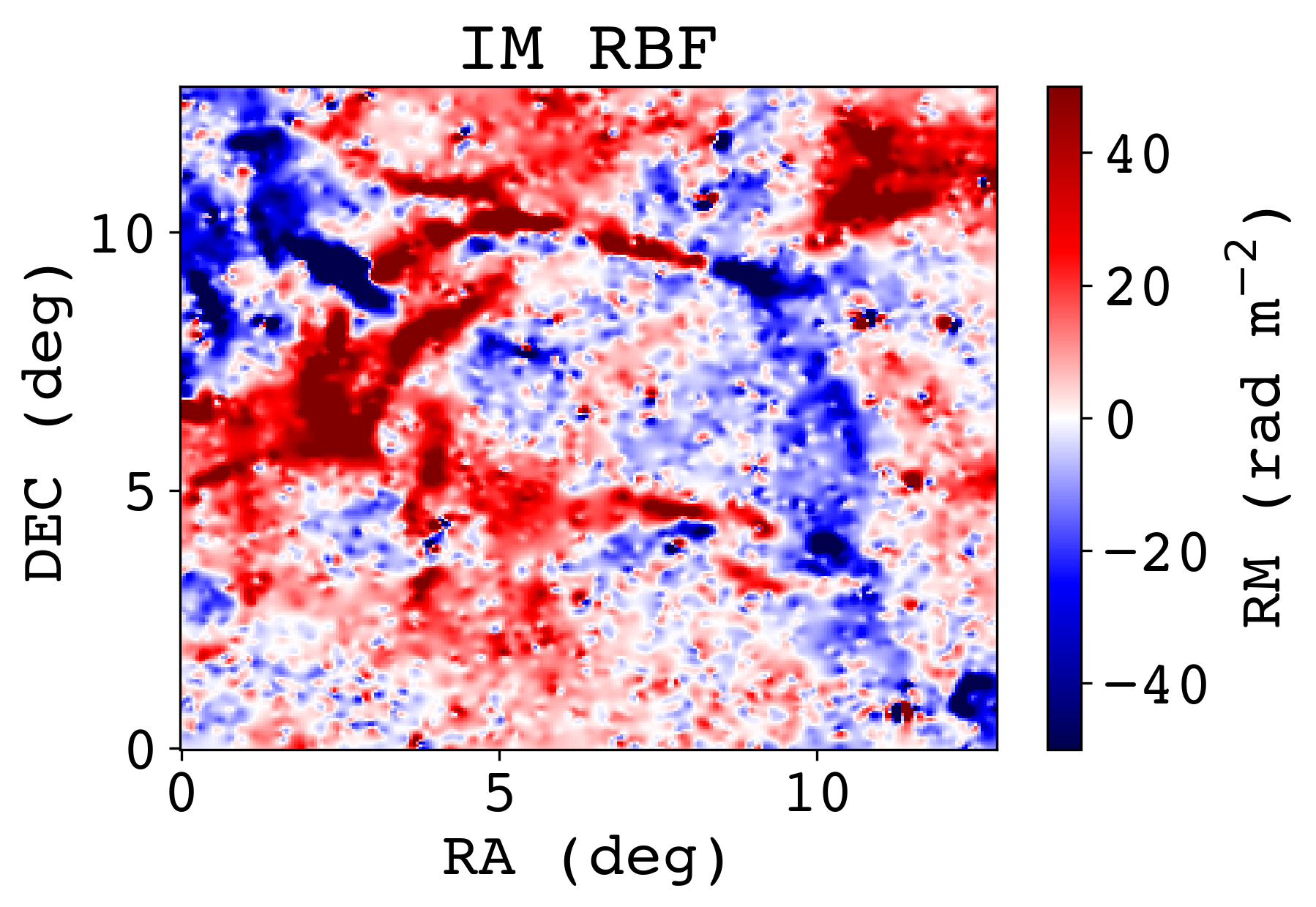}{0.33\textwidth}{}\fig{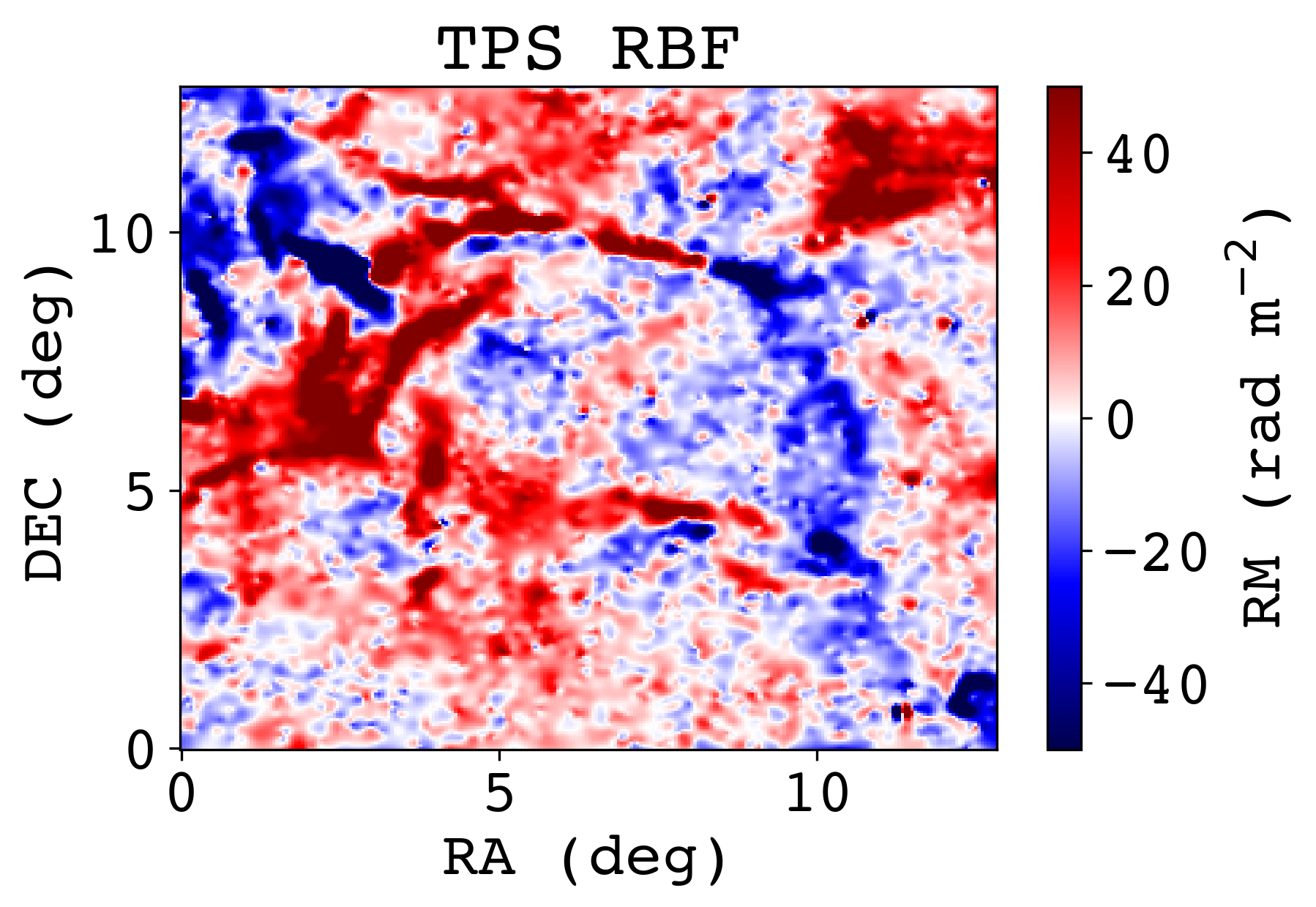}{0.33\textwidth}{}\fig{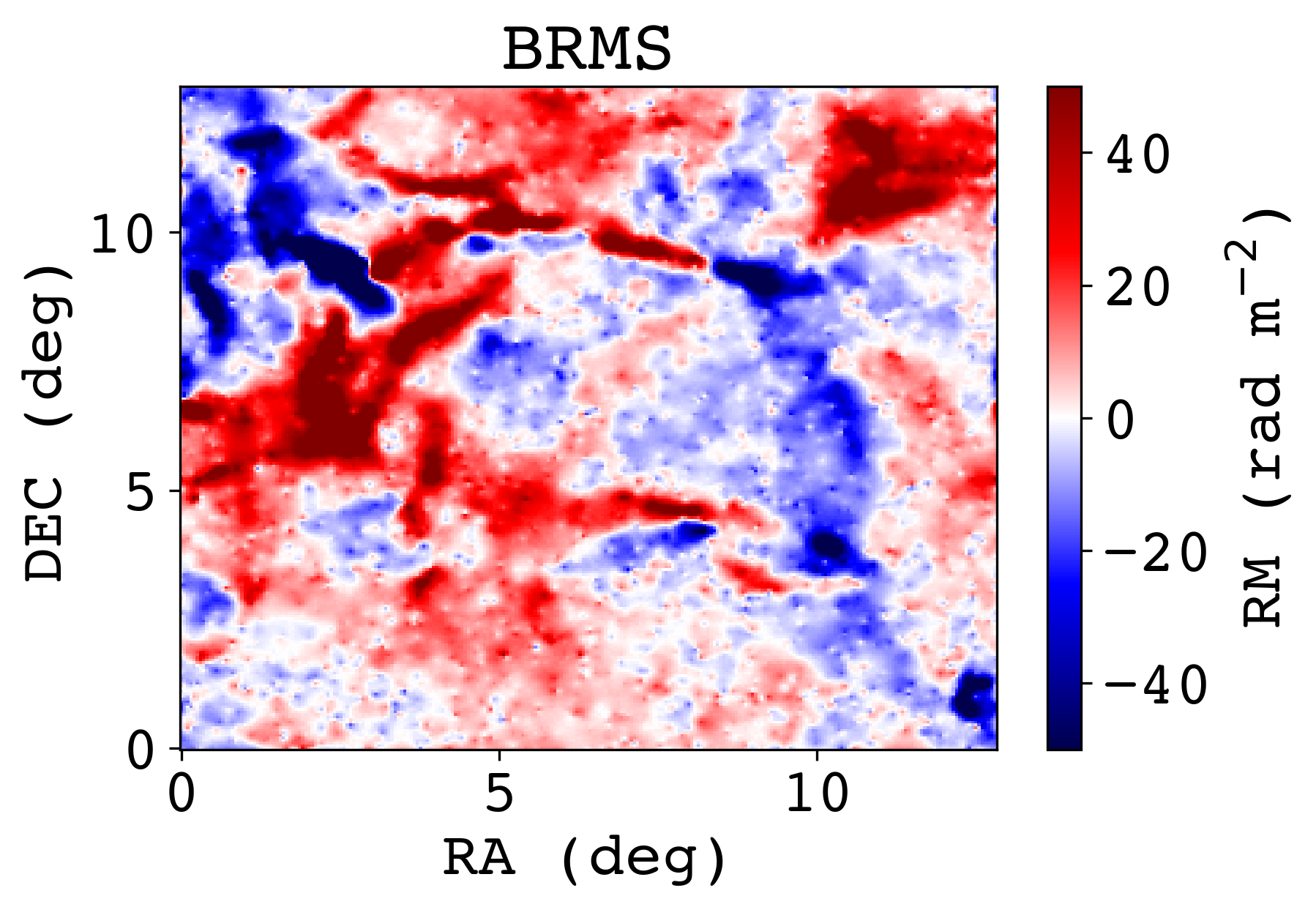}{0.33\textwidth}{}}
\caption{Reconstructed RM maps for each of the interpolation techniques for a foreground sky with filamentary structures, saturated from $-50$ to $+50$ rad m$^{-2}$. The color map represents RM, with red being positive and blue being negative. At the top-centre, we present the simulated foreground RM. From top-left to bottom-right, the reconstructions are as follows: IDW1, IDW2, NNI, IM RBF, TPS RBF, BRMS.}
\label{fig:6}
\end{figure*}

In this subsection, we present the reconstructions that were obtained for a filamentary RM sky, which was simulated using the high Mach simulated cubes. The results are portrayed in Figure \ref{fig:6}, which is saturated between $\pm 50$ rad m$^{-2}$. 

All of the interpolation techniques capture the correct sign for most RM structures, except IDW1, which fails to reconstruct most of the structure in the region $\alpha <8$ deg and $\delta <7$ deg. The other reconstructions do lead to an overall accurate reconstruction of the signs of the RM structures, however (as before) the amplitude is miscalculated for some of the structures. 

IDW1 constructs a map that has an RM amplitude that is far lower than that of the simulated foreground RM map and produces sharp peaks at the positions of input data points. The peaks at the positions of the input data points for IDW2 do still persist but they are far less pronounced when compared to the results of the patchy RM sky reconstruction.

NNI appears to accurately reconstruct almost all of the filamentary structure from the simulated foreground RM. The only small deviation that it has from the simulated foreground RM are the \pkg{nan} values (which, similar to the patchy RM reconstruction, appear at the edges of the interpolation region), and the small-scale RM structure of some of the smaller RM filaments that are mostly present at $\delta <5$ deg. There are far fewer deviations from the simulated foreground at lower $|\rm{RM}|$ values, when compared to the case of the patchy RM sky. 

Both RBF techniques (IM and TPS) seem to reconstruct most of the RM filaments from the simulated foreground. Notably, the sharp peaks in $|\rm{RM}|$ that were observed in the patchy RM sky are absent in the filamentary case. Furthermore, the fluctuations in RM near regions of sign reversal (in the RM) that were present in the patchy RM sky appear to persist; they are most apparent in regions of the sky with  low $|\rm{RM}|$.

BRMS does a good job of reconstructing the large RM structures and captures the correct signs of all RM filaments. It also accurately interpolates smaller structures within the RM filaments. 

\subsection{Residuals}
\label{sec:analysis}

\begin{figure*}[tp]

\gridline{\fig{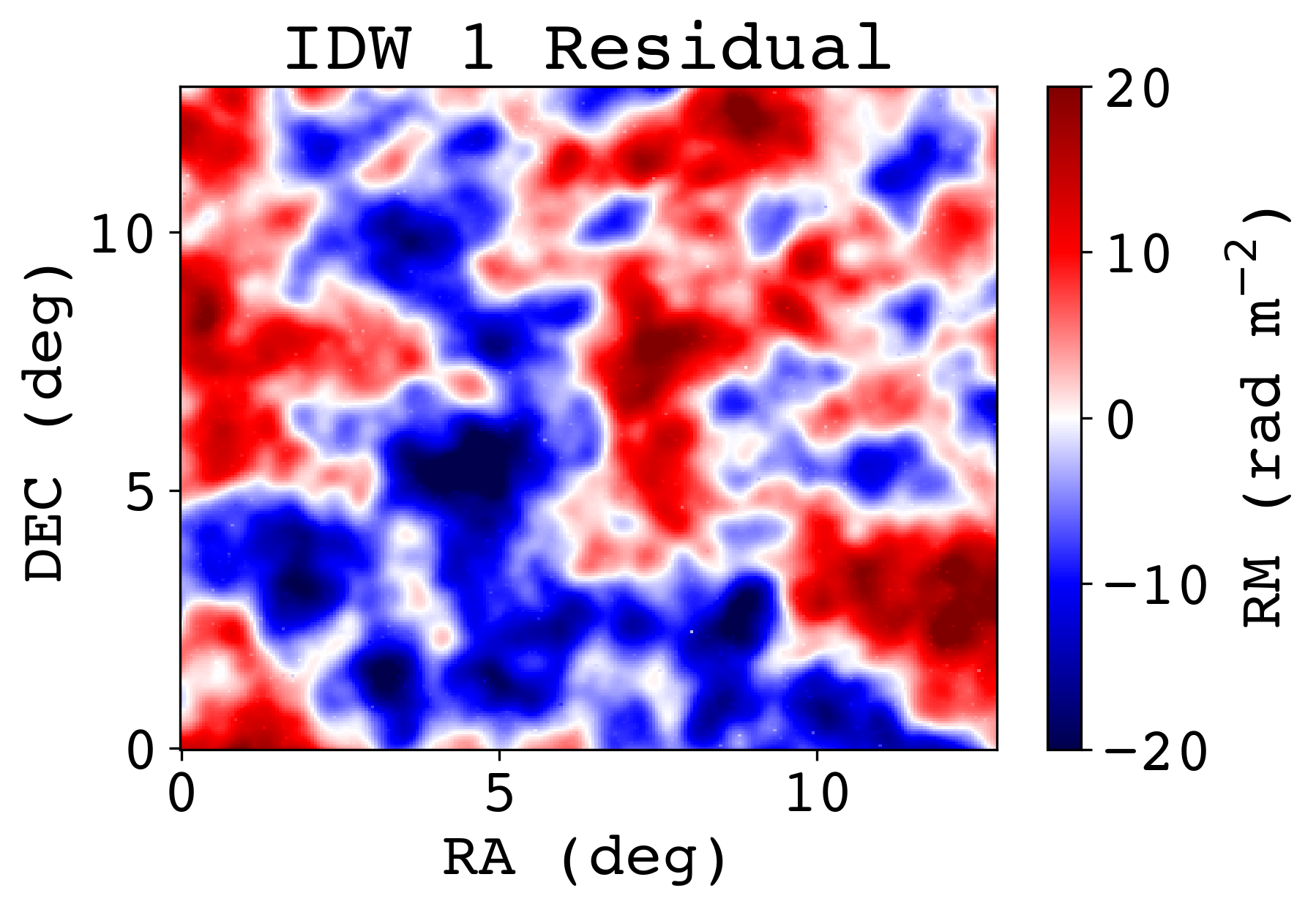}{0.33\textwidth}{}\fig{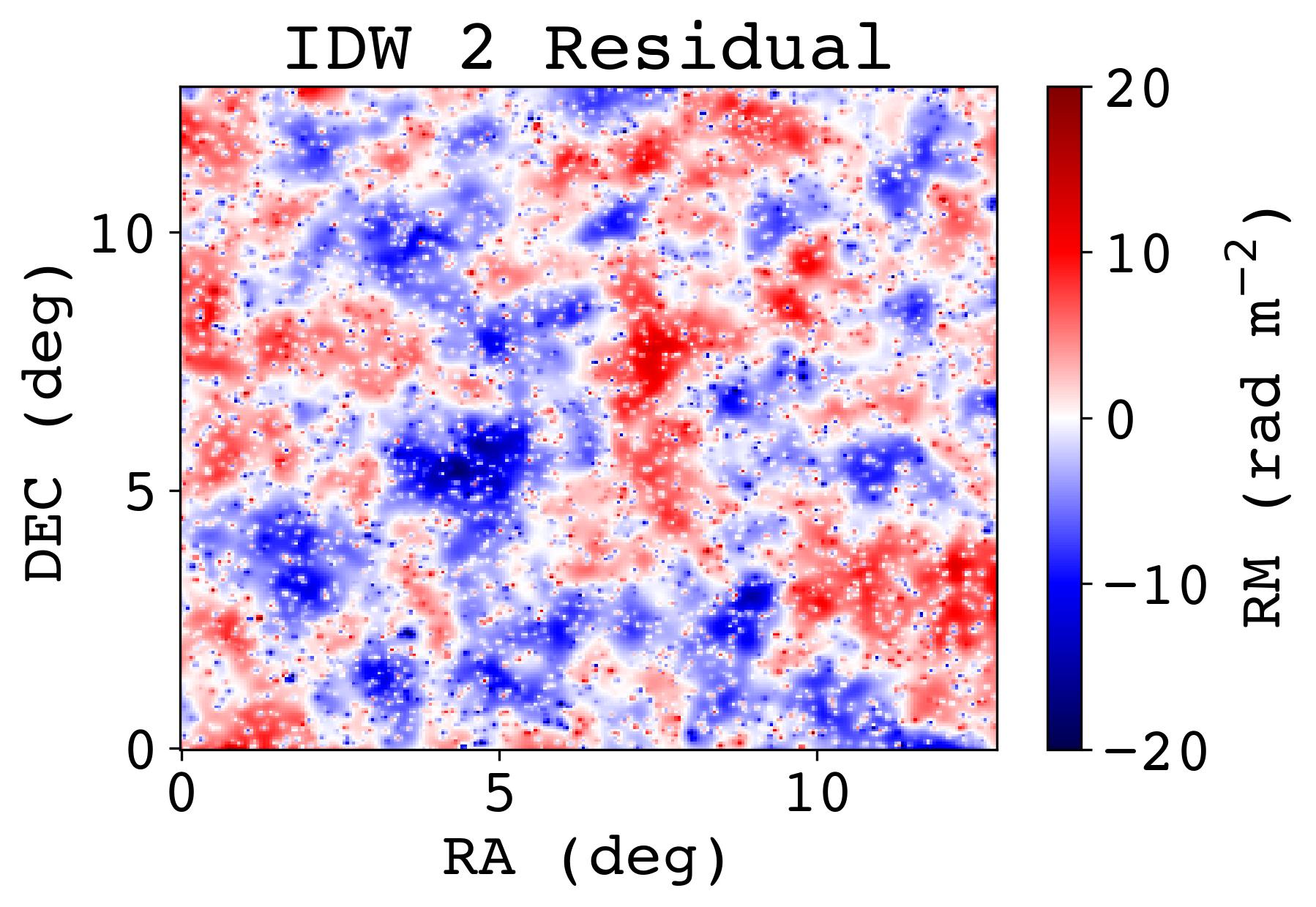}{0.33\textwidth}{}\fig{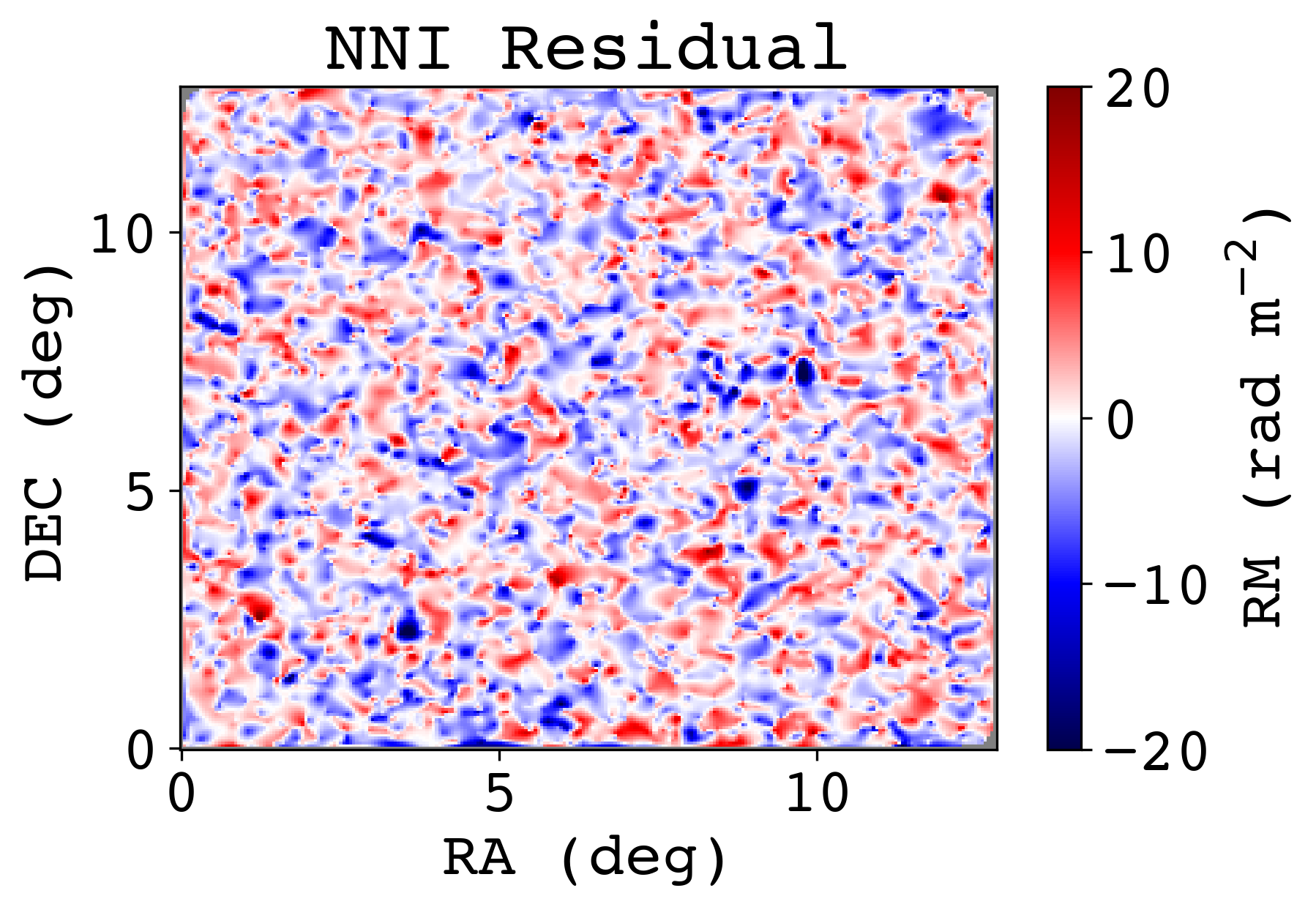}{0.33\textwidth}{}}
\gridline{\fig{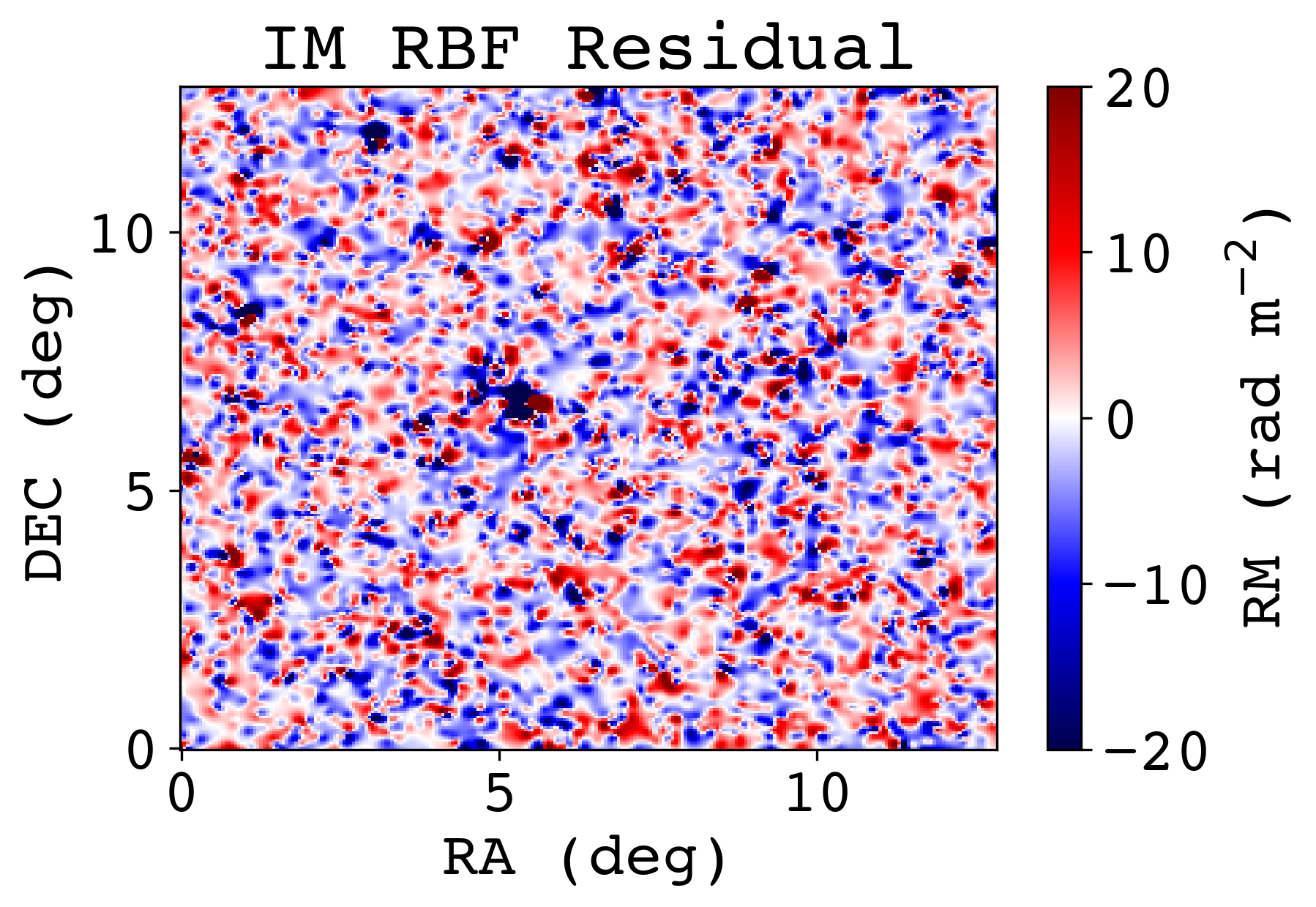}{0.33\textwidth}{}\fig{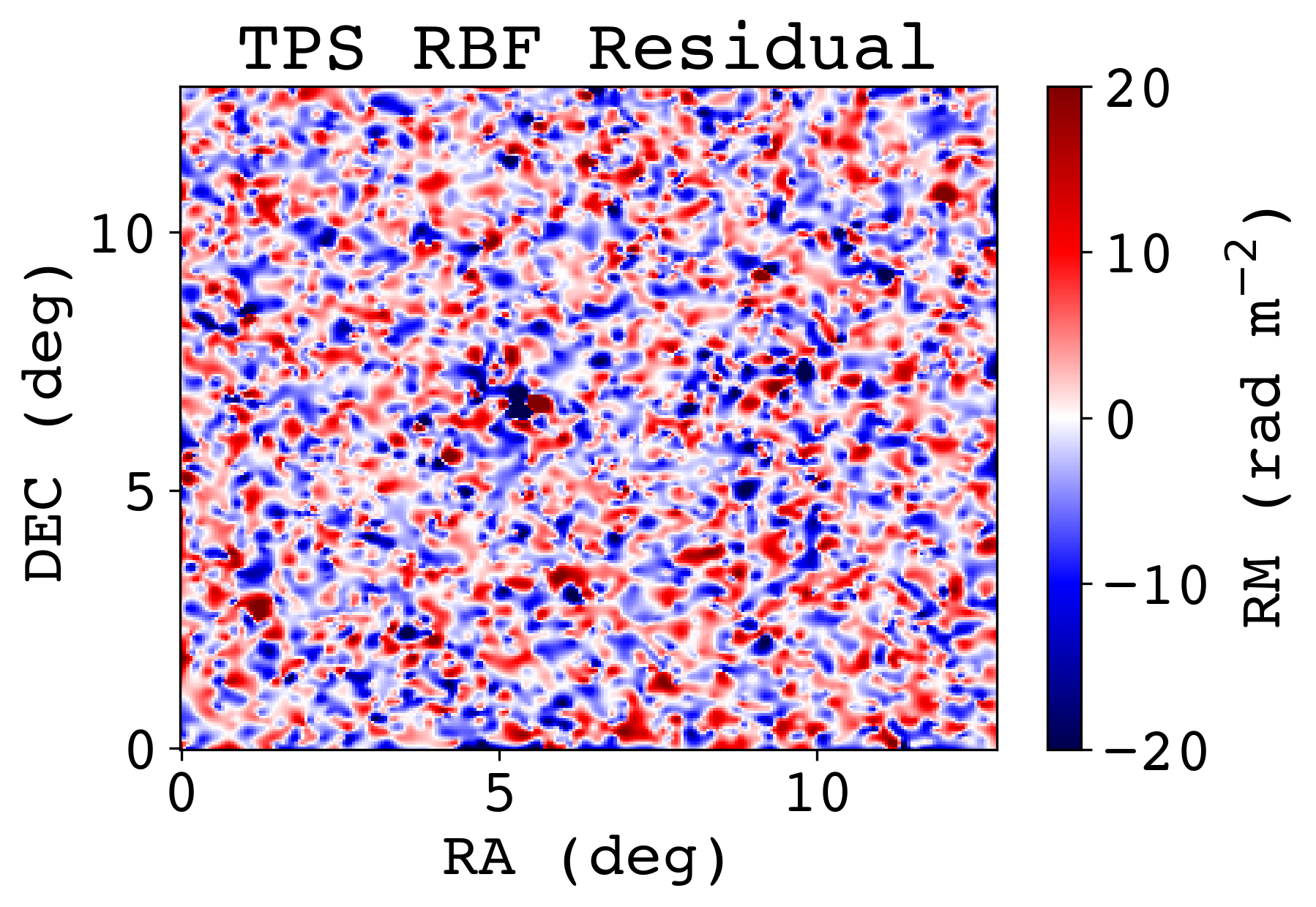}{0.33\textwidth}{}\fig{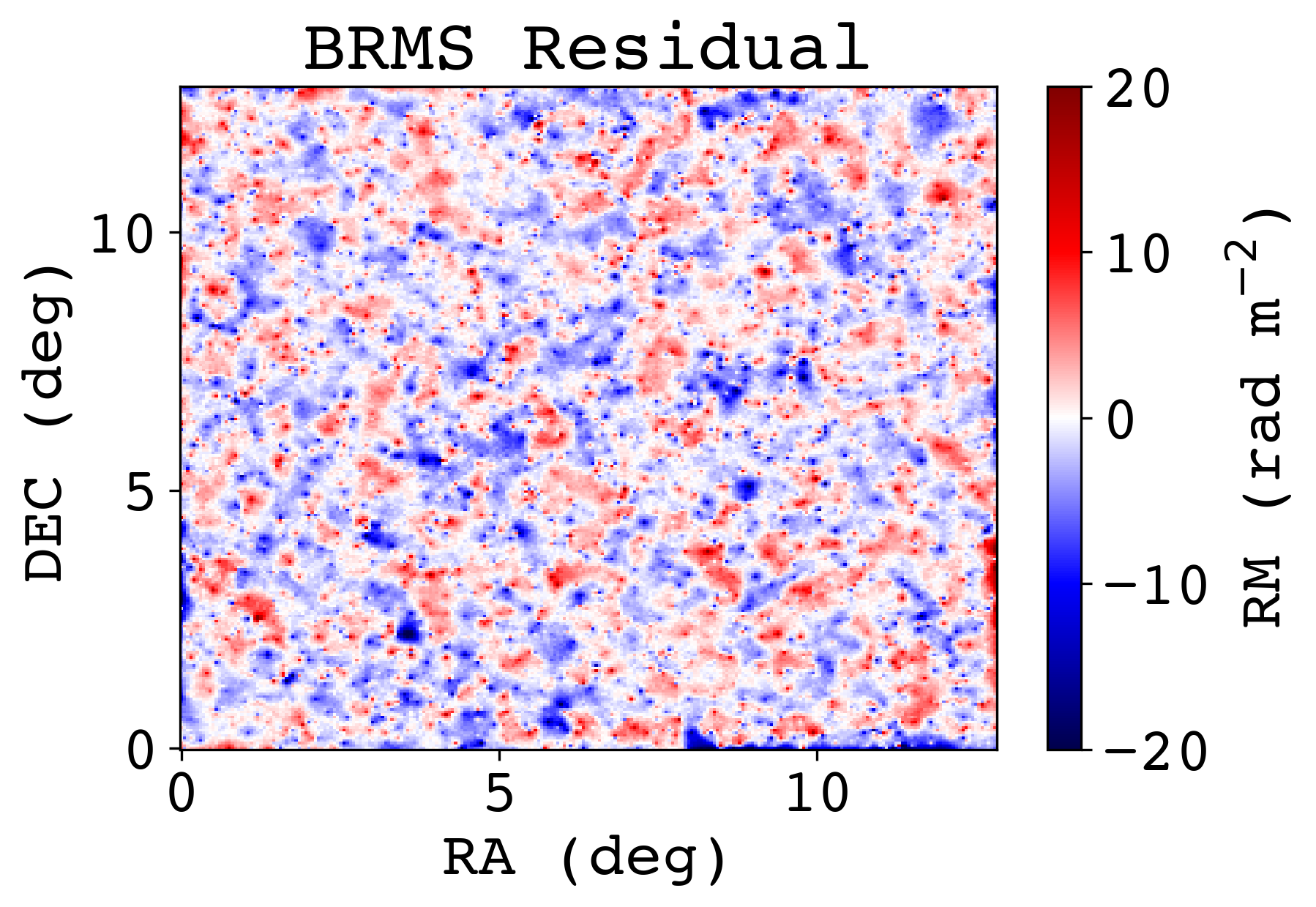}{0.33\textwidth}{}}

\caption{Residual RM maps for each of the interpolation techniques in the case of a patchy RM sky. In every map, the residual RM for each pixel was calculated by finding the difference in the RM value between the reconstructed pixel and the pixel from the simulated foreground RM. From top-left to bottom-right, the reconstruction are as follows: IDW1, IDW2, NNI, IM RBF, TPS RBF, and BRMS.}
\label{fig:12}
\end{figure*}

\begin{figure*}[tp]

\gridline{\fig{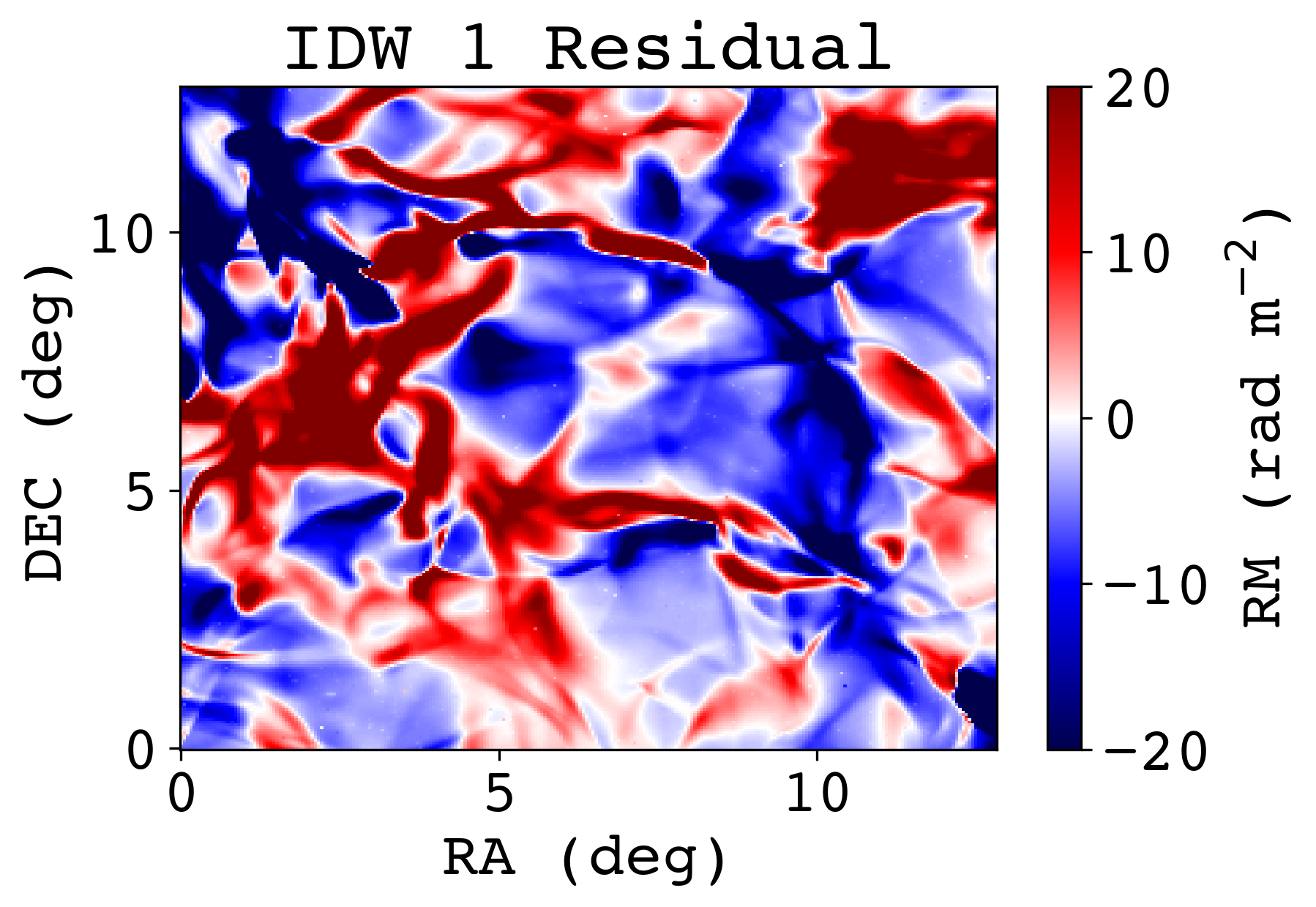}{0.33\textwidth}{}\fig{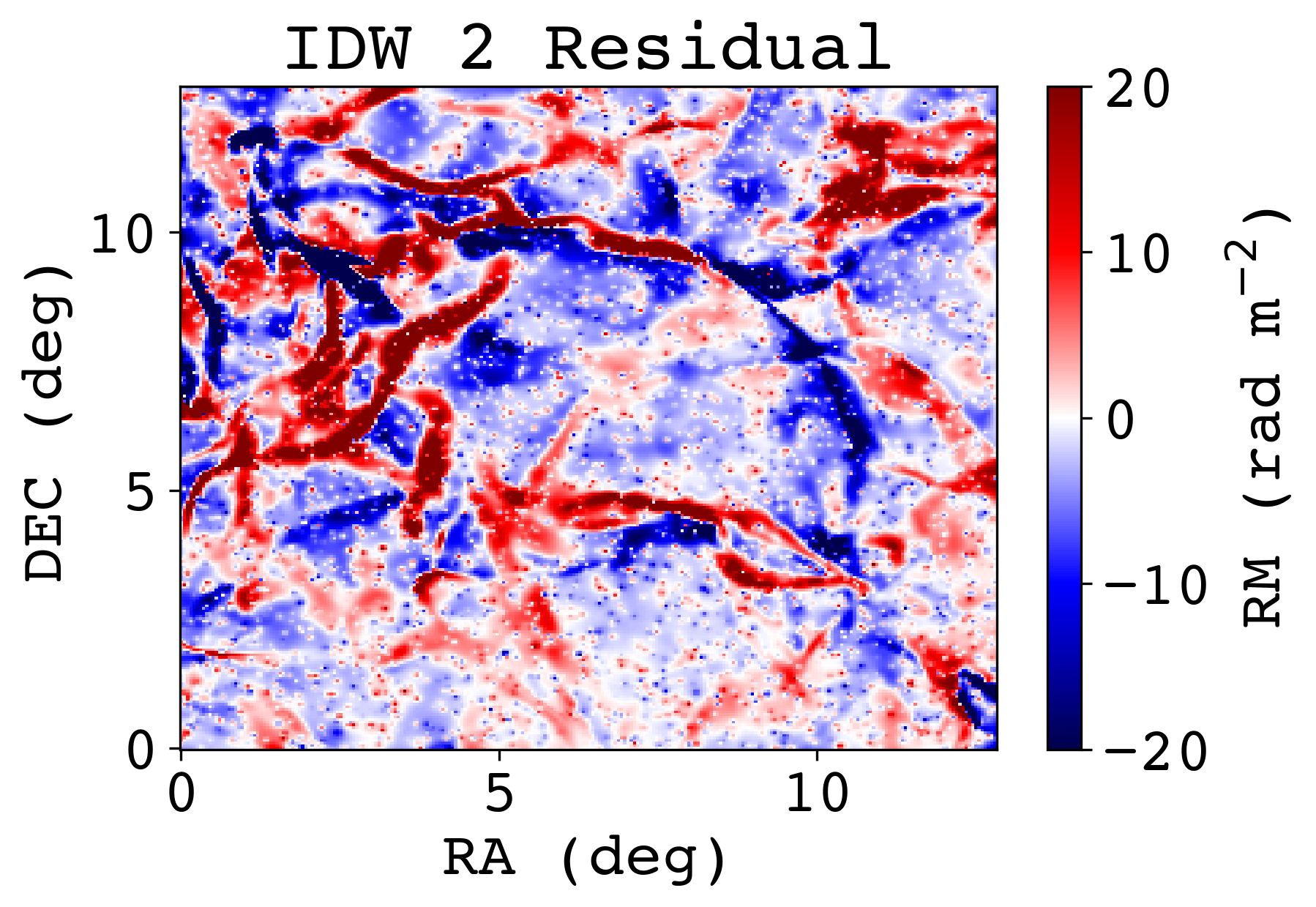}{0.33\textwidth}{}\fig{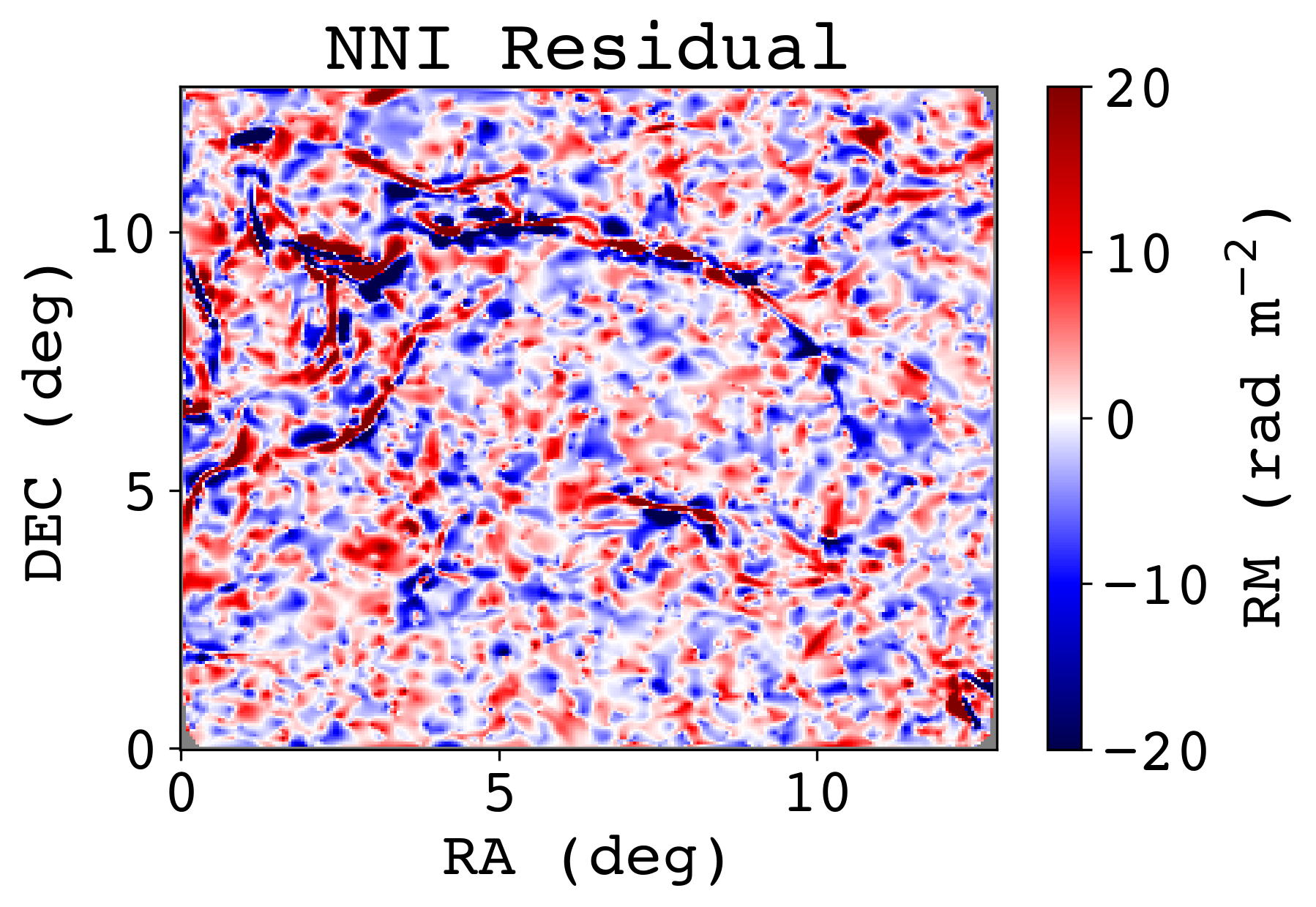}{0.33\textwidth}{}}
\gridline{\fig{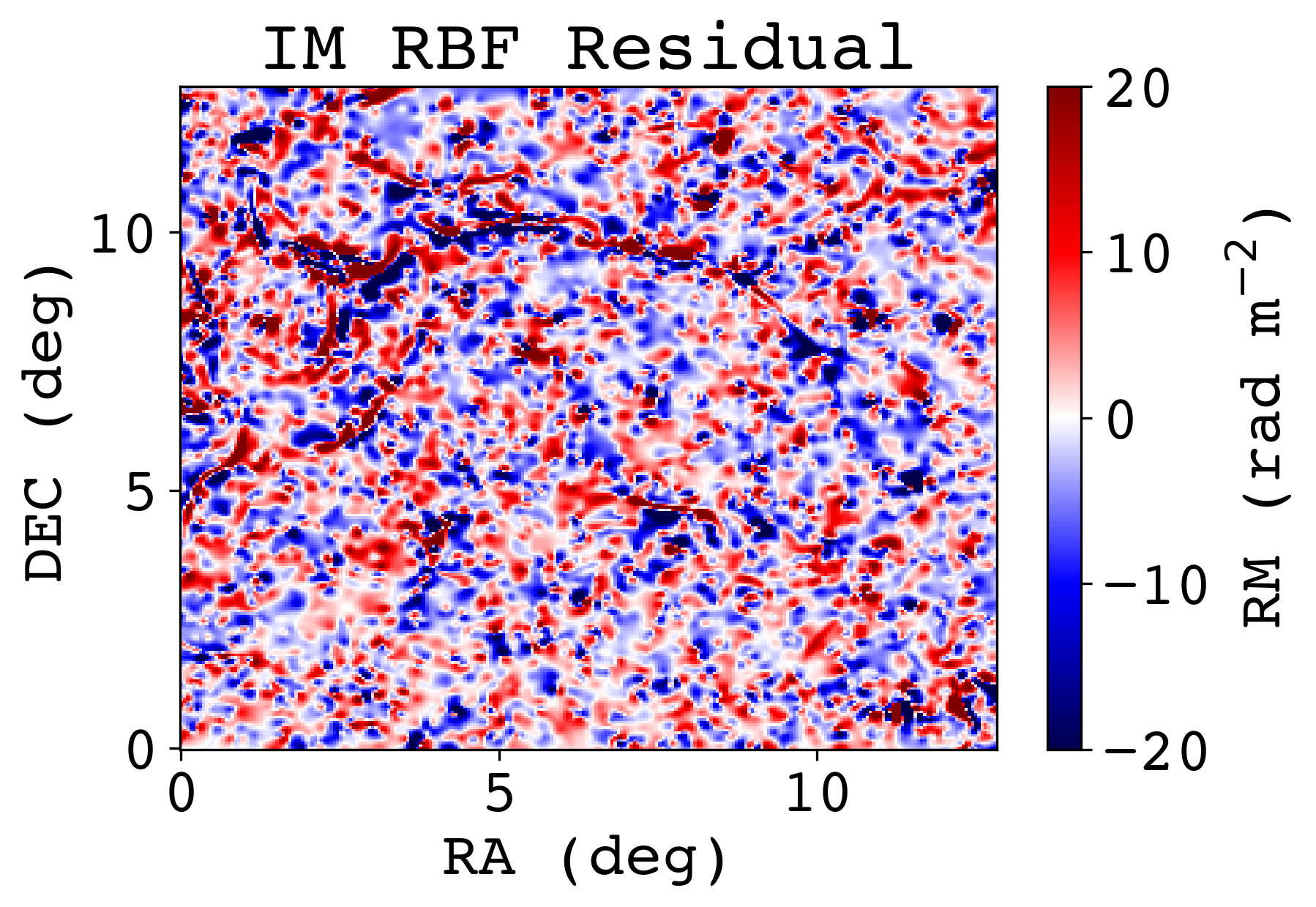}{0.33\textwidth}{}\fig{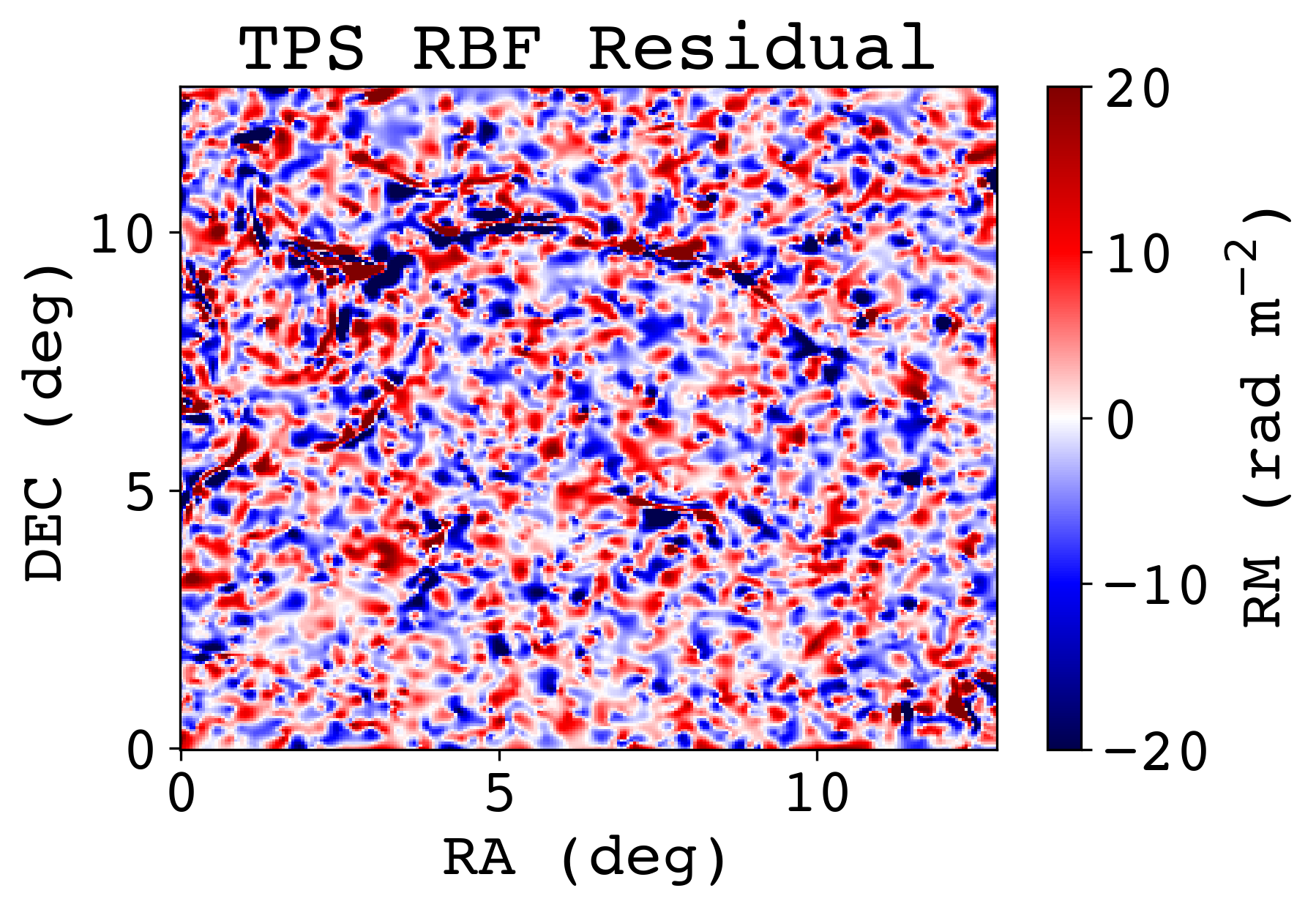}{0.33\textwidth}{}\fig{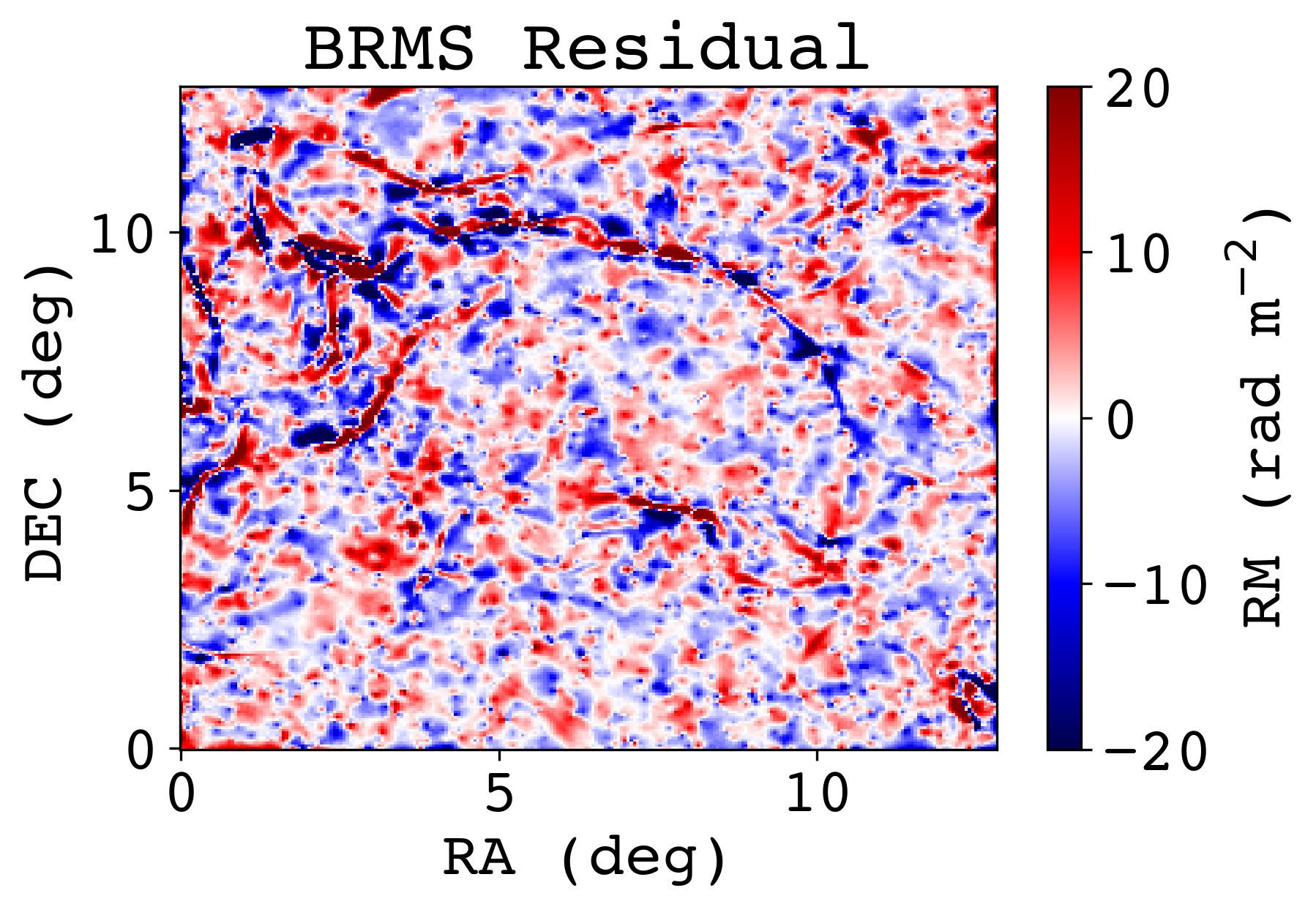}{0.33\textwidth}{}}

\caption{Residual RM maps for each of the techniques in the case of a filamentary RM sky. In every map, the residual RM for each pixel was calculated by finding the difference in the RM value between the reconstructed pixel and the pixel from the simulated foreground RM. From top-left to bottom-right, the reconstruction are as follows: IDW1, IDW2, NNI, IM RBF, TPS RBF, and BRMS.}
\label{fig:13}
\end{figure*}

The residuals for each of the techniques for the patchy and filamentary RM sky are presented in Figures \ref{fig:12} and \ref{fig:13}, respectively. 

In the patchy case, as described in the previous subsection, the residual values for both IDW algorithms form large RM structures, being indicative of the fact that it does not accurately reconstruct large RM structures. However, the mean residual for IDW2 is only bested by BRMS and NNI. For the rest of the techniques, we can see that the issues in reconstruction lie in the smaller scale RM structures. 

In the filamentary case, both IDW residuals again show that the algorithm has difficulty reconstructing the large-scale RM structures. All techniques have the most residual RM present along the high $|\rm{RM}|$ filaments. 

\begin{figure*}
\centering
\large \textbf{Residual Histograms}
    \gridline{\includegraphics[width = 0.49\textwidth]{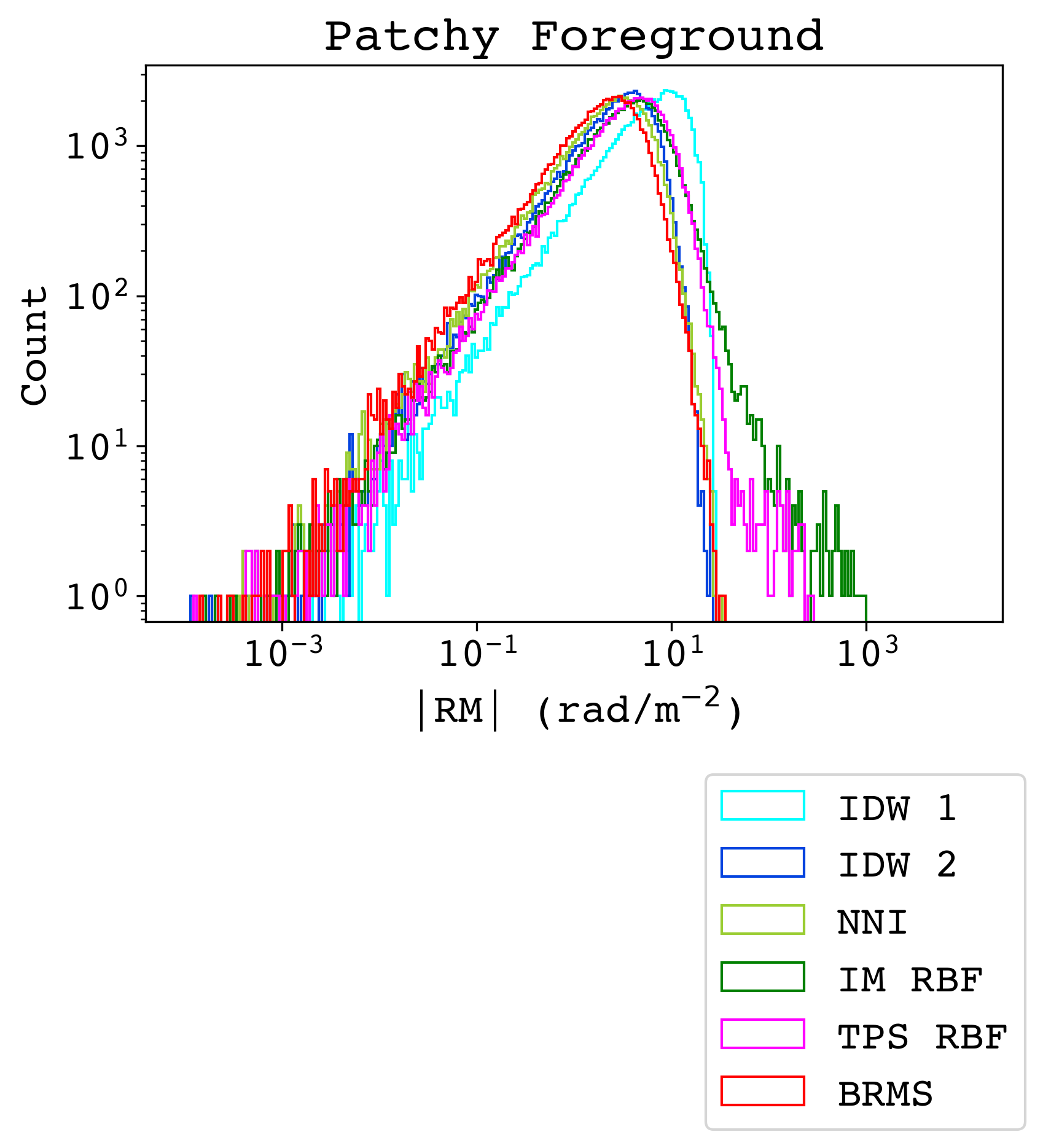} \includegraphics[width = 0.49 \textwidth]{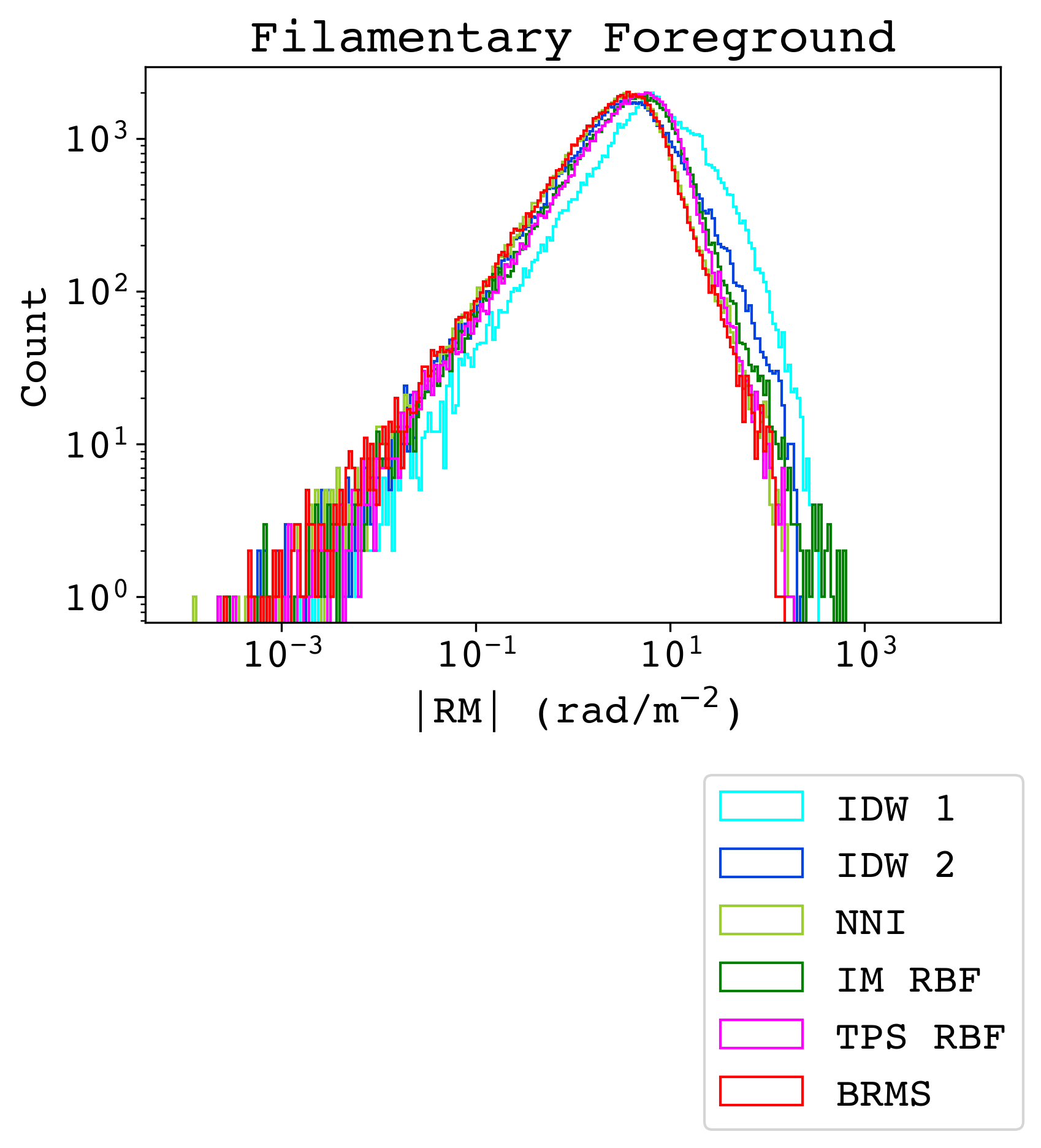}}
    \caption{A histogram of the magnitude of the residual RM for all the techniques applied for the patchy RM sky (left) and a histogram of the magnitude of the residual RM for all reconstructions applied to the filamentary RM sky (right). }
    \label{fig:14}
\end{figure*}

Figure \ref{fig:14} presents the histograms of the magnitudes of the residual RMs for all the interpolation techniques applied to both patchy and filamentary RM structures. In both RM skies, we see that BRMS performs the best, while IDW1 performs the worst. Contrary to what we expect, IDW2 seems to perform better than IM and TPS in the patchy sky. This will be discussed further in Section \ref{sec:discussion}. Further, qualitatively, some of the techniques seem to perform similarly. However, we will analyze the techniques quantitatively in Section \ref{sec:power}. Because of these reasons, we will use a different measure of the accuracy of the interpolation using their power spectra.

\subsection{Computational Costs}

We present a summary of the accuracy and costs of each of the techniques in Tables \ref{tab:1} and \ref{tab:2} for the patchy and filamentary RM structures, respectively. For all the reconstructions, the standard deviation of the absolute residual RM is greater in the case of the filamentary sky. 

Across all cases, the least computationally intensive algorithms are the RBF techniques, with TPS RBF being slightly less computationally intensive than IM RBF in the filamentary sky; both TPS and IM take $\sim$ 16s, compared to the other algorithms, which take $> 200$~s. However, the reconstruction accuracy of both techniques does suffer in the patchy case, having high mean residuals compared to NNI and BRMS. 

 In addition to the lack of accuracy, the computation times for the IDW algorithms are 2-3 orders of magnitude greater than those of the other algorithms ($\sim 10^3$~s compared to $\sim 10^2$~s and $\sim 10$~s).

In the case of the filamentary RM sky, both RBF techniques and NNI result in close mean residual RMs, and standard deviations as well. The only major difference is their computational cost; the RBF techniques take $\sim 10$~s while NNI takes $\sim 10^2$~s. 

\begin{table*}[!htb]
\centering
\textbf{Patchy Foreground Performance}
    \begin{tabular}{|c|c|c|c|}
        \hline
        \textbf{Interpolation Technique} & \textbf{$\langle|\text{Res RM}|\rangle$ / $\mathbf{\sigma_{\mathrm{\textbf{sim}}}}$}   & \textbf{$\sigma(|\text{Res RM}|)$ / $\mathbf{\sigma_{\mathrm{\textbf{sim}}}}$} & \textbf{Computational Time (s)}\\
        \hline 
        \hline
        IDW1 & 0.651 &0.475 & 6992\\
        IDW2 & 0.304 &  0.238 & 6311 \\
        NNI & 0.271 & 0.233 & 338\\
        IM RBF &0.476 & 1.29 & 16\\
        TPS RBF & 0.434 & 0.531 & 16\\
        BRMS &  0.228 & 0.206 & 272\\
        \hline
    \end{tabular}
\caption{Mean of the absolute value of the residual RMs (normalized by the standard deviation in the simulated foreground RM), their standard deviations (normalized by the standard deviation in the simulated foreground RM), and the computational times for each of the interpolation techniques we have tested. The number of data points used for each of the reconstructions was 6310 and the number of interpolation points on the patchy RM sky was $256^2$ (these were uniformly distributed throughout the interpolation region).}
    \label{tab:1}
\end{table*}

\begin{table*}[!htb]
\centering
\textbf{Filamentary Foreground Performance}
    \begin{tabular}{|c|c|c|c|}
        \hline
        \textbf{Interpolation Technique} & \textbf{$\langle|\text{Res RM}|\rangle$  / $\mathbf{\sigma_{\mathrm{\textbf{sim}}}}$}  & \textbf{$\sigma(|\text{Res RM}|)$  / $\mathbf{\sigma_{\mathrm{\textbf{sim}}}}$}  & \textbf{Computational Time (s)}\\
        \hline 
        \hline
        IDW1 & 0.496 & 0.767 & 6827\\
        IDW2 & 0.272 &  0.478 & 6634\\
        NNI  & 0.189 & 0.279 & 357\\
        IM RBF & 0.262 & 0.492 & 20\\
        TPS RBF & 0.240 & 0.296 & 14\\
        BRMS &  0.182 &0.258 & 546\\
        \hline
    \end{tabular}
\caption{Mean of the absolute value of the residual RMs (normalized by the standard deviation in the simulated foreground RM), their standard deviations (normalized by the standard deviation in the simulated foreground RM), and the computational times for each of the interpolation techniques we have tested. The number of data points used for each of the reconstructions was 6436 and the number of interpolation points on the filamentary RM sky was $256^2$ (these were uniformly distributed throughout the interpolation region).}
    \label{tab:2}
\end{table*}

\subsection{Power Spectra}
\label{sec:power}
\begin{figure*}
    \gridline{\includegraphics[width = 0.49\textwidth]{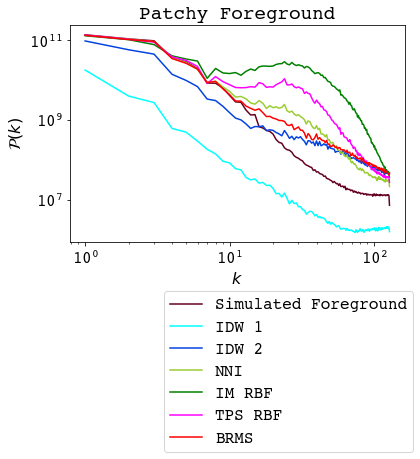} \includegraphics[width = 0.49 \textwidth]{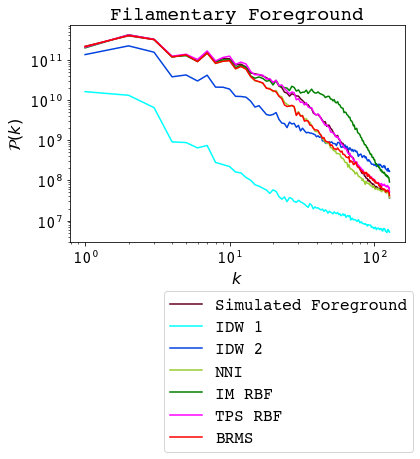}}
        \gridline{\includegraphics[width = 0.49\textwidth]{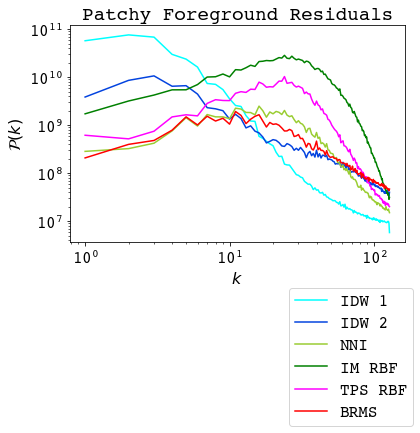} \includegraphics[width = 0.49 \textwidth]{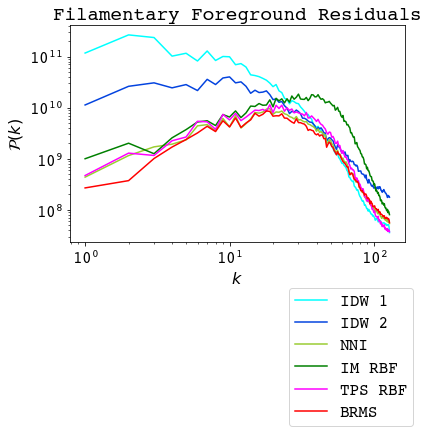}}
    \caption{The top figures display the power spectra of the reconstructed RM maps and the simulated foreground for the patchy sky (left) and the filamentary sky (right), and the bottom two figures display the power spectra of the residuals of the reconstructed RM maps for the patchy sky (left) and the filamentary sky (right).}
    \label{fig:power}
\end{figure*}

The Fourier power spectra were computed for the RM maps and their residuals under a flat sky approximation and are presented in Figure \ref{fig:power}. 

From the power spectra, we can see that IDW1 consistently underestimates the RM over large angular scales. From the power spectra of the residuals, IDW2 does not reconstruct large RM structures accurately in both skies, while IDW1 accurately reconstructs structure at smaller angular scales. All the other interpolation kernels perform well at large angular scales and at small angular scales. In the power spectra of the residuals, the largest power that is seen is in the intermediate angular scales, where the interpolation kernels perform from best to worst: BRMS, NNI, TPS and IM. 

We will be forming a goodness-of-fit parameter for the interpolation kernels using the power spectra of the residuals. From the power spectra of the residuals, we see that if the power is less, the reconstruction is more accurate. Additionally, we need to weight the power more for larger angular scales as reconstructing the general shape of a large RM structure is more important than the small scale features. Therefore, we will weight the parameter with the angular separation, i.e, by $1/k$ to define the following goodness-of-fit parameter: 

\begin{align}
    I_{\mathrm{res}} = \int \frac{1}{k} \frac{\mathcal{P}_{\mathrm{res}}(k)}{\mathcal{P}_{\mathrm{total}}} dk,
    \label{eq:integral}
\end{align}
where $\mathcal{P}_{\mathrm{res}}(k)$ is the power spectrum of the residual RM map and $\mathcal{P}_{\mathrm{total}}$ is defined as: 

\begin{align}
    \mathcal{P_\mathrm{total}} = \int{\mathcal{P}_\mathrm{sim}(k)}dk,
\end{align}
with $\mathcal{P}_{\mathrm{sim}} (k)$ being the power spectrum of the simulated foreground RM. Similarly, we can also define the correlation length of the RM reconstruction, $I$ as:

\begin{align}
    I = \int  \frac{1}{k} \frac{\mathcal{P}(k)}{\mathcal{P}_\mathrm{total}} dk,
\end{align}
where $\mathcal{P}(k)$ is the power spectrum of the reconstructed RM map. We can now convert the correlation length from $k-$space to a correlation separation in angular space, $\Lambda$:

\begin{align}
    \Lambda = \frac{\pi}{2 I}.
    \label{eq:scale}
\end{align}

This correlation separation gives us the information of the angular separation at which the power spectra undergo breaks, which can then be compared to the characteristic break scales obtained from the best-fit broken power laws for the structure functions of the simulated data from Section \ref{sec:data}.

Another set of quantities that are worth investigating are the minimum ($\Lambda_{\rm{min}}$) and maximum angular ($\Lambda_{\rm{max}}$) scales that each technique reconstructs accurately. For each technique the maximum angular separation simply corresponds to the maximum distance between two points on the grid which in both simulated foreground cases is $\sim 18$ deg. For the minimum scale, we will use the power spectra. In particular, we will define the minimum angular scale to correspond to the point in $k-$space where the reconstruction power is not within $20\%$ of the power of the simulated foreground (once the point in $k-$space is found, the minimum angular separation will be calculated using an equation analogous to Equation \ref{eq:scale}). Further, we will weight these minimum angular separations by the average angular separation between data points, $\bar{\Lambda}$:

\begin{align*}
    \bar{\Lambda} = \frac{1}{\sqrt{\bar{\rho}}},
\end{align*}
where $\bar{\rho}$ is the mean density of the data points.

The goodness-of-fit parameters (along with the correlation scales and minimum angular scales) for the patchy and filamentary sky are shown in Tables \ref{tab:3} and \ref{tab:4}; a lower $I_\mathrm{res}$ value indicates a more accurate reconstructed RM map. In the patchy simulated foreground, the accuracies from best to worst are: BRMS, NNI, TPS, IDW2, IM and IDW1. In the filamentary simulated foreground, the accuracy is: BRMS, NNI, TPS, IM, IDW2, and IDW1. 

Further, we can compare the correlation length of the reconstructed RM maps, $\Lambda$, to the break scales of the SFs, $\Delta \theta_{\rm{break}}$. For the patchy simulated foreground, we see that the techniques from best to worst in capturing the break scale of the SF are: BRMS, NNI, TPS, IM, IDW2, and IDW1. In the filamentary simulated foreground, the reconstruction best capturing the break-scales are: TPS, IM, NNI, BRMS, IDW2, and IDW1. Again, note that the correlation scales for the best techniques are skewed to around 6 deg (similar to the SF in Section \ref{sec:data}), likely due to the increase in structure at that angular scale.

From the minimum angular scales, we see that the techniques that reconstruct information on the smallest scales from best to worst for the patchy foreground are: BRMS, NNI, TPS, IM, IDW2 and IDW1; and the best technique in the filamentary foreground is TPS, the worst are IDW2 and IDW1, and IM, NNI and BRMS are all mediocre.

\begin{table*}[!htb]

\centering
\textbf{Patchy Foreground}

    \begin{tabular}{|c|c|c|c|}
        \hline
        \textbf{Technique} & $I_\mathrm{res}$ & $\Lambda$ \textbf{(deg)} & $\Lambda_{\rm{min}}/\bar{\Lambda}$\\
        \hline 
        \hline
        IDW1 & $2.71 \times 10^{-1}$ & 51.5 & 9.77\\
        IDW2 & $3.85 \times 10^{-2}$& 6.28 & 9.77\\
        NNI & $(1.107 \pm 0.005) \times 10^{-2}$& $3.598 \pm 0.008$ & $1.40\pm 0.02$\\
        IM RBF  & $(3.10 \pm 0.48)\times 10^{-2} $& $3.45 \pm 0.03$ & $1.10 \pm 0.06$\\
        TPS RBF & $(3.03 \pm 0.02) \times 10^{-2}$& $3.43 \pm 0.01$ & $1.22 \pm 0.03$\\
        BRMS & $(0.82 \pm 0.14) \times 10^{-2}$& $3.66\pm 0.04$ & $0.75\pm 0.05$\\
        \hline
    \end{tabular}
\caption{The goodness-of-fit parameters, the correlation separation and the weighted minimum angular scales that were accurately reconstructed for the interpolation kernels for a patchy simulated foreground RM. Lower values of $I_\mathrm{res}$ indicate a more accurate reconstruction. We decided to exclude calculation of the error for IDW techniques due to the excessive computational cost.}
    \label{tab:3}
\end{table*}

\begin{table*}[!htb]
\centering
{\textbf{Filamentary Foreground}}

    \begin{tabular}{|c|c|c|c|}
        \hline
        \textbf{Technique} & $I_\mathrm{res}$ & $\Lambda$ (\textbf{deg}) & $\Lambda_{\rm{min}}/\bar{\Lambda}$\\
        \hline 
        \hline
        IDW1 & $1.78 \times 10^{-1}$ & 214 & 9.77\\
        IDW2 & $3.52 \times 10^{-2}$ & 13.8 & 9.77\\
        NNI & $(0.66 \pm 0.01) \times 10^{-2}$ & $6.83\pm 0.02$ & $0.75 \pm 0.03$\\
        IM RBF & $(1.40 \pm 0.18) \times 10^{-2}$ & $6.67\pm 0.03$ & $1.40 \pm 0.10$\\
        TPS RBF & $(0.77 \pm 0.02) \times 10^{-2}$ & $6.26 \pm 0.02$ & $0.41 \pm 0.03$\\
        BRMS & $(0.55 \pm 0.02)\times 10^{-3}$ & $6.64\pm 0.04$ & $0.65\pm 0.04$\\
        \hline
    \end{tabular}
\caption{The goodness-of-fit parameters , the correlation separation and the weighted minimum angular scales that were accurately reconstructed for the interpolation kernels in a filamentary simulated foreground RM. Lower values of $I_\mathrm{res}$ indicate a more accurate reconstruction. We decided to exclude calculation of the error for IDW techniques due to the excessive computational cost.}
    \label{tab:4}
\end{table*}

\subsection{RM structures without extragalactic RM}
\label{sec:noegal}
It is also instructive to look at the results of the interpolation techniques in the ideal case, where we have a perfect algorithm to remove all extragalactic RM contribution from the data. We obtained similar RM maps as above (see Appendix \ref{app:a}). 

\begin{figure*}
\centering
    \large\textbf{Residual Histograms (No Extragalactic RM)}
\gridline{\includegraphics[width = 0.49\textwidth]{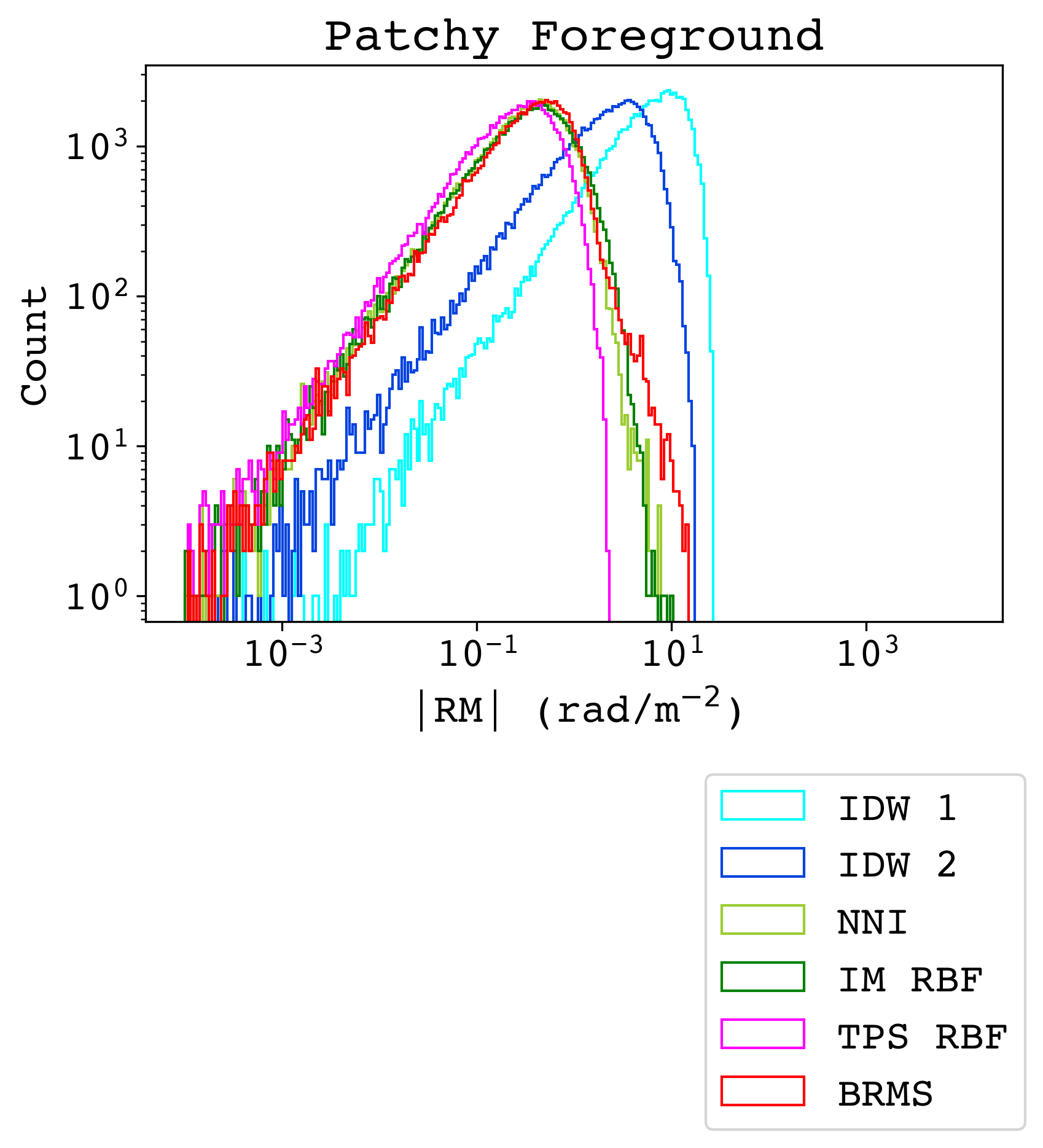} \includegraphics[width = 0.49 \textwidth]{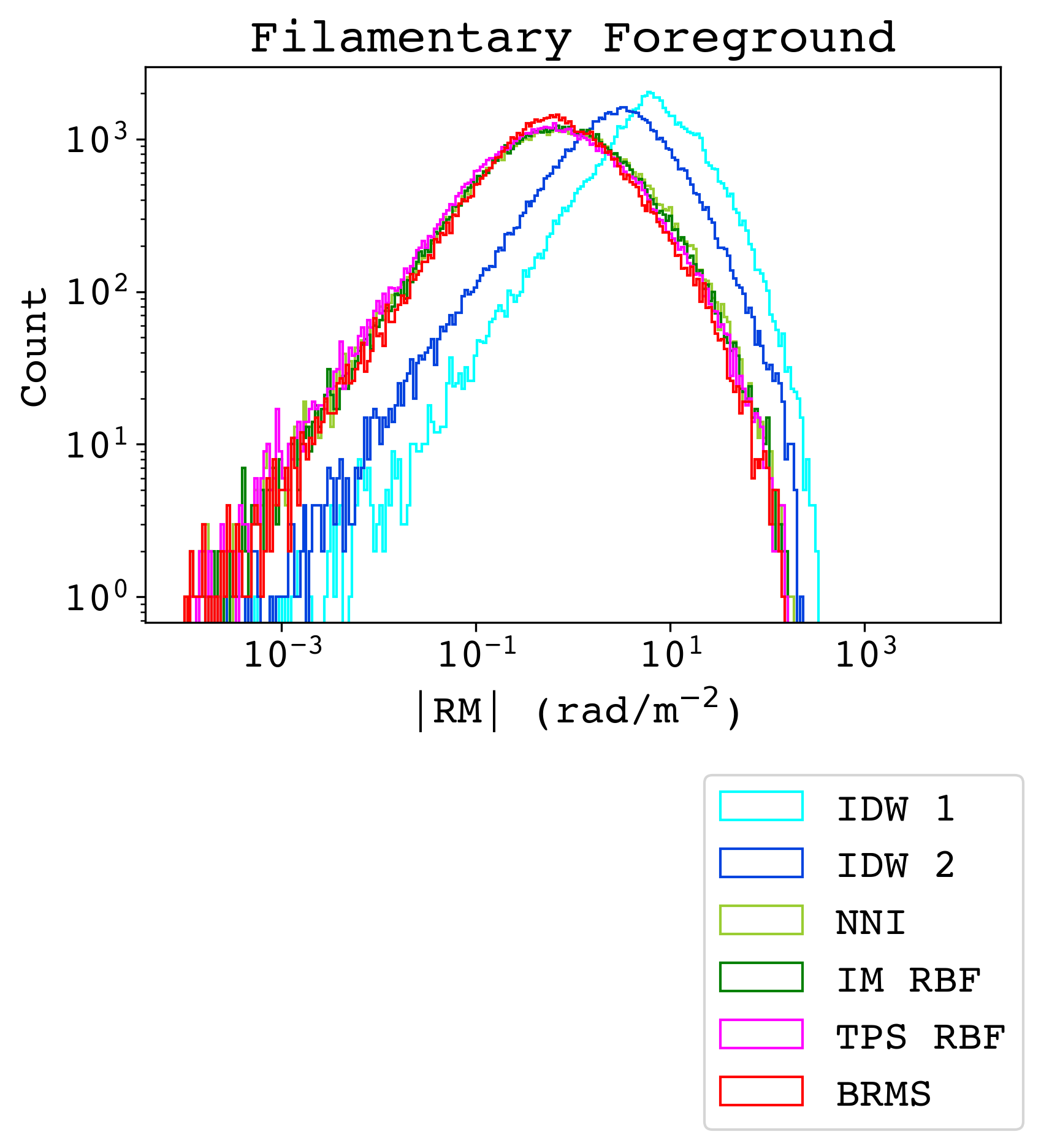}}
    \caption{A histogram of the magnitude of the residual RM for all the techniques applied for the patchy RM sky (left) and a histogram of the magnitude of the residual RM for all reconstruction applied to the filamentary RM sky (right) with the data sets having no extragalactic RM contributions. }
    \label{fig:hist_gal}
\end{figure*}

The histograms of the residuals are displayed in Figure \ref{fig:hist_gal}. Both IDW techniques seem to perform similar to before. The biggest difference appears in BRMS, NNI and the RBF maps, all of which significantly improve, specifically in the filamentary foreground RM.

\begin{figure*}
    \centering \large\textbf{No Extragalactic RM}\gridline{\includegraphics[width = 0.49\textwidth]{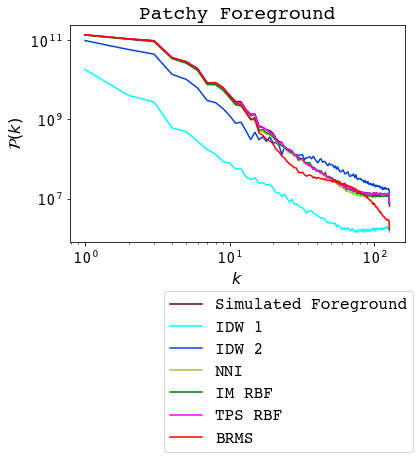} \includegraphics[width = 0.49 \textwidth]{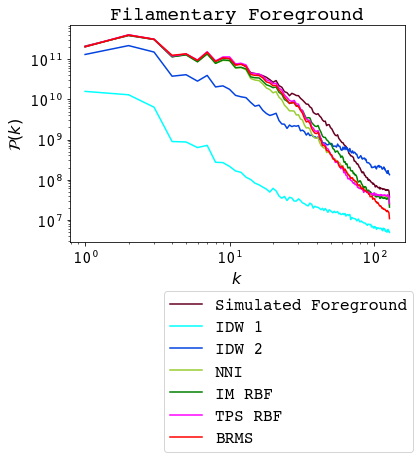}}
        \gridline{\includegraphics[width = 0.49\textwidth]{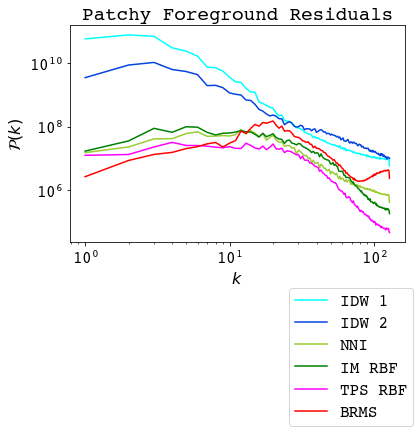} \includegraphics[width = 0.49 \textwidth]{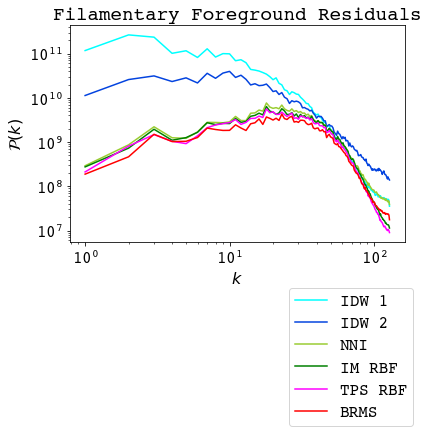}}
    \caption{The top figures display the power spectra of the reconstructed RM maps and the simulated foreground for the patchy sky (left) and the filamentary sky (right), and the bottom two figures display the power spectra of the residuals of the reconstructed RM maps for the patchy sky (left) and the filamentary sky (right), with both data sets having no extragalactic contributions.}
    \label{fig:power_gal}
\end{figure*}

Figure \ref{fig:power_gal} displays the power spectra of the reconstructed RM maps when the data sets have no extragalactic contributions. From the power spectra of the residuals, we see that as the angular scale decreases the accuracy of all the RM maps increases, except for BRMS which has a decreasing accuracy for very small angular scales. 

Tables \ref{tab:patchy_gal} and \ref{tab:fil_gal} present the accuracy of the interpolation kernels for the data sets without extragalactic contributions. Using the goodness-of-fit parameter, $I_\mathrm{res}$, we see that the accuracy of the techniques in patchy RM structures from best to worst is: TPS, BRMS, NNI, IM, IDW2, and IDW1; while in the filamentary structures it is: BRMS, TPS, IM, NNI, IDW2, and IDW1.

Plots of the computational time against the accuracy of the interpolation kernel for the different RM structures (with and without extragalactic contributions to the data points) is included in Appendix \ref{app:b}.

\begin{table*}[!htb]
\centering
\textbf{Patchy Foreground (No Extragalactic RM)}
    \begin{tabular}{|c|c|c|c|c|}
        \hline
        \textbf{Technique} & $|\langle \mathbf{Res RM} \rangle |$/$\mathbf{\sigma_{\mathrm{\textbf{sim}}}}$ & $I_\mathrm{res}$ & $\Lambda$ (\textbf{deg}) & $\Lambda_{\rm{min}}/\bar{\Lambda}$\\
        \hline 
        \hline
        IDW1 & $0.65\pm 0.47$  & $2.70 \times 10^{-1}$ & 51.0 &  9.77\\
        IDW2 & $0.26 \pm 0.21$ & $3.54 \times 10^{-2}$ & 6.26 & 9.77\\
        NNI & $0.040\pm 0.039$ & $(3.680\pm 0.005) \times 10^{-4}$ & $3.62\pm 0.06$ & $0.814\pm 0.01$\\
        IM RBF & $0.045 \pm 0.040$ & $(5.08\pm 0.9) \times 10^{-4}$ & $3.68\pm 0.01$ & $0.698\pm 0.030$\\
        TPS RBF & $0.030\pm0.026$ & $(1.87\pm 0.08) \times 10^{-4}$ & $3.571\pm 0.009$ & $0.0763\pm 0.03$\\
        BRMS  & $0.00\pm 0.06$ & $(3.65\pm 0.03) \times 10^{-4}$ & $3.57\pm$ & $0.698\pm 0.020$\\
        \hline
    \end{tabular}
\caption{The mean of the absolute value of the residual RMs (normalized by the standard deviation in the simulated foreground RM), the goodness-of-fit parameters, the correlation separations, and the weighted minimum angular scales that were accurately reconstructed for the interpolation kernels for a patchy simulated foreground RM with data sets having no extragalactic contributions. We decided to exclude calculation of the error for IDW techniques due to the excessive computational cost.}
    \label{tab:patchy_gal}
\end{table*}

\begin{table*}[!htb]
\centering
\textbf{Filamentary Foreground (No Extragalactic RM)}
    \begin{tabular}{|c|c|c|c|c|}
        \hline
        \textbf{Technique} & $|\langle \mathbf{Res RM} \rangle |$/$\mathbf{\sigma_{\mathrm{\textbf{sim}}}}$  & $I_\mathrm{res}$ & $\Lambda$ (deg) & $\Lambda_{\rm{min}}/\bar{\Lambda}$\\
        \hline 
        \hline
        IDW1 & 0.497 & $1.78 \times 10^{-1}$ & 215 & 9.77\\
        IDW2 & 0.254 & $3.48 \times 10^{-2} $ & 13.9 & 9.77 \\
        NNI & 0.105 & $(4.57\pm 0.05) \times 10^{-3}$ & $6.64\pm 0.01$ & $0.70\pm 0.040$\\
        IM RBF & 0.0980 & $(4.04\pm 0.5) \times 10^{-3}$ & $6.67 \pm 0.02$ & $0.70\pm 0.030$\\
        TPS RBF & 0.0892 & $(3.63\pm 0.06) \times 10^{-3}$ & $6.36 \pm 0.02$ & $0.39 \pm 0.01$\\
        BRMS  & 0.0814 & $(2.93\pm 0.01) \times 10^{-3}$ & $6.45 \pm 0.02$ & $0.54 \pm 0.02$\\
        \hline
    \end{tabular}
\caption{The mean of the absolute value of the residual RMs (normalized by the standard deviation in the simulated foreground RM), the goodness-of-fit parameters, the correlation separations, and the weighted minimum angular scales that were accurately reconstructed for the interpolation kernels for a filamentary simulated foreground RM with data sets having no extragalactic contributions. We decided to exclude calculation of the error for IDW techniques due to the excessive computational cost.}
    \label{tab:fil_gal}
\end{table*}

\section{Discussion}
\label{sec:discussion}
The different interpolation techniques yielded varied degrees of accuracy and computational efficiency. The interpolation techniques that yielded the most accurate RM maps in both patchy and filamentary simulated foregrounds were BRMS and NNI.

\subsection{IDW}
For IDW, the interpolation took an average time of  $\sim$ 6000 s across all the tests, with the reconstruction in the filamentary RM sky taking a few more minutes than the patchy sky for both values of the $p$ parameter. The reason for this incredibly large computational time is because IDW needs to independently calculate the weights of all the data points at each interpolation point (and repeat this process for all the interpolation points). For $N$ data points and $M$ interpolation points, the time complexity is $\mathcal{O}(N\times M)$. Because of this, simply adding an additional data point increases the computational time by a factor of $M$, and adding an additional interpolation point increases it by a factor of $N$. For these reasons, IDW is inadvisable for an interpolation over a large number of data points or for a high resolution reconstruction.

The most apparent feature of the IDW1 reconstructions is the incredibly low $|\rm{RM}|$ throughout the interpolation region for both filamentary and patchy foregrounds. The reason for this is that for each interpolation point IDW1, weighs the RM of every data point as $1/r$. Because of this relatively small dependence on $r$, data points far away from the interpolation point are also given a significant weight. And this results in all the data points have weights of the same order of magnitude. Since the regions chosen has roughly equal positive and negative RMs, this leads the sum of all the weighted data point RMs to be very close to zero at every interpolation point. This issue is no longer present for IDW2 because the weights now go as $1/r^2$ so far away data points are weighted far less than before. It must also be noted that if a high enough $p$ is being used, we may ignore points outside a critical radius where the weight is approximately zero to increase computational efficiency. 

From the residuals presented in Section \ref{sec:analysis}, IDW more accurately models the RM sky with a power parameter $p = 2$. The reason for this is because data points farther away are weighted to have less effect on an interpolation point with a higher power parameter.  Increasing the power parameter, causes IDW to transition from a global interpolation kernel to a local interpolation kernel. This might be useful to model filamentary RM structures because of their sharp, small scale RM variations.  

An artefact that is quite visible on both of the IDW techniques is that the RM map produced has sharp peaks at the data points, as can be seen in Figure \ref{fig:4}. This is likely because the algorithm calculates the norm $||x_i-x||$ for interpolation points close to $x_i$ to be equal to zero. Because of this, all of the points in some neighbourhood around a data point are set to have a constant RM value. A possible way to improve this is to use arrays that can store larger numbers. However, this would still produce a similar large scale RM map that is just simply smoother, while also requiring greater computation power and time.

\subsection{NNI}

NNI had a run-time of roughly 350 seconds for both RM skies and reconstructs almost all the large scale RM structures accurately, but faces some issues with the smaller scale RM structures as displayed in Figures \ref{fig:12} and \ref{fig:13}. 

\begin{figure}[!htb]
    \includegraphics[width = 0.47\textwidth]{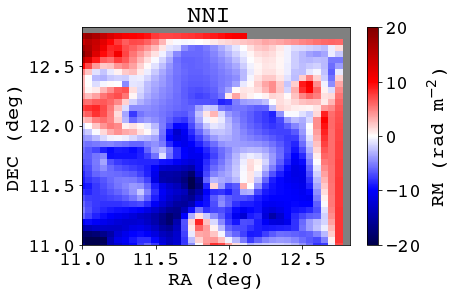}
    \caption{Zooming into a corner of the NNI RM reconstruction, highlighting the \pkg{nan} values present at the edges of the interpolation region. The \pkg{nan} values in this case are the grey pixels along the top-right corner.}
    \label{fig:8}
\end{figure}

As noted in Section \ref{sec:results}, at the edges of the interpolation region, NNI produces \pkg{nan} values, as can be seen in Figure \ref{fig:8}.

\begin{figure}[!htb]
    \includegraphics[width = 0.47\textwidth]{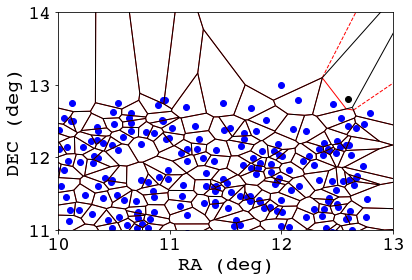}
    \caption{A plot of the Voronoi diagram for an interpolation point (in black) at the edges of the grid. The edges of the Voronoi polygons of the data points (in blue) are shown with black lines, and the Voronoi polygon for the interpolation point is shown in red. A dashed line represents an edge that extends to infinity.}
    \label{fig:9}
\end{figure}

This artefact is caused by the structuring of the Voronoi diagrams. For an interpolation point at the edges of the interpolation region, the Voronoi diagram is shown in Figure \ref{fig:9}, with the unbounded Voronoi polygon of an interpolation point shown in red.

\begin{figure}[!htb]
    \includegraphics[width = 0.47\textwidth]{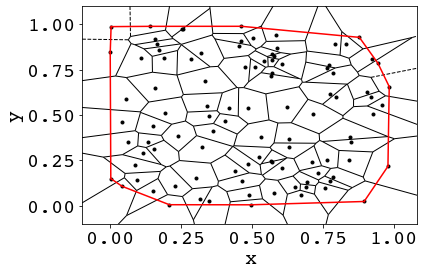}
    \caption{A plot of the convex hull (in red) of a random set of points. Voronoi edges that are unbounded are illustrated with dotted lines.}
    \label{fig:10}
\end{figure}

To classify where and why these unbounded polygons occur, we introduce the \textit{convex hull} of the data points to be the intersection of all \textit{convex sets} that contain the data points \citep{10.1093/biomet/52.3-4.331}. Here, we define a \textit{convex set} to be a set such that a straight line between any two points in this set is contained in the set  \citep{klee1971convex}. Then, we note that points that lie inside the convex hull have bounded Voronoi polygons, while points outside the convex hull have unbounded Voronoi polygons \citep{Liang2010ASA}, as illustrated in Figure \ref{fig:10}.

Therefore, interpolation points outside this convex hull have an infinite captured area $A_i$, so we cannot use Equation \ref{eq:6} to calculate the weights. Therefore, NNI fails at such points. This issue occurs with similar spatial interpolation kernels. To have an accurate comparison between the RM maps, we restrict ourselves to the section where NNI does not obtain \pkg{nan} values.

For real data, a possible way to avoid \pkg{nan} values in the reconstruction from NNI is to reduce the size of the interpolation region, so that the interpolation region lies inside the convex hull of the data points, or conversely to pad the data set with additional data points such that the new convex hull completely contains the interpolation region. In regions of the sky that are close to the edge of the POSSUM field or near POSSUM tiles that are yet to be observed, an alternate approach can be applied using `ghost points' \citep{afe98d6abb5c4e01bc5c4ddcecad657f} but further study of this for RM data remains out of the scope of this paper.

Another artefact that was noted with NNI (and with both RBF techniques) in the patchy sky is that at low magnitude RMs these algorithms present fluctuations in the RM that were not present in the input foreground RM. These fluctuations arise due to extragalactic RM. They are not very distinguishable at regions with high RM magnitudes because the extragalactic RM is comparatively smaller; however, in regions with RM close to zero, the extragalactic RM is prevalent. This can be resolved by having a more robust algorithm to deal with extragalactic RM. Further, this artefact is only present in the NNI and the RBF techniques because these techniques are local (i.e they only take into consideration points in a local neighbourhood such as in NNI or they strongly down-weight more distant points, as in TPS). We can see this artefact arising in Figure \ref{fig:4} in IDW as the power parameter is increased from 1 to 2. 

Being a local kernel, NNI works particularly well in the case of a filamentary RM sky. The rapidly fluctuating RM over small scales between the filaments is particularly difficult for a more global technique, such as IDW, to accurately reconstruct. In addition, the reason why the aforementioned artefact does not show up in the filamentary case is because in regions containing LOS magnetic field reversals, the large RM magnitude of the filaments far outweighs any effects that the extragalactic RM has on the reconstruction (a similar effect occurs for both the RBF techniques).

\subsection{RBF}
Each of the RBF interpolation techniques took roughly 20 seconds to run and are, therefore, the most computationally efficient interpolation techniques out of the ones we consider.

\subsubsection{IM} The largest artefact that is present in the IM RBF approach are the peaks in the reconstruction in regions where the RM is relatively uniform, as displayed in Figure \ref{fig:4}. It is also these same peaks that also heavily skew the accuracy of the IM reconstruction (even when using the accuracy metric from Equation \ref{eq:integral} that takes into account the angular scales), making its accuracy worse than IDW2. Once again, these peaks mostly occur near regions of RM sign reversal. This is due to the use of a constant shape parameter (the shape parameter depends on what the expected RM map looks like; if the map has varying structure, then the shape parameter should also reflect this). While out of the scope of this work, one could explore the use of Variably Scaled Kernels (VSKs) \citep{rossini2018interpolating} to resolve this issue, where a scaling function $c(x, y)$ is introduced and the interpolation problem is now solved on the space $\{(x, y, c(x, y)):  (x, y) \in \Omega \}$, where $\Omega$ is the interpolation region being considered. This artefact is not prevalent in the filamentary case for the same reasons as outlined previously: the large RM magnitude of the filaments far outweighs any irregularities that occur near RM sign reversals. 

It is also these artefacts in the local interpolation kernels (RBF, as well as NNI) that lead to the residual histogram and the mean residual being inaccurate measures of the accuracy, as a few erroneous interpolated values that are far from the simulated foreground value skew the histogram. For this reason, the power spectrum and its $1/k$ weighted integral are the better ways to measure the accuracy, as they also contain information about the angular scales.

\subsubsection{TPS} 
Compared to IM RBF, TPS produces far fewer peaks in the case of the patchy RM sky. The peaks that are formed are likely once again due to the choice of the smoothing parameter. Using VSKs would help resolve this issue as well. In the filamentary RM sky, TPS produces the most accurate reconstruction.

Additionally, TPS performs particularly well in the filamentary foreground (which is further accentuated in the case without extragalactic RM). In particular, it must be noted that it most accurately captures the break scales of the SF and reconstructs the RM map accurately to the smallest angular scales across all techniques in both RM structures. This makes it a good candidate for the reconstruction of RM structures that are known to be filaments. However, even for filamentary structures, we recommend using it in conjunction with another technique such as BRMS or NNI. This is because if there are any patchy RM structures present, a reconstruction that purely relies on TPS would likely loose a lot of information, particularly if there is some extragalactic contamination. In addition to this, from Table \ref{tab:patchy_gal}, we see that TPS in the patchy foreground without extragalactic contamination, reconstructs the RM accurately at the smallest possible scales; in fact it must be noted that this is the angular scale that corresponds to the highest possible $k$ value for the grid we have used and therefore TPS was mostly accurate down to each individual pixel.

\subsection{BRMS}

BRMS took an average time of $\sim$ 600~s to run on each of the RM structures. However, this is while using the extragalactic RM removal technique described in Section \ref{sec:data}. When using the built-in extragalactic RM removal in BRMS the run-time increases by roughly 3 times. In addition to this, there are hyperparameters in the optimization process involved in BRMS and adjusting these will also greatly affect the computational cost. It must also be noted that with the newer version of \pkg{NIFTy}, \pkg{NIFTy.re} \citep{niftyre}, the computational cost will decrease significantly.

Of all the techniques, BRMS gives the most accurate results in the patchy RM sky and the filamentary RM sky and deals well with the presence of extragalactic RM. It must also be noted that based on Tables \ref{tab:3} and \ref{tab:4}, BRMS has a wider margin to the other interpolation techniques in the case of the patchy foreground RM. The reason for this is likely that a turbulent ISM with a Mach number of 0.1 has been shown to be Gaussian by \citet{2021MNRAS.502.2220S} and the Gaussian priors used for BRMS favour such RM structures. For a filamentary RM structure, it is possible that the performance can be improved by using a non-Gaussian prior and is something that can be explored in future work.

\subsection{Local vs. Global Interpolation Kernels}

The biggest difference between local (techniques that significantly weigh data points within a neighbourhood of around $\lesssim$0.5 deg of the interpolation point, e.g NNI and RBF) and global (techniques that significantly weigh all data points) is that global interpolation kernels take far longer to run. The primary reason for this is that global interpolation kernels take into account every data point to estimate the RM at the interpolation point; while local interpolation kernels only take data points within a close neighbourhood of the interpolation point.

Another difference between the two types of kernels is that local interpolation kernels, on average, seem to improve in accuracy far more than global interpolation kernels when extragalactic contributions are not present (with the RBF techniques improving the most and accuracy of IDW techniques being relatively unchanged). The reason for the dramatic improvement in the accuracy of local kernels is because points with significant extragalactic contributions are given far more weight than in global kernels; when extragalactic contributions are removed these points no longer skew the estimated RM of the interpolation point. This improvement is most apparent in the lack of `peaks' in the reconstructed maps for the RBF techniques (see Appendix \ref{app:a}), when compared to the previous case containing extragalactic RMs.

\subsection{Other Interpolation Kernels}

A similar approach can be applied to test other interpolation kernels. We've avoided investigating the other interpolation algorithms that have been used in previous works for various reasons, for the few we've mentioned before, the reasoning is as follows: the median technique developed by \citet{taylor2009rotation} is more rudimentary than IDW and would likely give a worse accuracy (however, the computational time would be considerably less); the wavelet interpolation \citep{Frick_2001} relies on a similar concept of basis functions as RBF and also uses the angular separation between points to set the basis functions. The spherical harmonics technique used by \citet{Dineen} would likely yield similar results to BRMS as it also uses Gaussian priors. 

\subsection{Drawbacks of Interpolation}
The greatest inherent drawback of any interpolation technique is the density of the data points (refer to Appendix \ref{app:e} for further discussion regarding data density) and the accuracy of the data. For example, consider a grid with two data points that have the same magnitude RM but different signs. In this case, all interpolation techniques would predict that the RM at a point equidistant from the two data points has an RM of 0 rad m$^{-2}$. However, this might not necessarily be the case as the underlying RM structure might be more complicated than this. For this reason, having an increased RM density is crucial for accurate RM maps, which is why large radio surveys like POSSUM are crucial. It must also be noted that the data density can be used to weigh the errors in the reconstructed RM, as done by \citet{Hutschenreuter_2022}.

\subsection{Quantifying filamentariness of RM structures}

It must be noted that in this work we have categorized the RM structures to be patchy or filamentary based on visual inspection. There have been attempts to quantify how filamentary magnetic fields are using Minkowski functionals \citep{2020PhRvF...5d3702S}. This might be of interest in future works as the accuracy of some of the interpolation techniques is similar for both RM structures, therefore these structures that appear to be visually distinct might have similarities in their statistical properties. 

\section{Conclusions}
\label{sec:conclusion}
In this paper, we have tested different priors for interpolating RM data. To do this, we have implemented the following interpolation algorithms: IDW, NNI, IM RBF, TPS RBF, and BRMS.

We tested these techniques over simulated data sets with an area of $\sim$ 144 deg$^{2}$; the first data set modelled a patchy RM sky (turbulent Mach number of 0.1), and the second modelled a filamentary RM sky (turbulent Mach number of 10). Initially, the patchy RM sky contained 6459 data points and the filamentary RM sky contained 6556. We used these data sets under two conditions: the first with extragalactic RM of mean 0 rad m$^{-2}$ and standard deviation $7$ rad m$^{-2}$, and in the second without any extragalactic contribution. To refine the data with extragalactic contributions, we excluded all data points that were not within 3$\sigma$ of the mean of their ten closest data points. We note that the coefficient that is used for $\sigma$ in this removal of extragalactic RM is difficult to determine for real data and making the correct choice will greatly impact the interpolation result.  

In the case that the patchy foreground had extragalactic contributions, the accuracy from best to worst was: BRMS, NNI, TPS RBF, IDW2, IM RBF, and IDW1. For the filamentary simulated foreground with extragalactic RM, the performance of the techniques was BRMS, NNI, TPS RBF, IM RBF, IDW2 and IDW1. 

When there was no extragalactic contribution, the accuracy in the simulated patchy foreground from best to worst was: TPS RBF, BRMS, NNI, IM RBF, IDW2 and IDW1; while in the filamentary foreground it was: BRMS, TPS RBF, IM RBF, NNI, IDW2 and IDW1

The least computationally intensive algorithms were the RBF algorithms (both algorithms took $\sim 10$  seconds to run). However, both RBF techniques do not produce ideal results for data sets with extragalactic contributions. The most computationally intensive algorithm was IDW (taking $\sim 10^3$ seconds to run). 

The biggest issue with NNI is that it fails to produce a result outside the convex hull of the data set. On an all-sky data set (when considering a patch of the sky), this can be resolved by padding the FOV with data points outside the convex hull. 

An issue that was faced by NNI and both RBF techniques is that in regions with low RM magnitude (which usually occur in regions with line of sight magnetic field reversals), extragalactic RM contributions greatly affect the result. An effective solution to this would be to construct a more robust algorithm to remove data points dominated by extragalactic RM. 

We conclude by noting that the choice of the prior greatly affects the reconstructed RM map, as well as the computational cost that is required. Local interpolation kernels (NNI, IM RBF, TPS RBF, IDW with a high $p$ value) produce more accurate results in regions where there are sharp, small scale variations in the RM and there is relatively little extragalactic contribution. In addition, a robust algorithm to remove extragalactic contributions is essential for accurately reconstructing Galactic foreground RM. It must also be noted that over small scales, it is difficult to distinguish between extragalactic RM and the foreground RM. More global priors (as used in BRMS, and in IDW with low $p$ values) produce better results in regions of large scale RM variations and deal with extragalactic contributions better.

Large upcoming RM sky surveys such as POSSUM \citep{2024AJ....167..226V} will provide us with much more RM data over the entire southern sky. Large sky surveys such as the Low Frequency Array Two-metre Sky Survey \citep[loTSS;][]{2023MNRAS.519.5723O} and the MeerKAT Large Area Synpotic Survey \citep[MeerKLASS;][]{2016mks..confE..32S} either have provided or will also provide dense RM grids, though with smaller sky coverage than POSSUM. The work in this paper is done in advance of the analysis of data from these sky surveys that will provide us with a better understanding of the GMF, particularly on smaller scales. Again, we note that these interpolation techniques can be applied to any other 2 dimensional astronomical data with signed scalars. 

\section*{Acknowledgements}
The authors would like to thank the anonymous reviewers for their constructive comments that helped further develop more detail in this manuscript. The University of Toronto operates on the traditional land of the Huron-Wendat, the Seneca, and most recently, the Mississaugas of the Credit River; we are grateful to have the opportunity to work on this land. The Dunlap Institute is funded through an endowment established by the David Dunlap family and the University of Toronto. The POSSUM project (\url{https://possum-survey.org}) has been made possible through funding from the Australian Research Council, the Natural Sciences and Engineering Research Council of Canada (NSERC), the Canada Research Chairs Program, and the Canada Foundation for Innovation. 
A.P. is funded by the NSERC Canada Graduate Scholarships--Doctoral program.
S.H. acknowledges funding from the European Research Council (ERC) under
the European Union’s Horizon 2002 re-
search and innovation programme (grant agreement No. 101055318). B.M.G. acknowledges the support of the NSERC through grant RGPIN-2022-03163, and of the Canada Research Chairs program. SPO acknowledges support from the Comunidad de Madrid Atracción de Talento program via grant 2022-T1/TIC-23797, and grant PID2023-146372OB-I00 funded by MICIU/AEI/10.13039/501100011033 and by ERDF, EU.

\appendix

\section{Further discussion on interpolation techniques}
\label{app:discussion}
Most of the interpolation algorithms (aside from BRMS) that have been discussed in this paper, have been applied or been developed to study geoprocessing \citep{sambridge1995geophysical, yamamoto1998review, zimmerman1999experimental}. Most of the data used in these studies can be quantified as being `patchy' based on the standards that have been used in this paper. Therefore, the study of the performance of interpolation techniques (which is of particular interest in astronomy) for filamentary data in this paper is important.

For larger data sets, the processing time naturally grows for all the interpolation algorithms. It grows particularly fast with IDW, which we do not recommend using. After IDW, the cost grows fastest for NNI, as computing the Delaunay triangulations and Voronoi diagrams will become increasingly computationally intensive. \citet{Fan2005HardwareAssistedNN} introduced a GPU-based algorithm for NNI that interpolates across several interpolation points simultaneously, and according to their results can perform up to 10 times as quickly as a traditional NNI algorithm for large enough data sets. A later work by \citet{10.1145/1869790.1869817} further developed this GPU-based algorithm by avoiding interpolating over regions where there is a lack of data; this might be useful in regions where there is significant depolarization due to Faraday screens or other reasons \citep{1966MNRAS.133...67B, 1998MNRAS.299..189S, 2012MNRAS.421.3300O} in order to obtain a more accurate reconstruction. In addition to this, for their purposes, they have also increased the number of interpolation points that can be simultaneously computed to $10^6$, compared to \citep{Fan2005HardwareAssistedNN} who's algorithm could perform simultaneous computations of $\lesssim10^{2}$ points.
\citet{10.1145/1869790.1869817} have also utilized CUDA, allowing most of the computation to take place locally on GPU. For this reason, we suggest exploring a GPU algorithm for computationally expensive data sets; further investigating this remains out of the scope of this work and will be studied in a future work. For BRMS, the newest versions of Nifty \citep{niftyre} (the framework BRMS was
developed in) are trivially portable to the GPU, but this development
was too recent to be take into account in this work.

\section{Reconstructed RM Maps with no extragalactic RM}
\label{app:a}

\subsection{Patchy Foreground}
\begin{figure}[!htb]
\gridline{\fig{patchy.png}{0.33\textwidth}{}}
\gridline{\fig{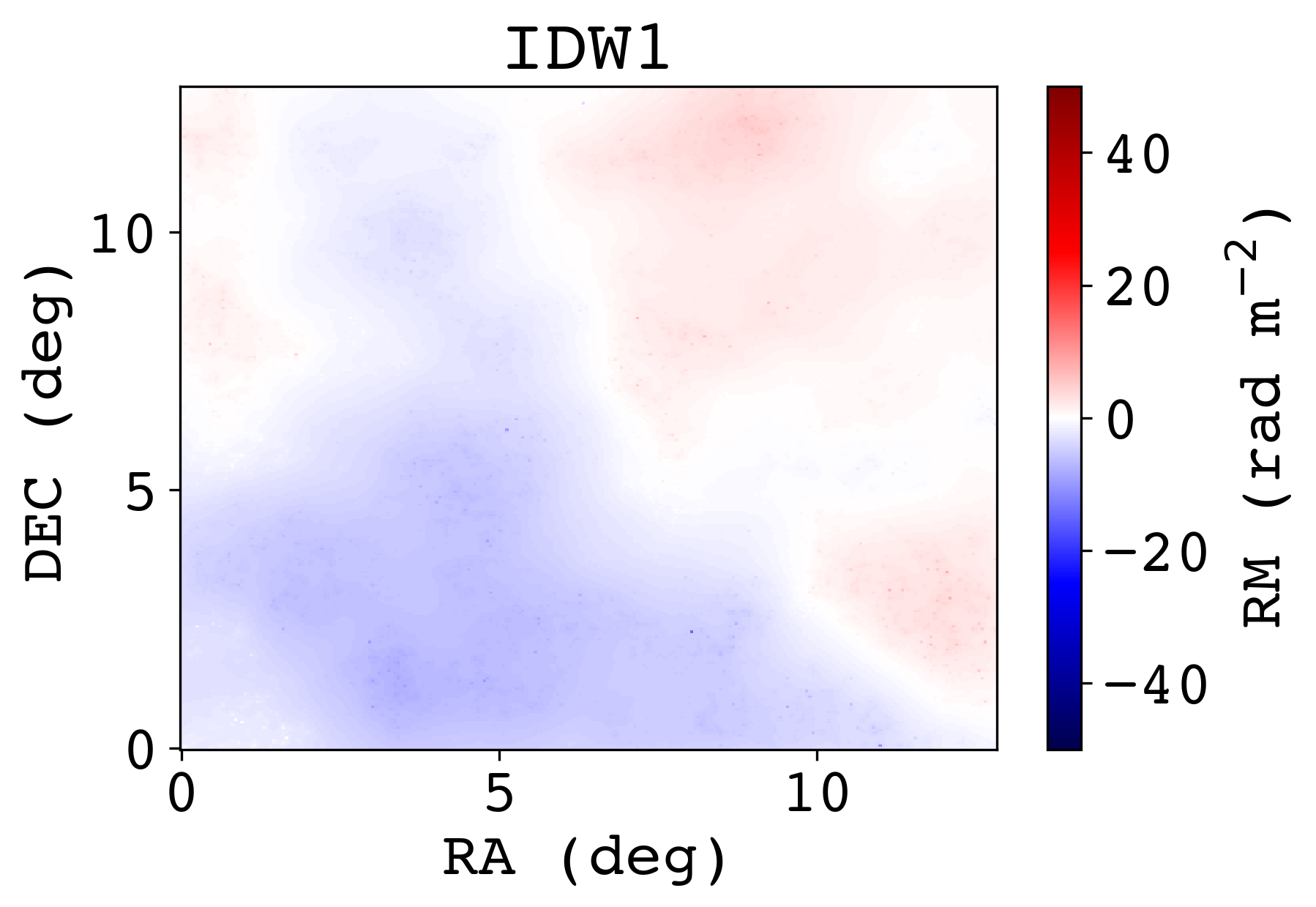}{0.33\textwidth}{}\fig{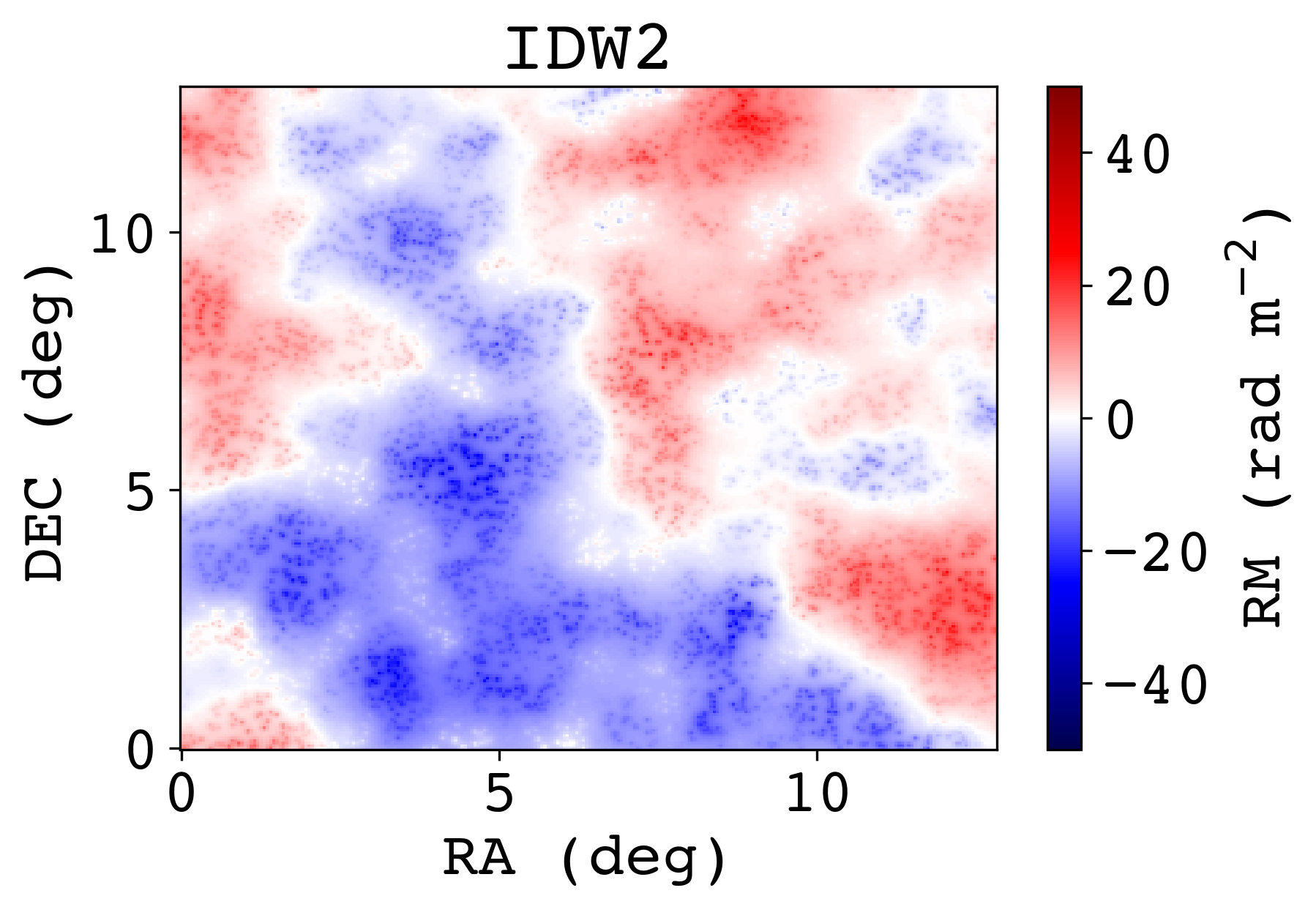}{0.33\textwidth}{}\fig{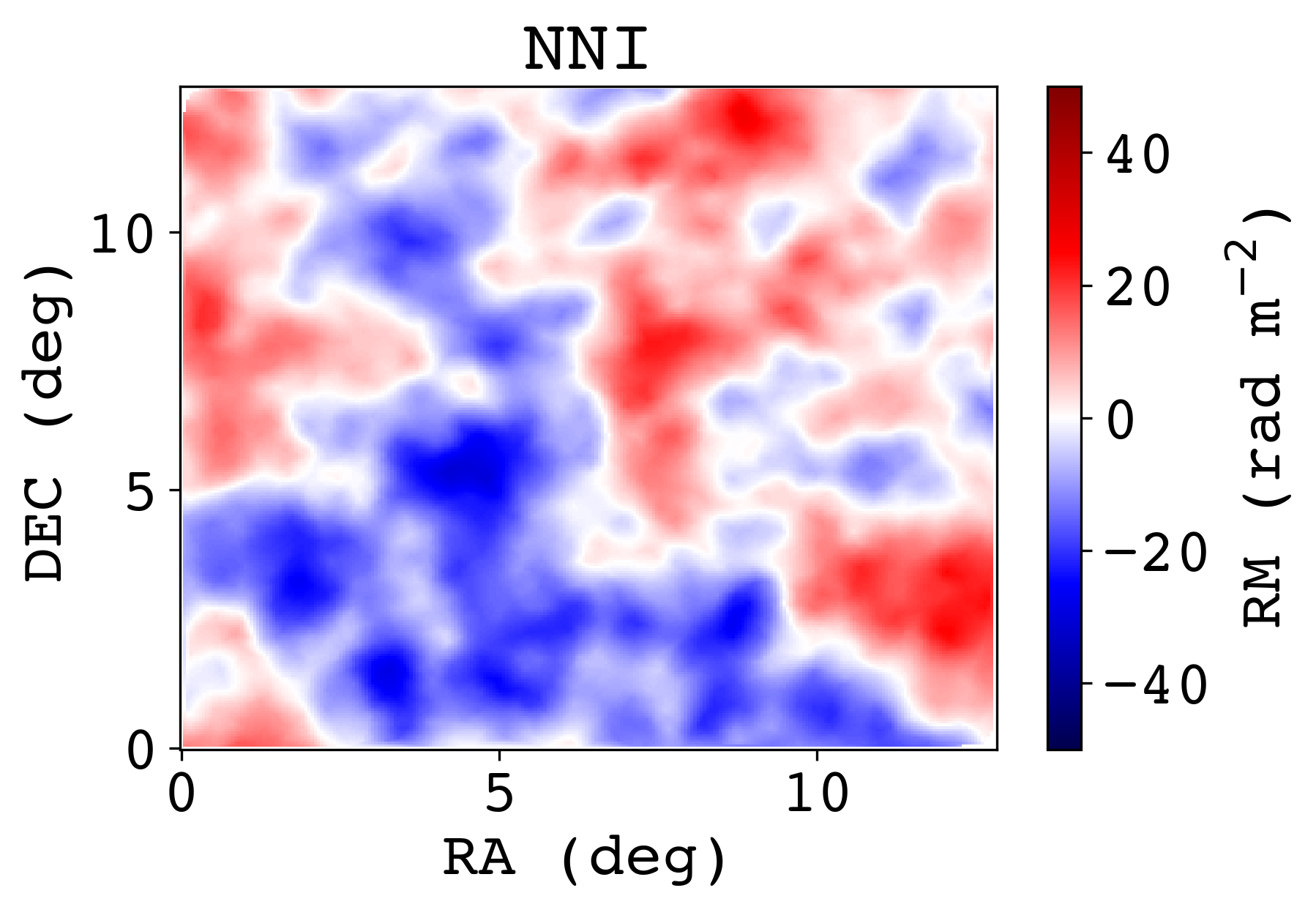}{0.33\textwidth}{}}
\gridline{\fig{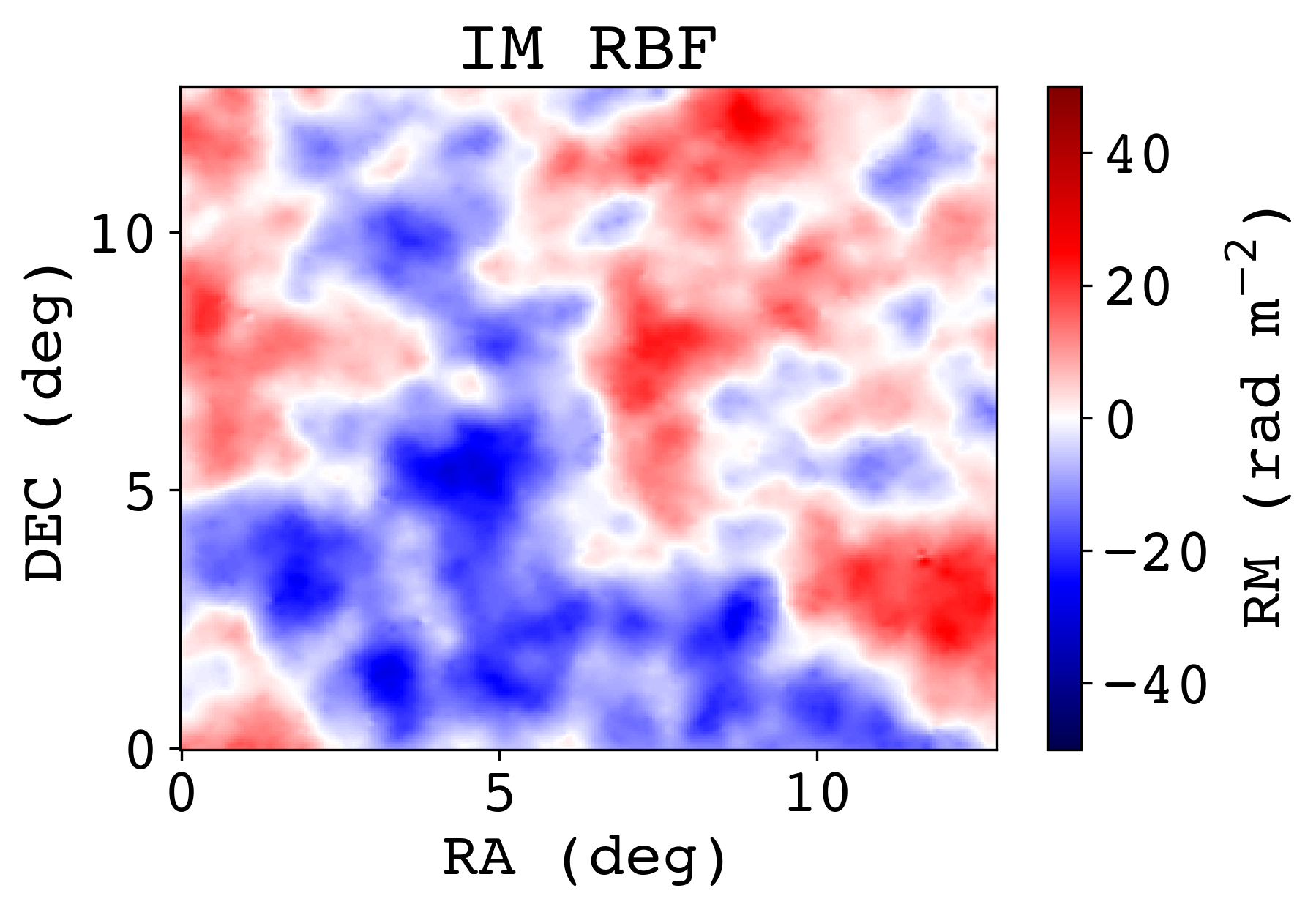}{0.33\textwidth}{}\fig{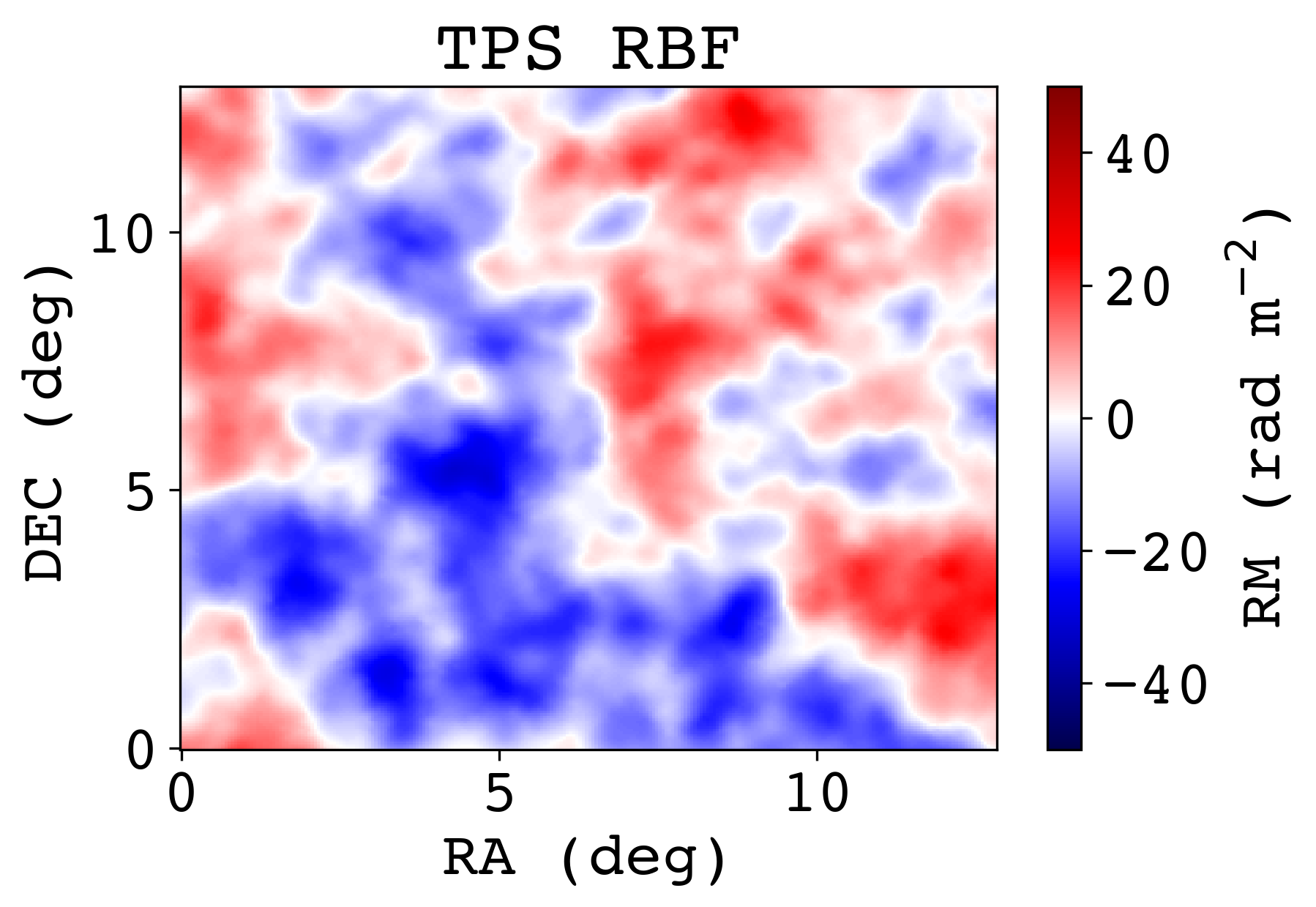}{0.33\textwidth}{}\fig{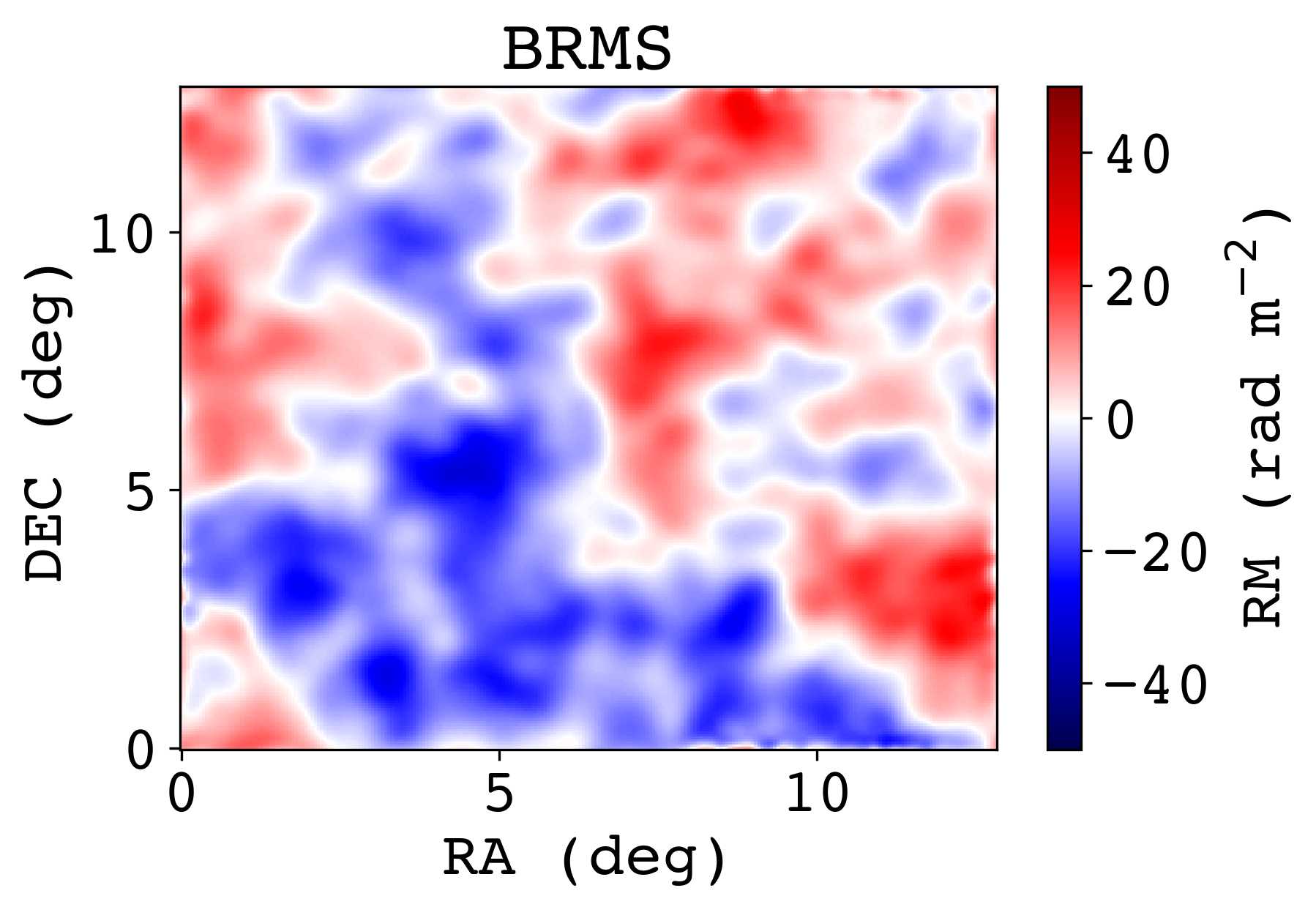}{0.33\textwidth}{}}
\caption{Reconstructed RM maps for each of the interpolation techniques for the patchy structures (with the data points having no extragalactic contributions), saturated from $-50$ to $+50$ rad m$^{-2}$. The color map represents RM, with red being positive and blue being negative. At the top-centre, we present the simulated foreground RM. From top-left to bottom-right, the reconstructions are as follows: IDW1, IDW2, NNI, IM RBF, TPS RBF, BRMS.}
\label{fig:app1}
\end{figure}

Figure \ref{fig:app1} display the reconstructed RM maps for a patchy simulated foreground without any extragalactic contribution, saturated  between $- 50$ and $+50$ rad m$^{-2}$. 

Comparing these RM maps to the ones from Section \ref{sec:results}, the maps produced by BRMS and both IDW techniques do not have any notable differences. The maps that do significantly improve are the ones produced by NNI and the RBF techniques. Most notably, the sharp peaks that had been earlier noted are far fewer in the RBF RM maps.

\begin{figure}[!htb]

\gridline{\fig{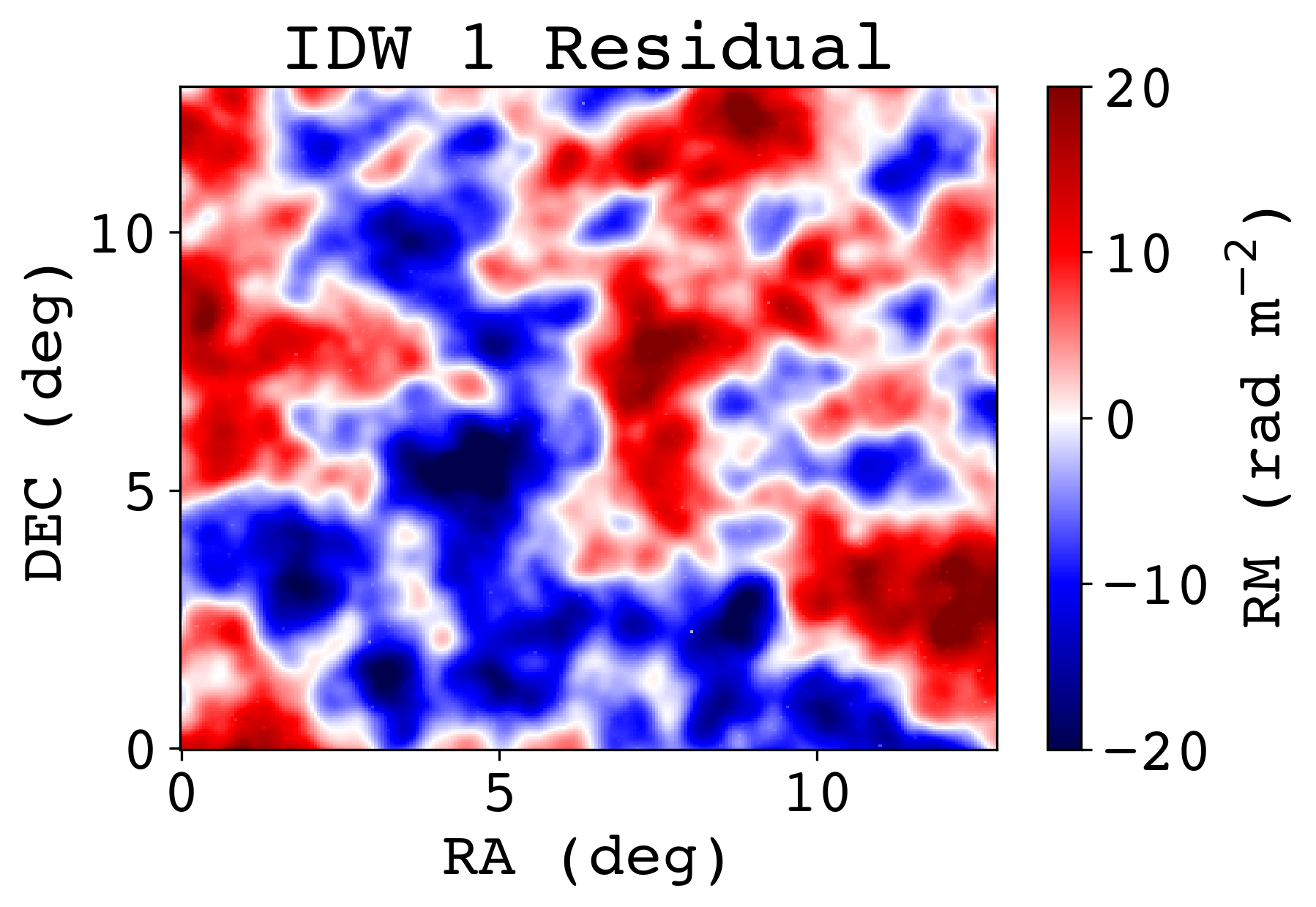}{0.33\textwidth}{}\fig{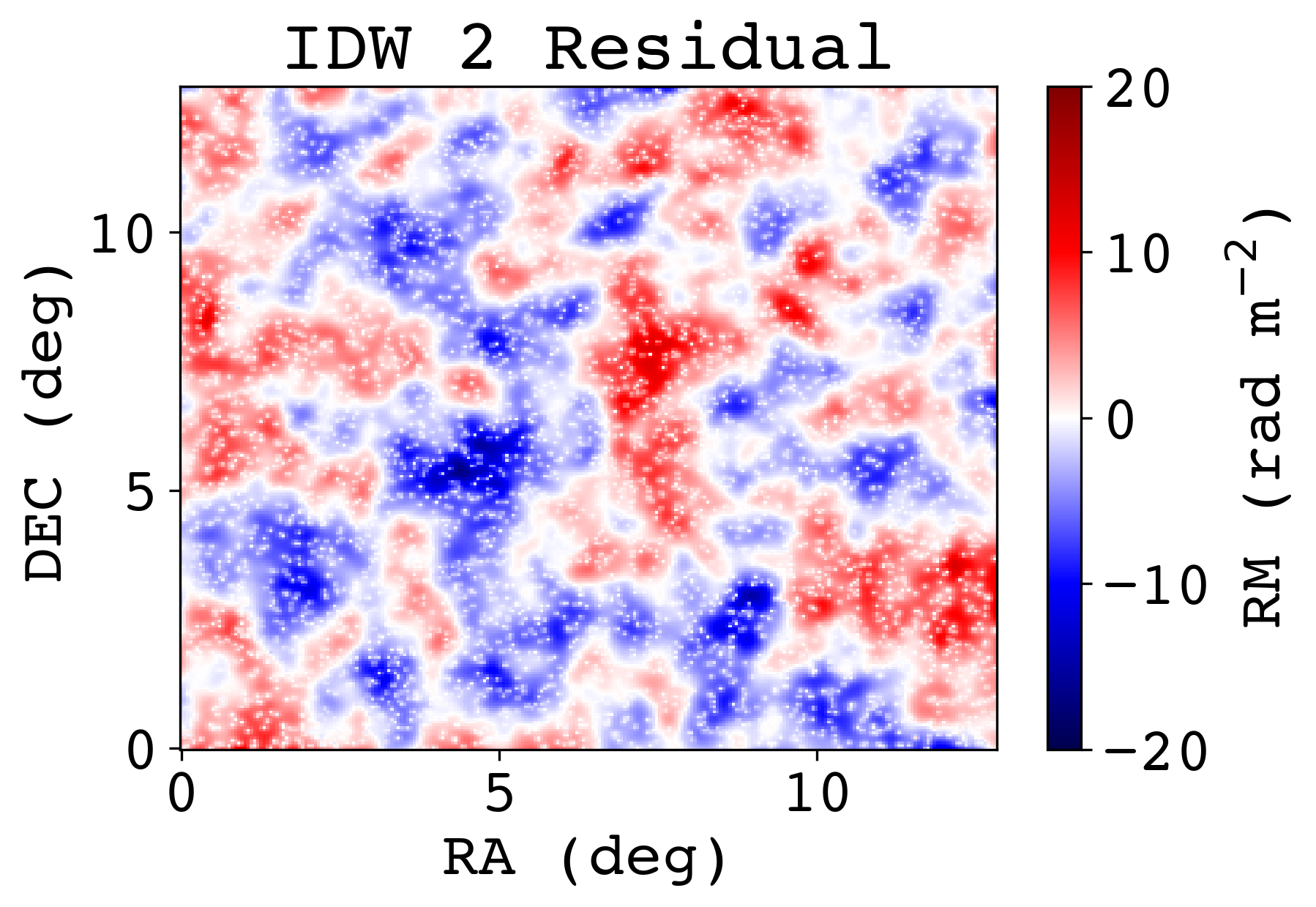}{0.33\textwidth}{}\fig{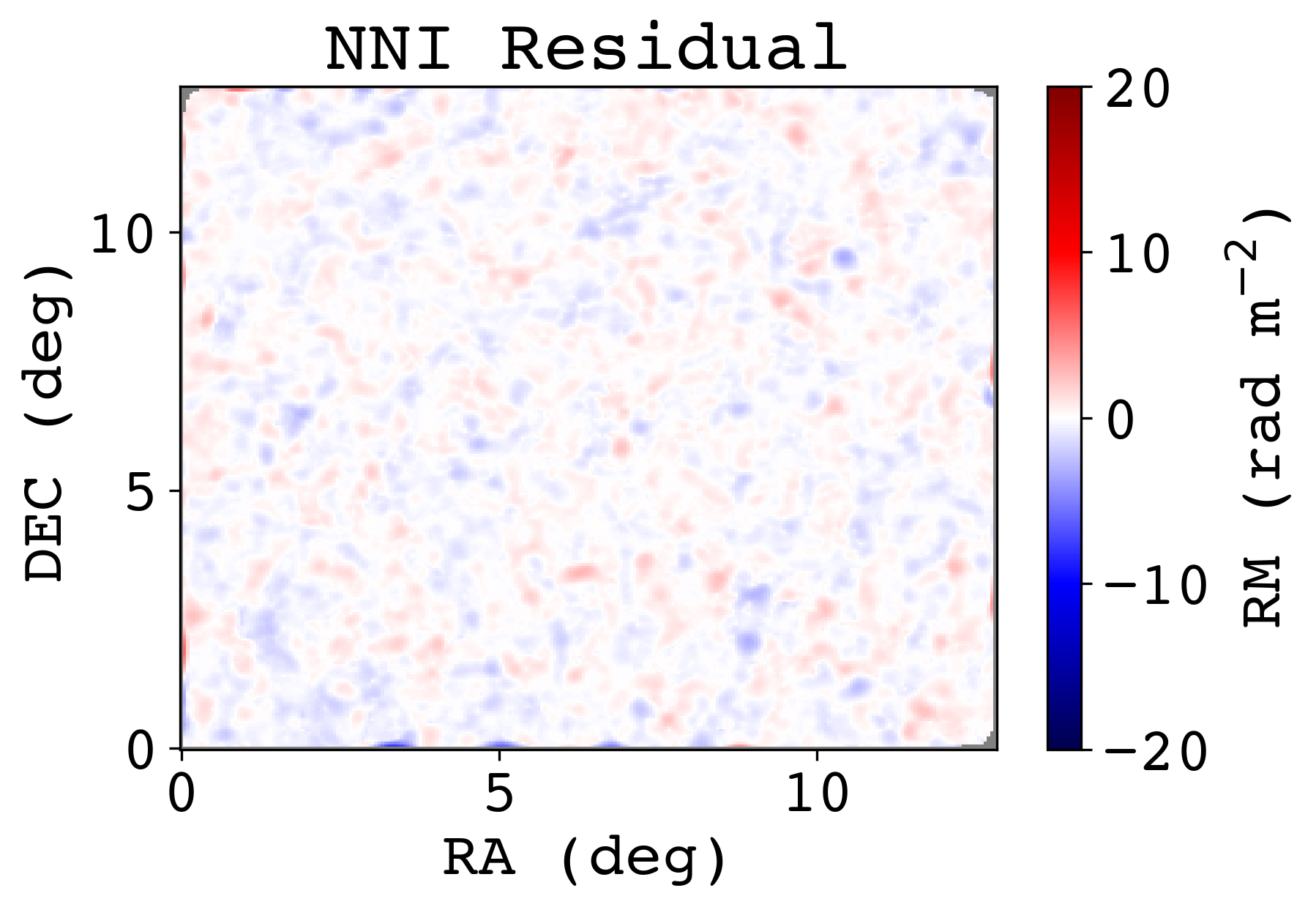}{0.33\textwidth}{}}
\gridline{\fig{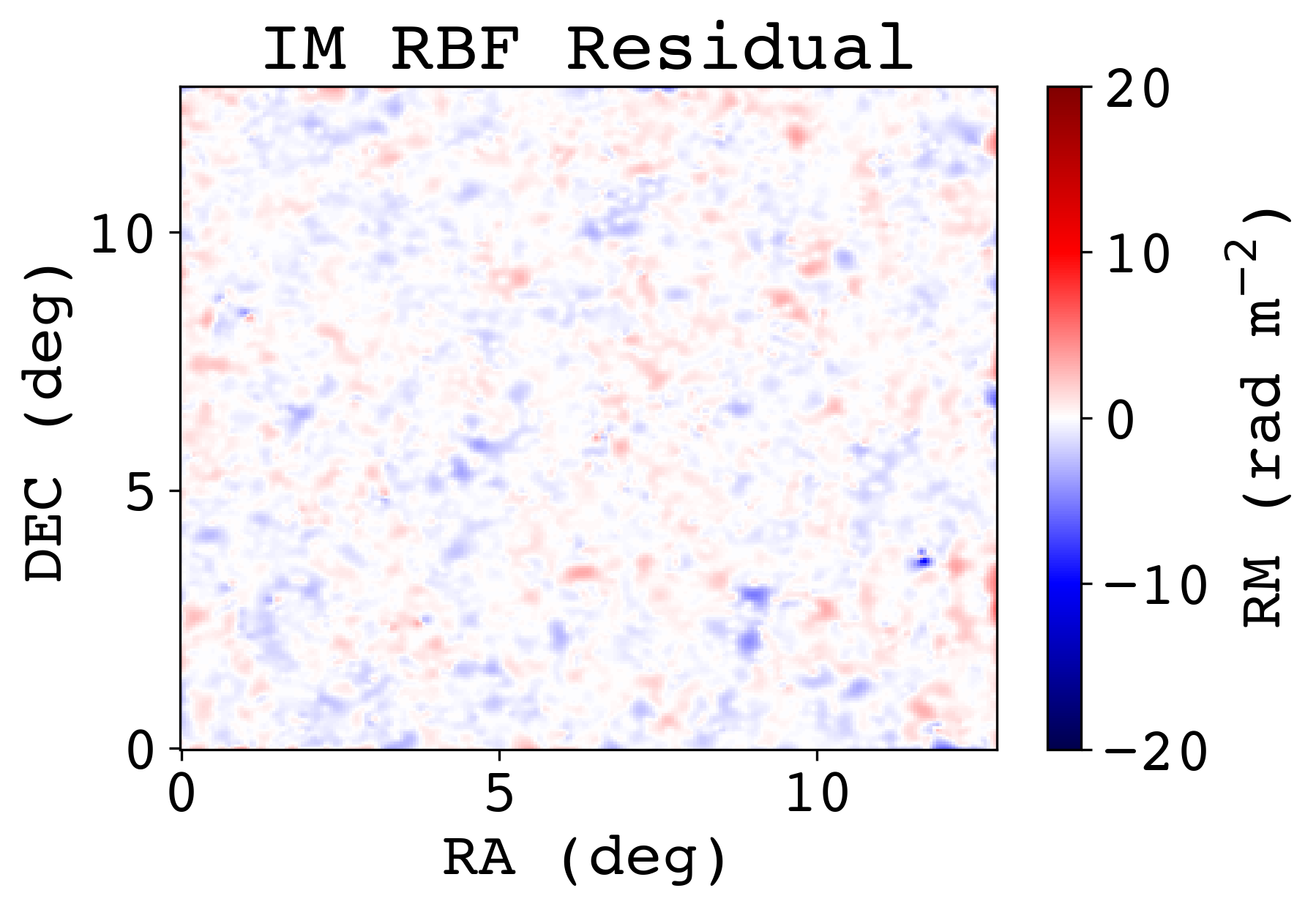}{0.33\textwidth}{}\fig{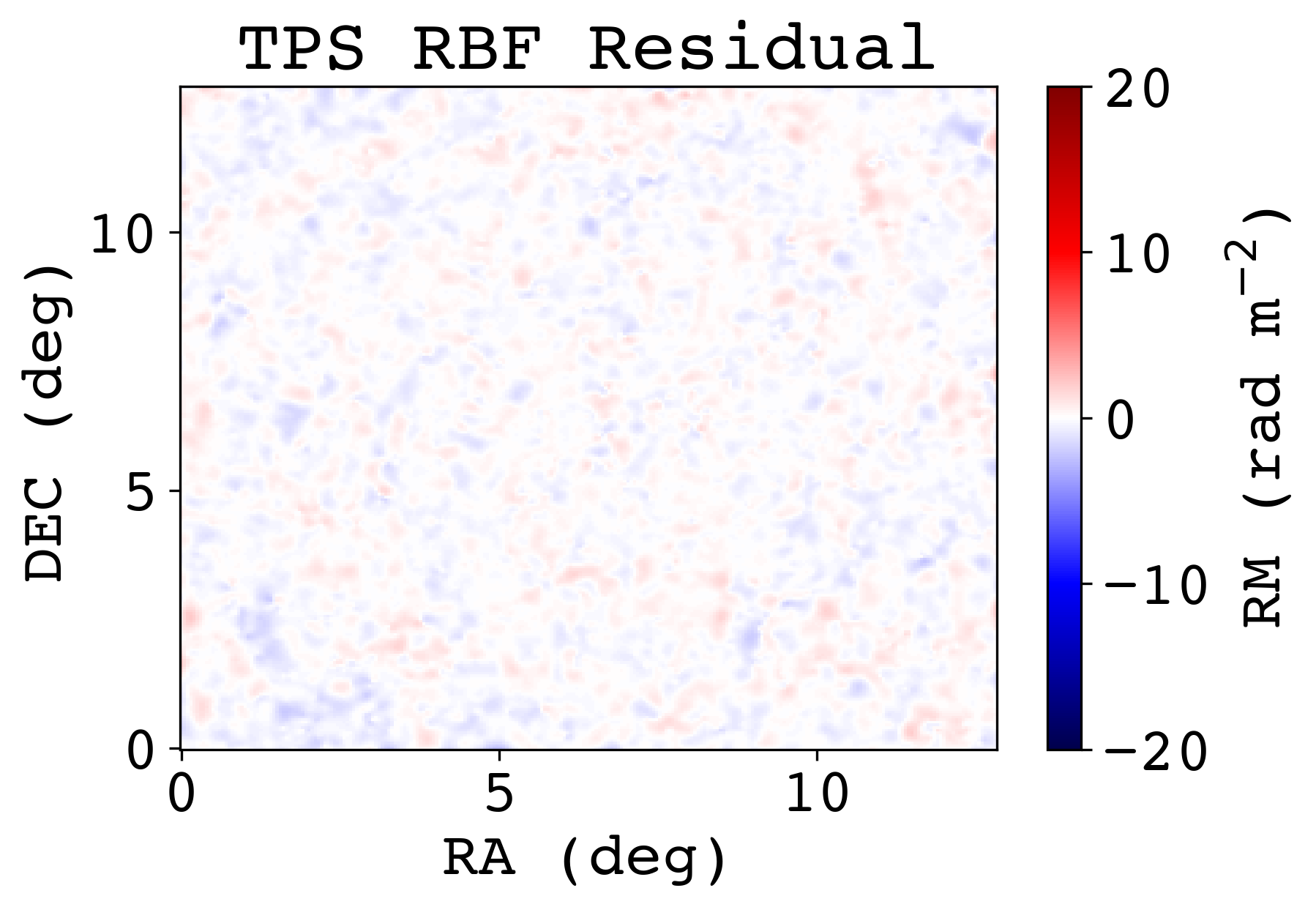}{0.33\textwidth}{}\fig{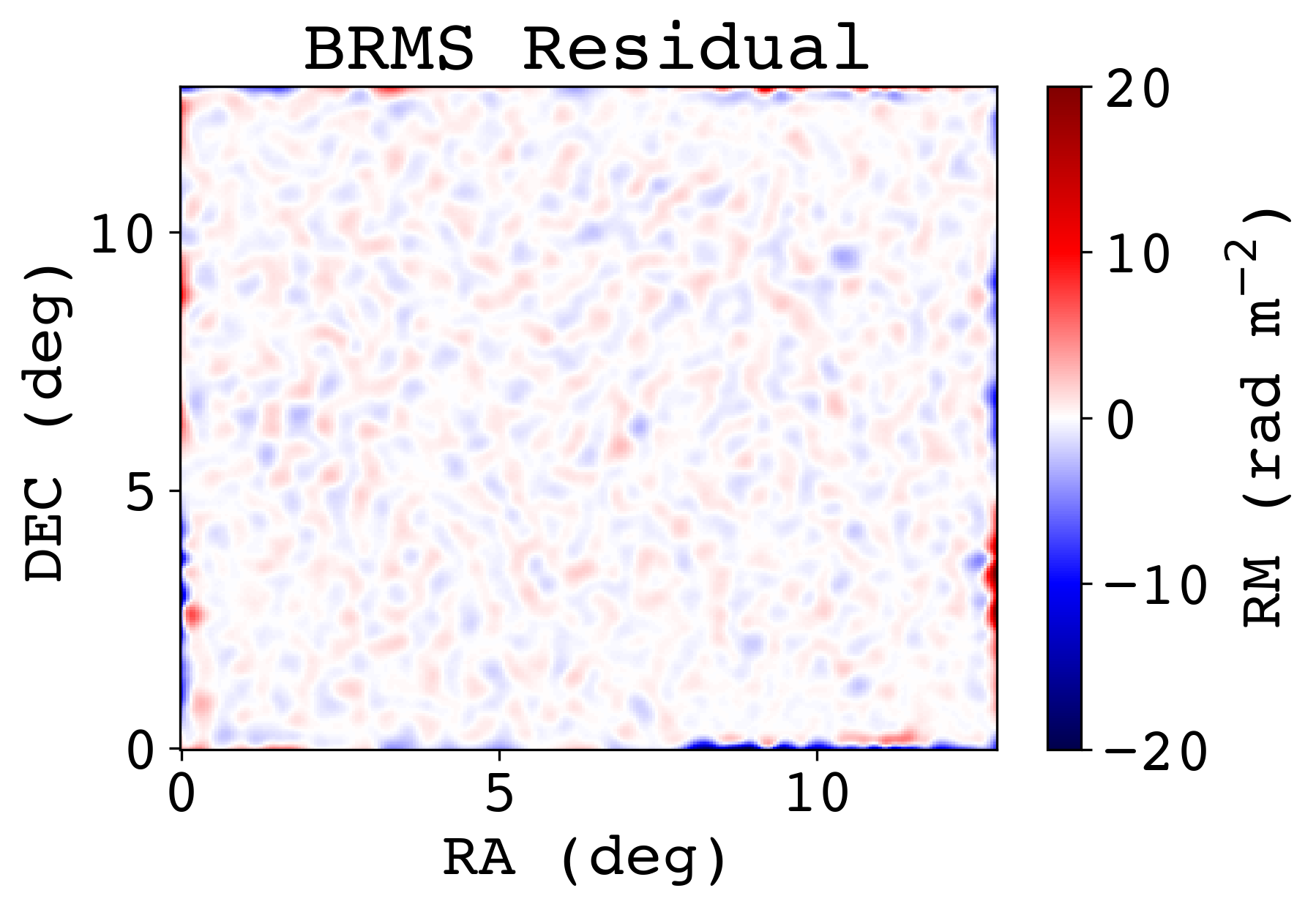}{0.33\textwidth}{}}

\caption{Residual RM maps for each of the interpolation techniques in the case of a patchy RM sky (with the data points having no extragalactic contributions). In every map, the residual RM for each pixel was calculated by finding the difference in the RM value between the reconstructed pixel and the pixel from the simulated foreground RM. From top-left to bottom-right, the reconstruction are as follows: IDW1, IDW2, NNI, IM RBF, TPS RBF, and BRMS.}
\label{fig:app3}
\end{figure}

The improvement in the RM maps of NNI and RBF techniques is far more obvious when compared to the IDW and BRMS maps, as displayed in Figure \ref{fig:app3}. The residuals for NNI and RBF are almost identically zero except for some small perturbations, and the peaks in RM for the RBF techniques.

\FloatBarrier
\subsection{Filamentary Foreground}

\begin{figure}[!htb]
\gridline{\fig{fil.png}{0.33\textwidth}{}}
\gridline{\fig{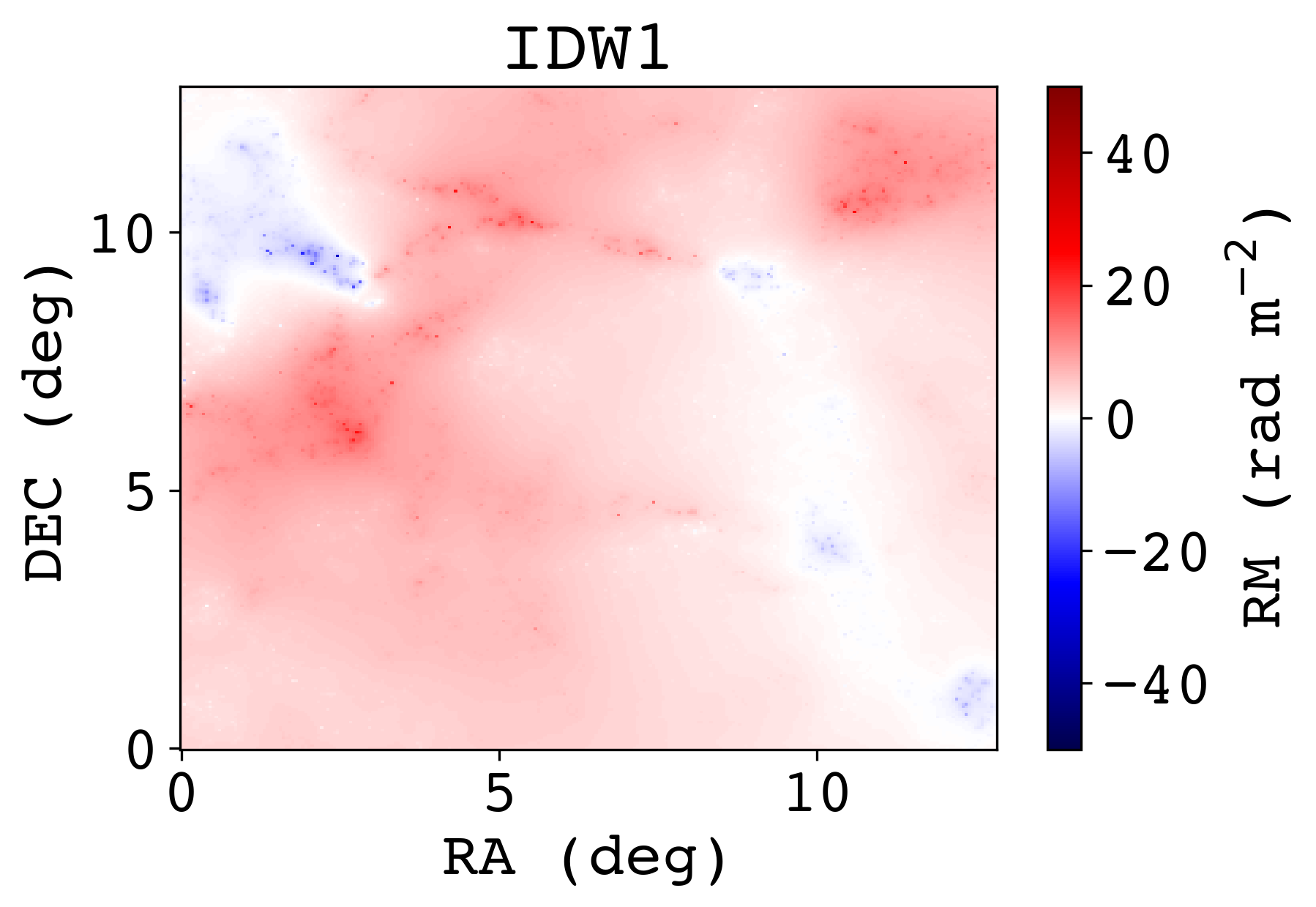}{0.33\textwidth}{}\fig{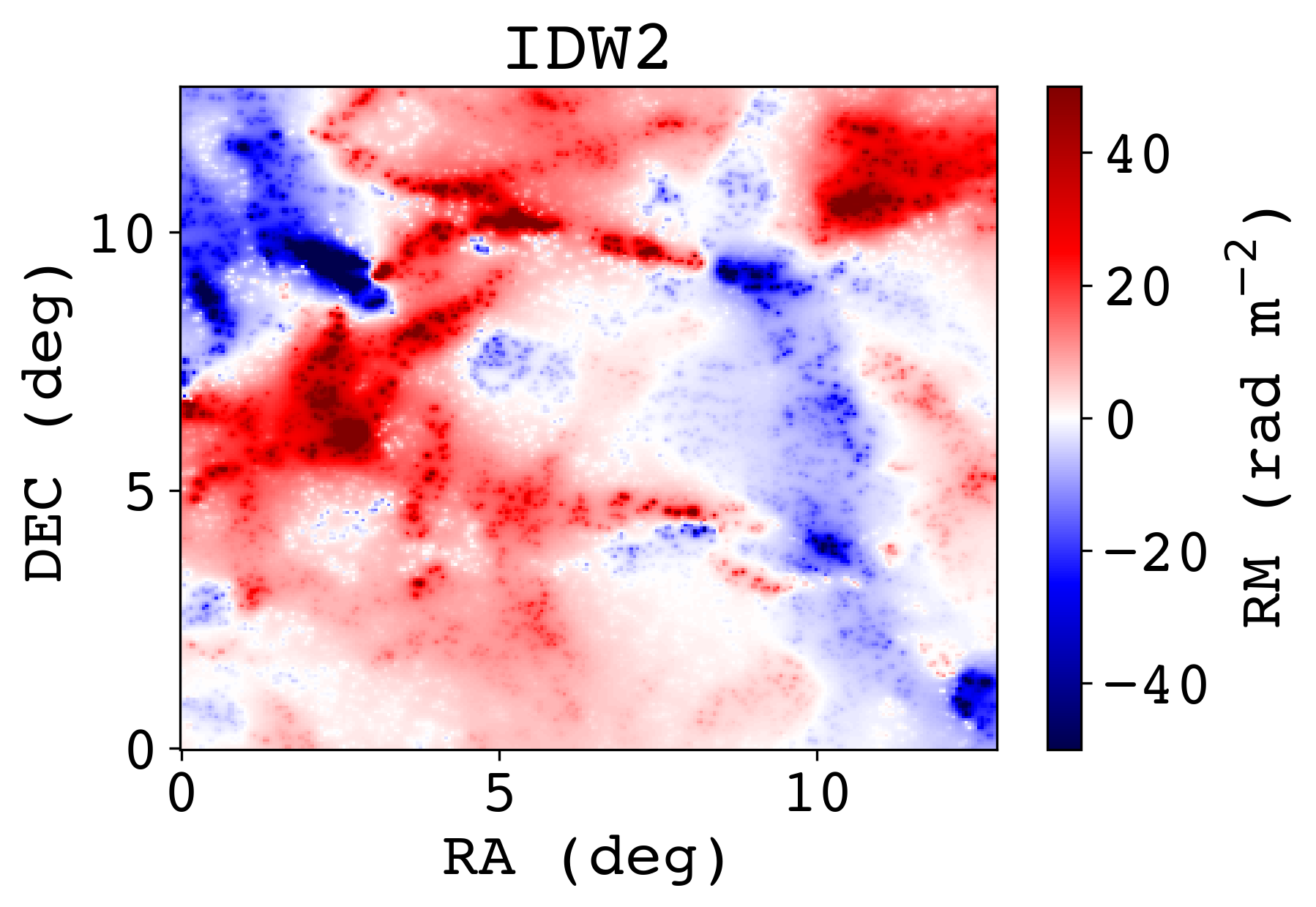}{0.33\textwidth}{}\fig{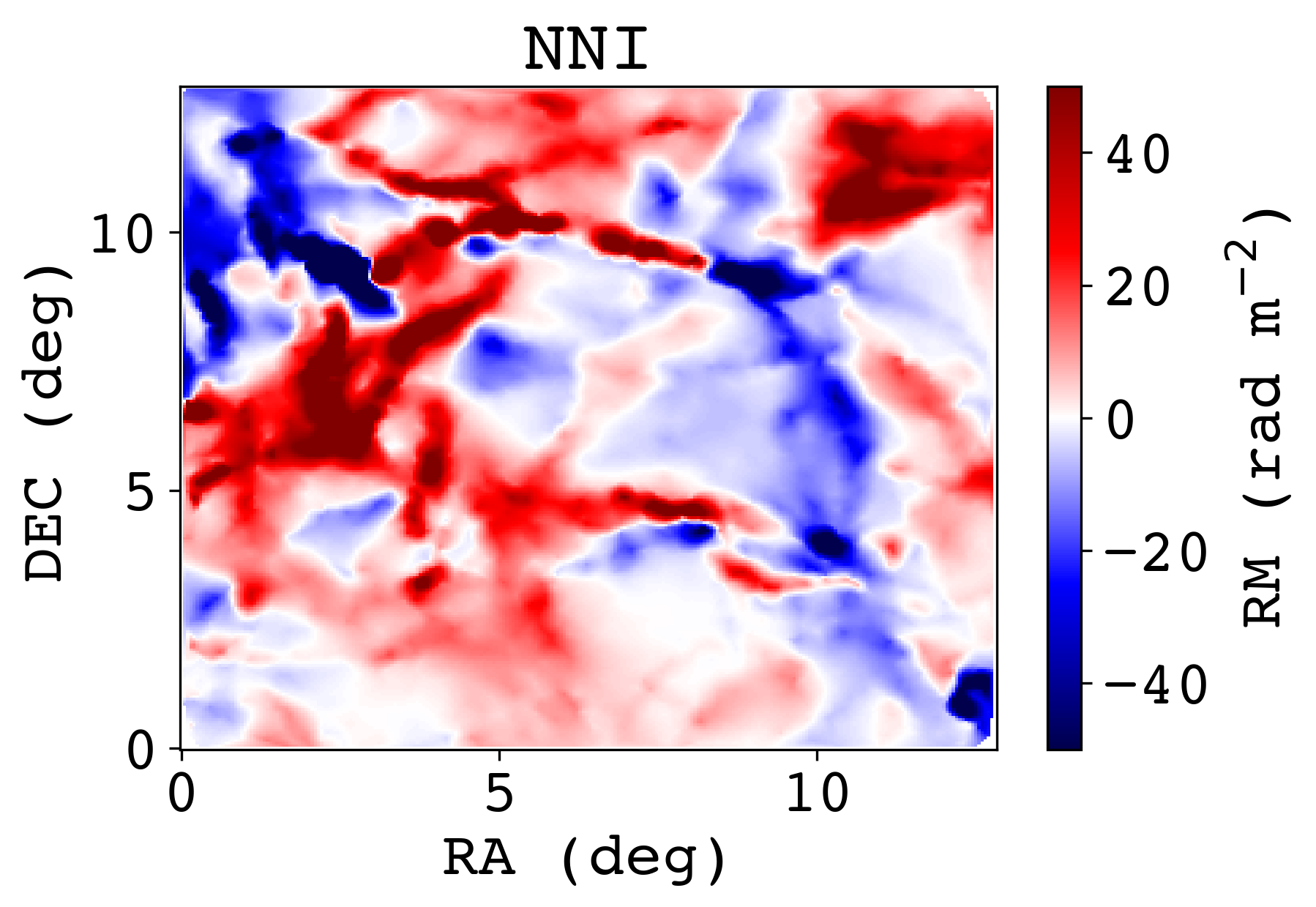}{0.33\textwidth}{}}
\gridline{\fig{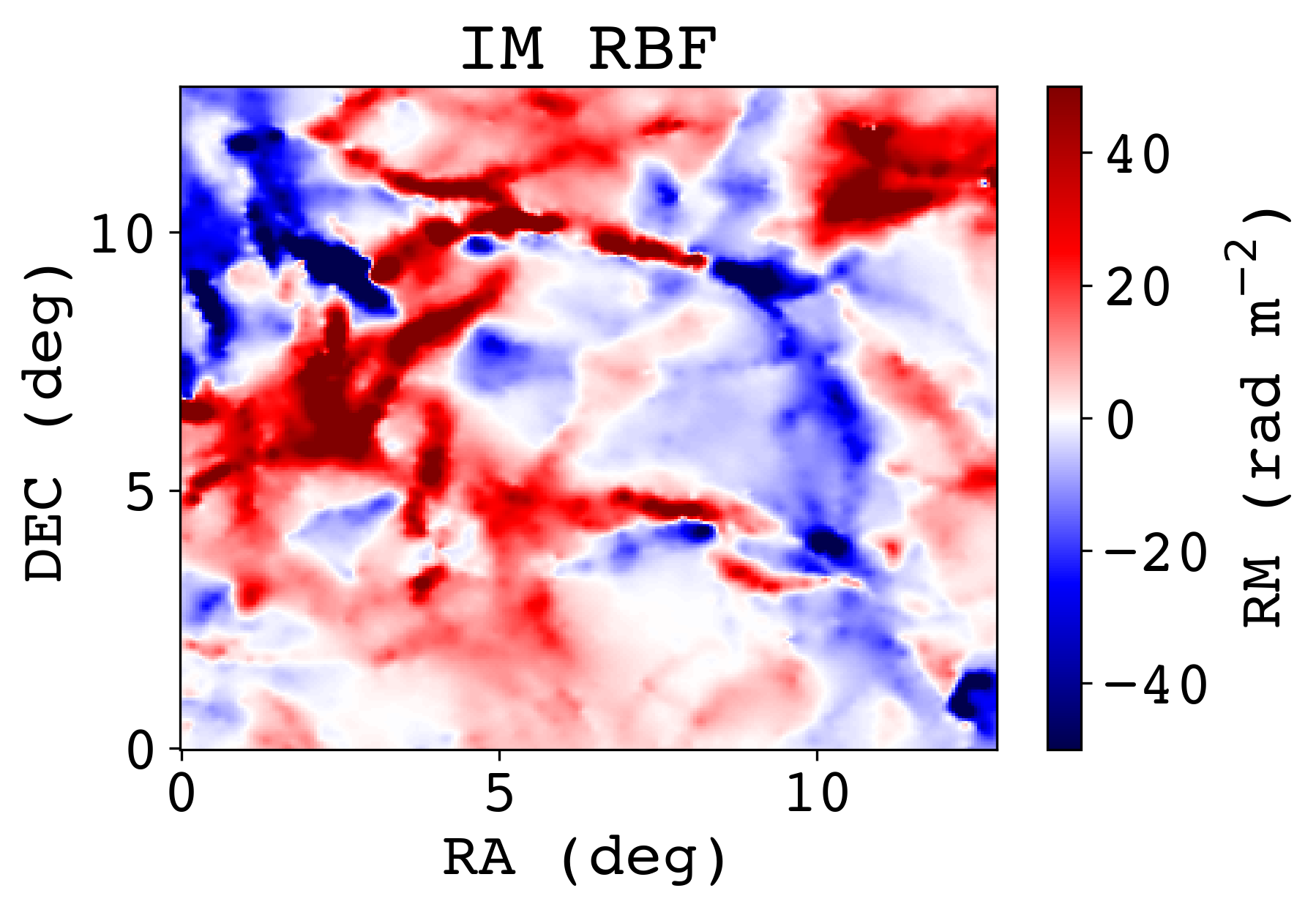}{0.33\textwidth}{}\fig{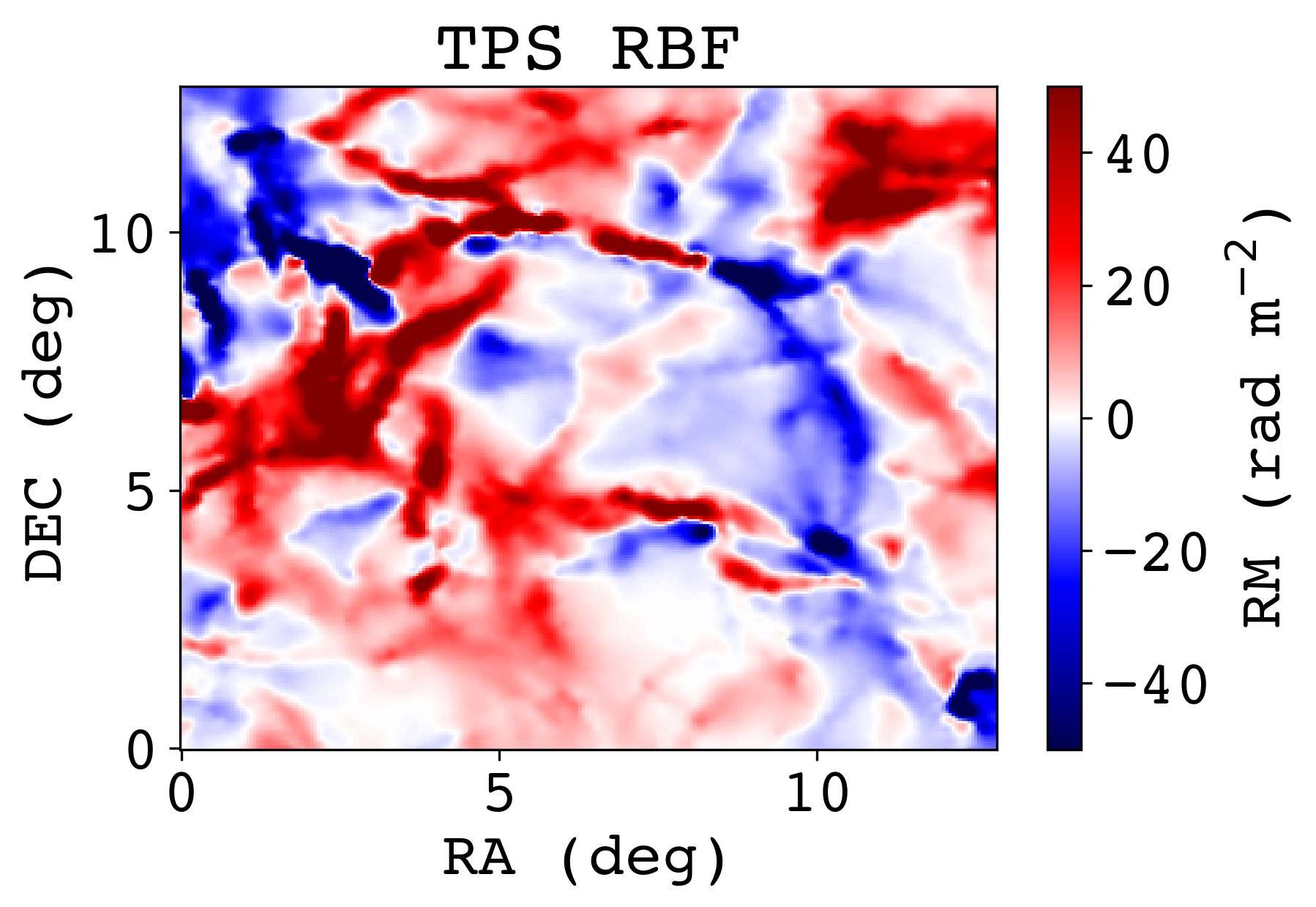}{0.33\textwidth}{}\fig{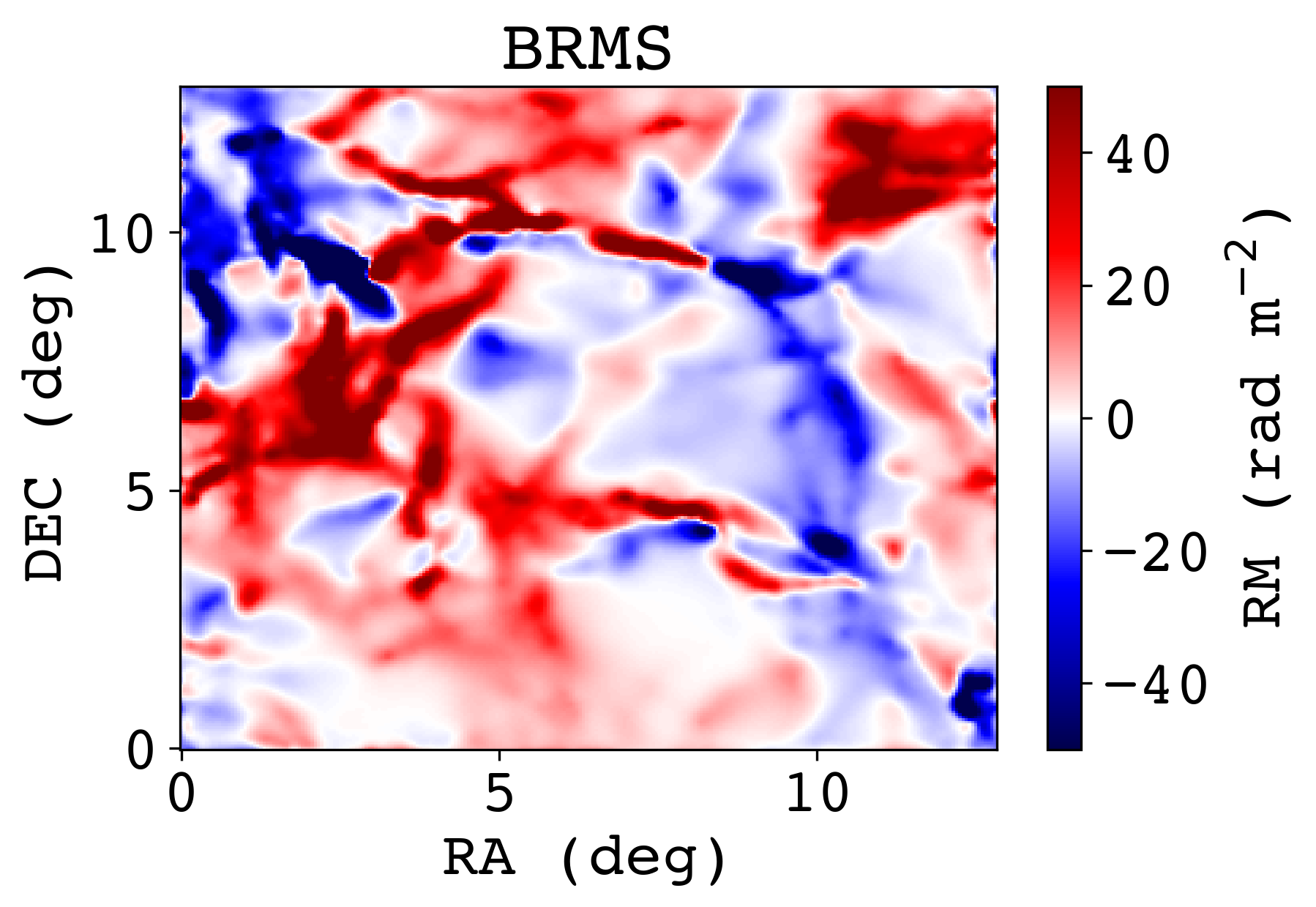}{0.33\textwidth}{}}
\caption{Reconstructed RM maps for each of the interpolation techniques for a simulated foreground with filamentary structures (with the data points having no extragalactic contributions), saturated from $-50$ to $+50$ rad m$^{-2}$. The color map represents RM, with red being positive and blue being negative. At the top-centre, we present the simulated foreground RM. From top-left to bottom-right, the reconstructions are as follows: IDW1, IDW2, NNI, IM RBF, TPS RBF, BRMS.}

\label{fig:app4}
\end{figure}

The reconstructed RM maps for the filamentary simulated foreground (with the data points having no extragalactic contributions) are displayed in Figure \ref{fig:app4}. As before, the maps produced from RBF techniques have far fewer jumps in RM than before. In addition to this, the filaments produced by NNI and RBF are far smoother; before, the filaments had small clumps forming on their edges.

\begin{figure*}[tp]

\gridline{\fig{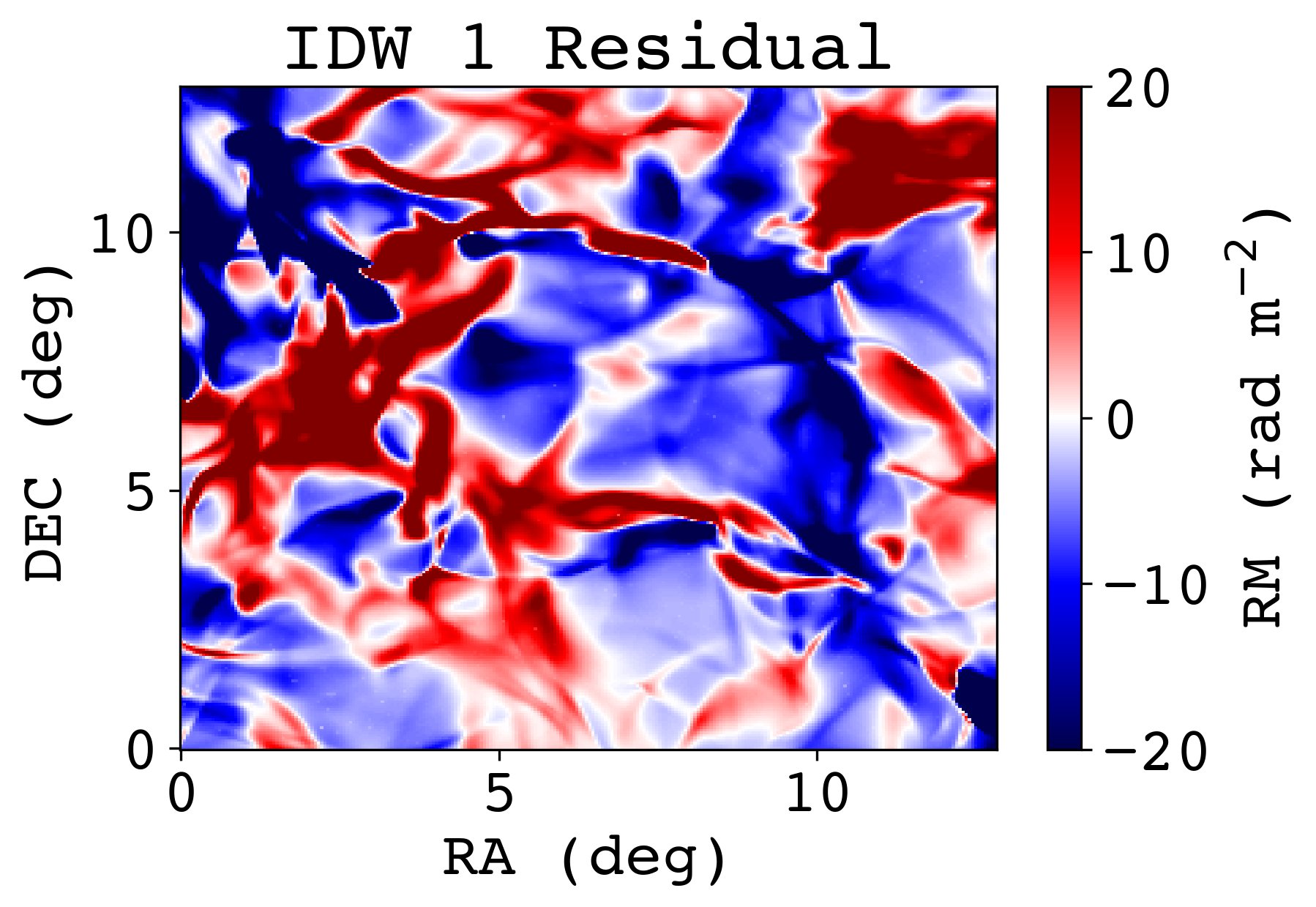}{0.33\textwidth}{}\fig{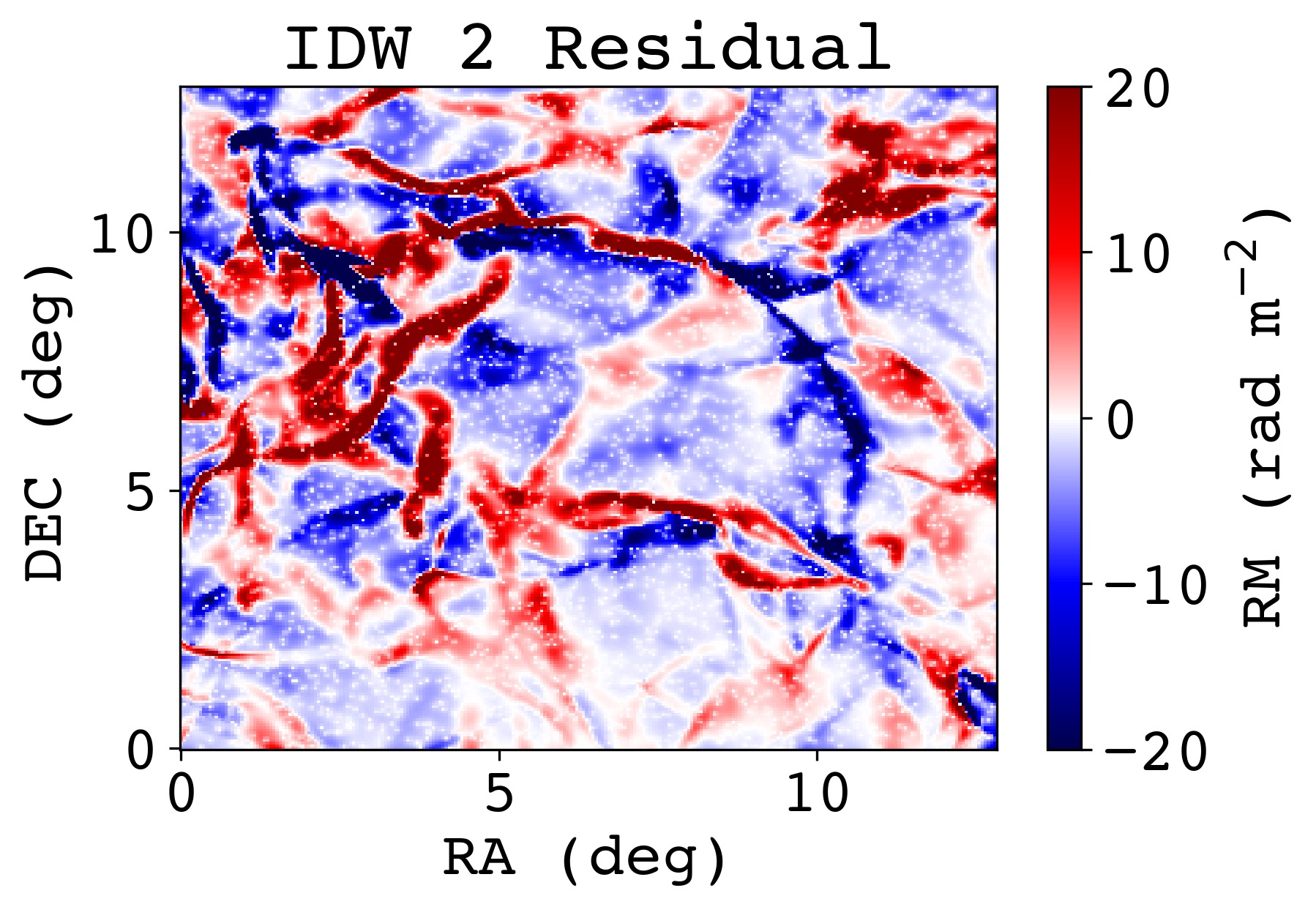}{0.33\textwidth}{}\fig{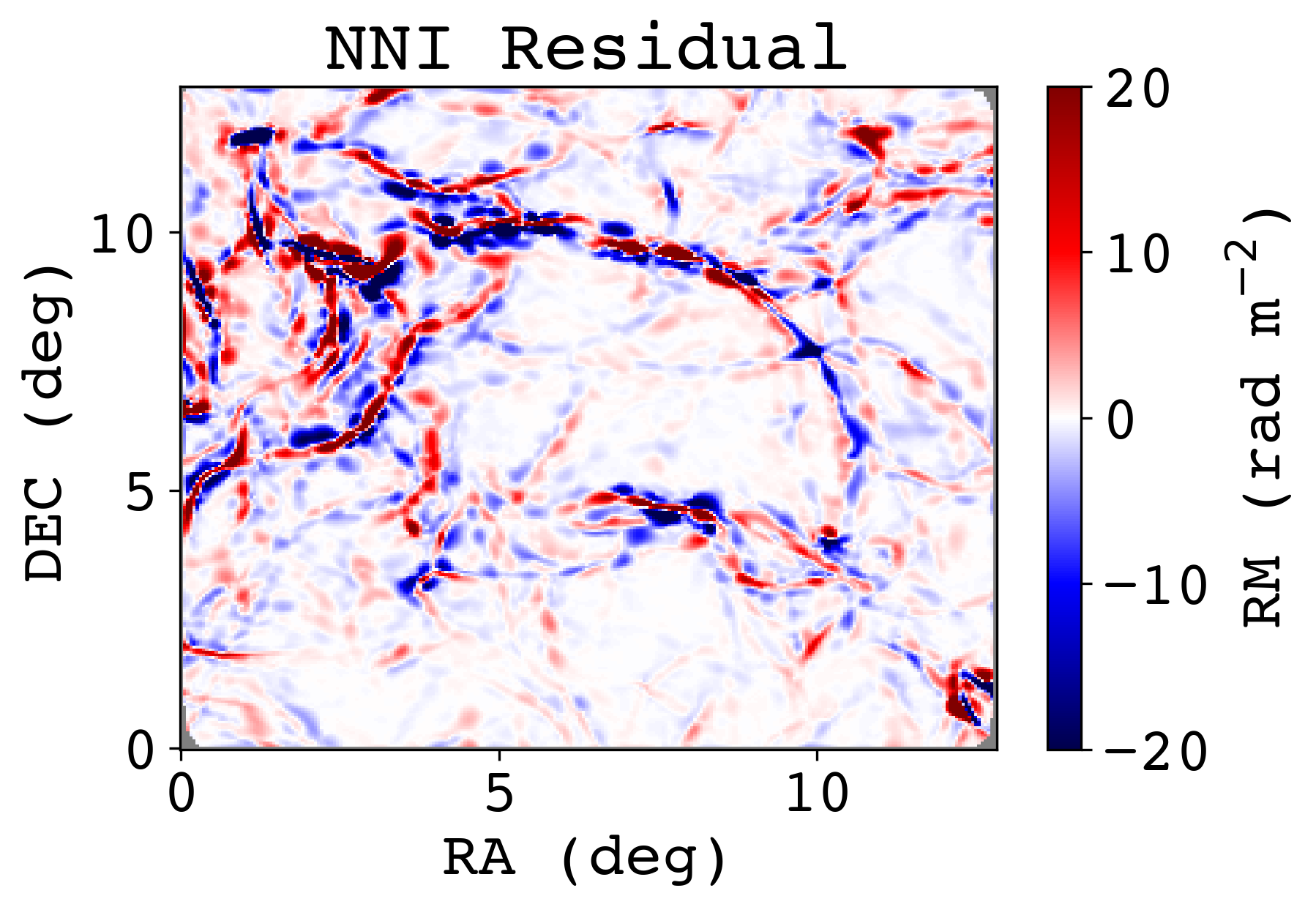}{0.33\textwidth}{}}
\gridline{\fig{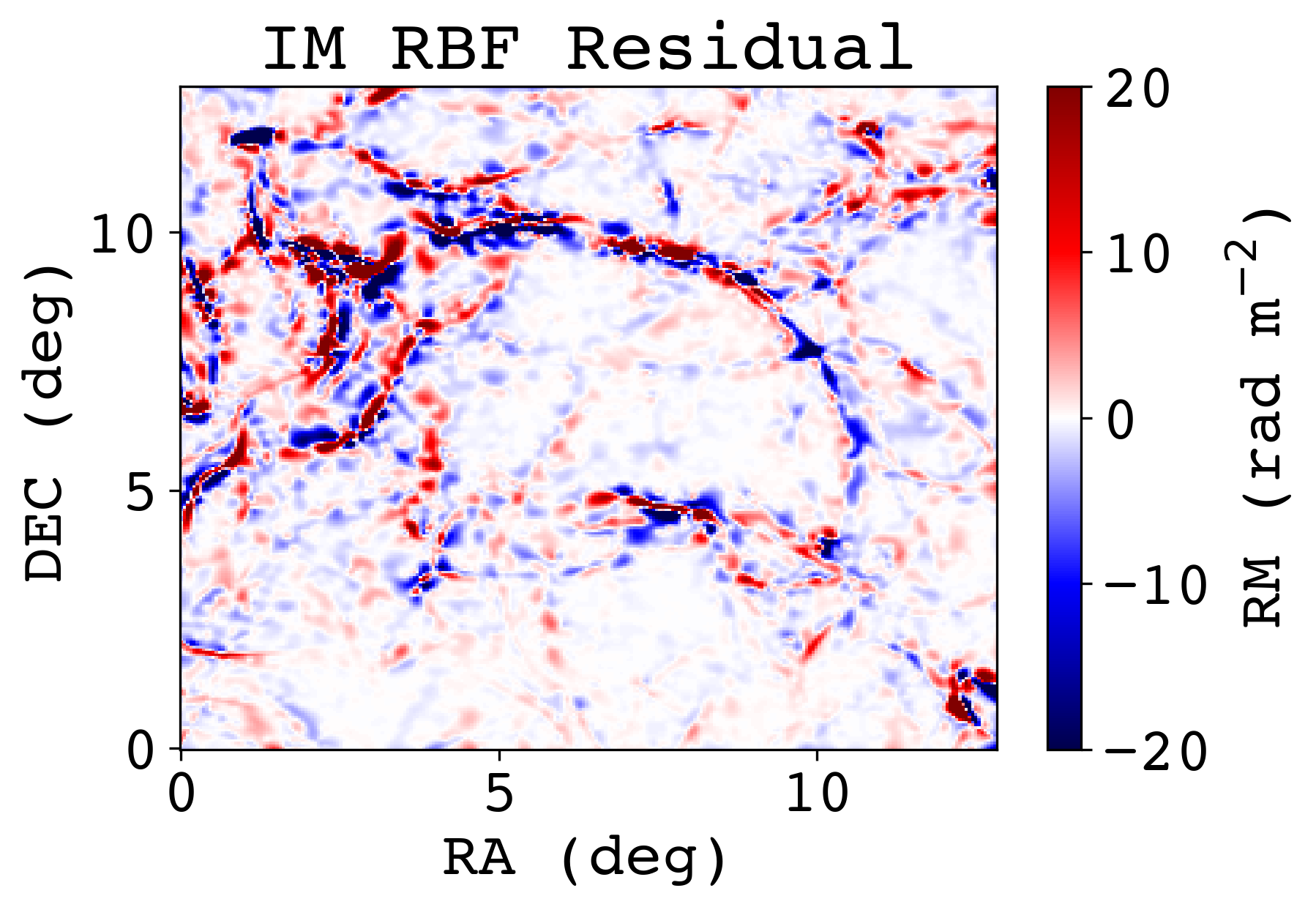}{0.33\textwidth}{}\fig{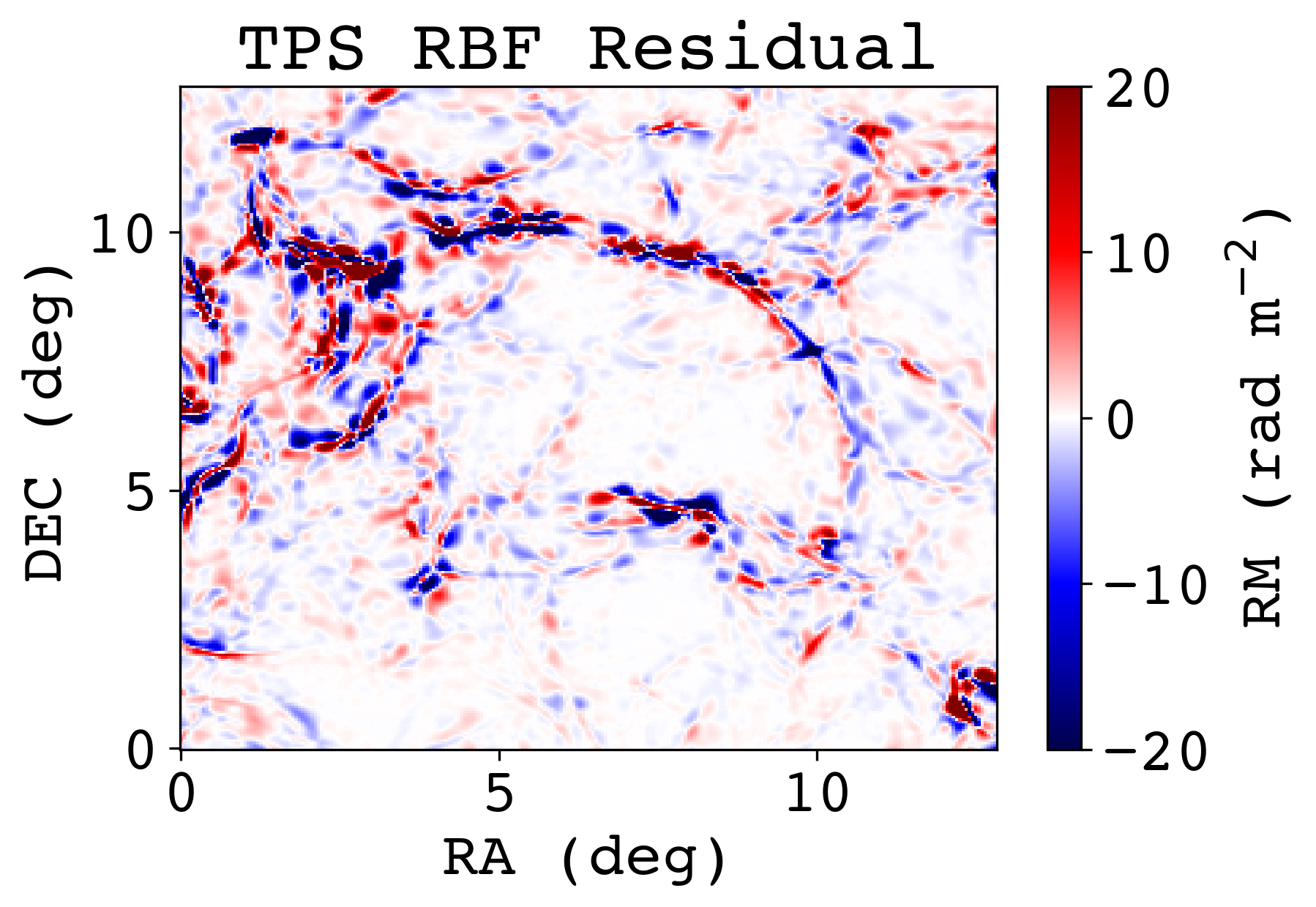}{0.33\textwidth}{}\fig{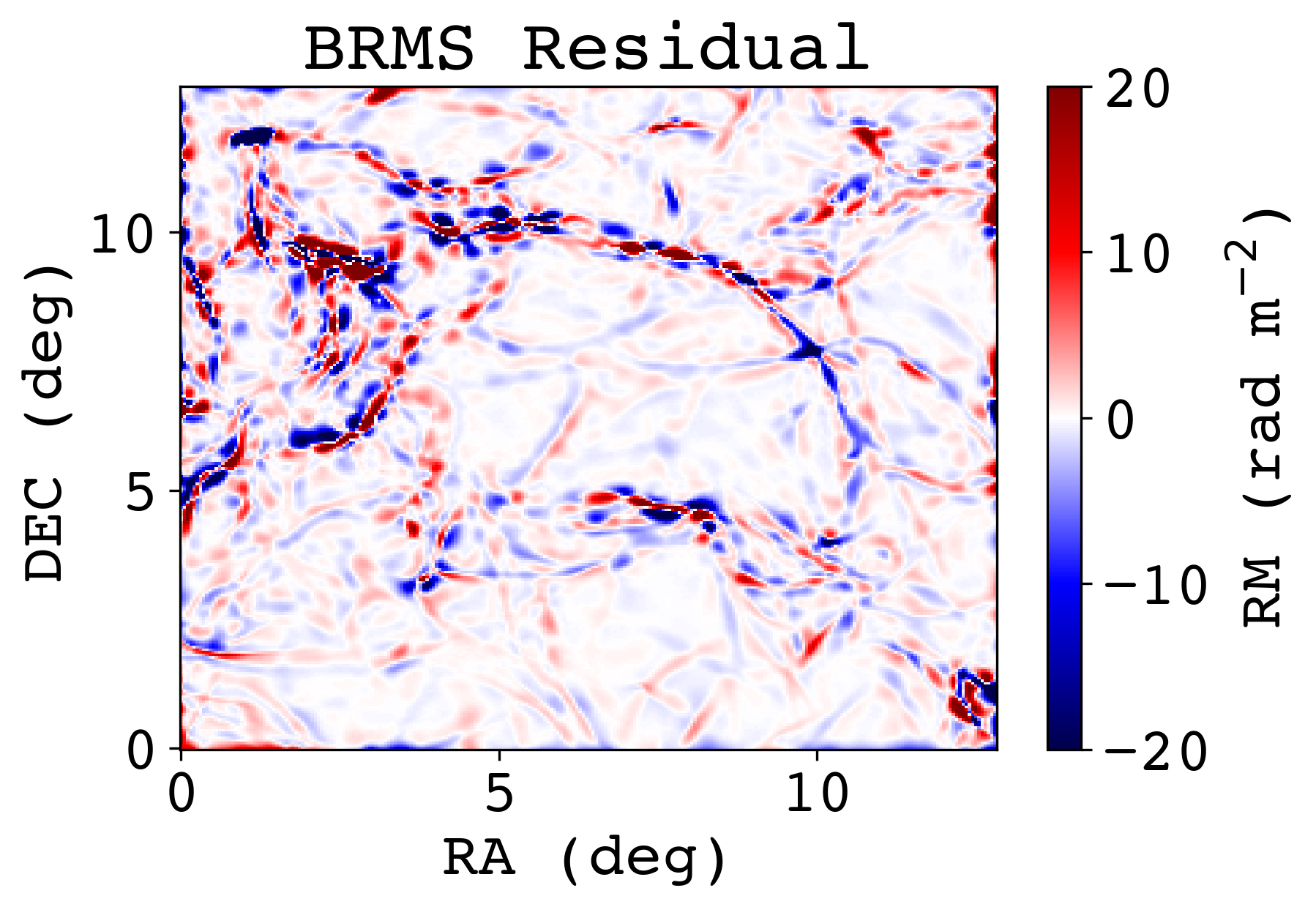}{0.33\textwidth}{}}

\caption{Residual RM maps for each of the techniques in the case of a filamentary foreground (with the data points having no extragalactic contributions). In every map, the residual RM for each pixel was calculated by finding the difference in the RM value between the reconstructed pixel and the pixel from the simulated foreground RM. From top-left to bottom-right, the reconstruction are as follows: IDW1, IDW2, NNI, IM RBF, TPS RBF, and BRMS.}
\label{fig:app6}
\end{figure*}

Figure \ref{fig:app6} portrays the residual RM maps saturated at $\pm 20$ rad m$^{-2}$. As before, NNI and RBF show the biggest improvement (with the residuals having the most RM at the filaments).

\FloatBarrier
\section{Accuracy v. Computation Time}
\label{app:b}

Figure \ref{fig:comp} shows a plot of the accuracy (in terms of the goodness-of-fit parameter $I_\mathrm{res}$) against the computation time for the interpolation kernels for all the RM skies. Note that a lower $I_\mathrm{res}$ value indicates better accuracy. 

\begin{figure}[!htb]
    \centering
    \gridline{\includegraphics[width = 0.49\textwidth]{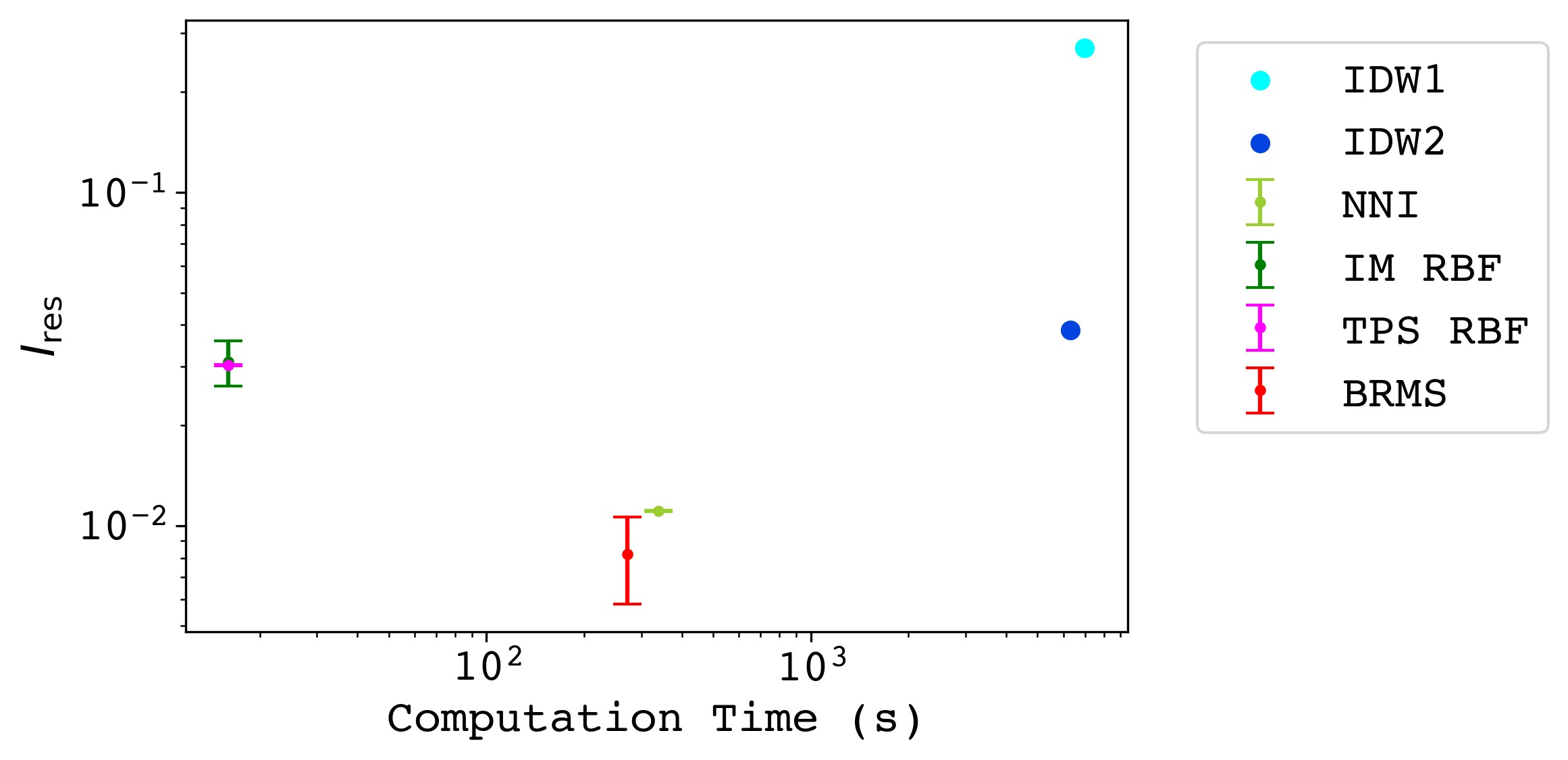} \includegraphics[width= 0.49\textwidth]{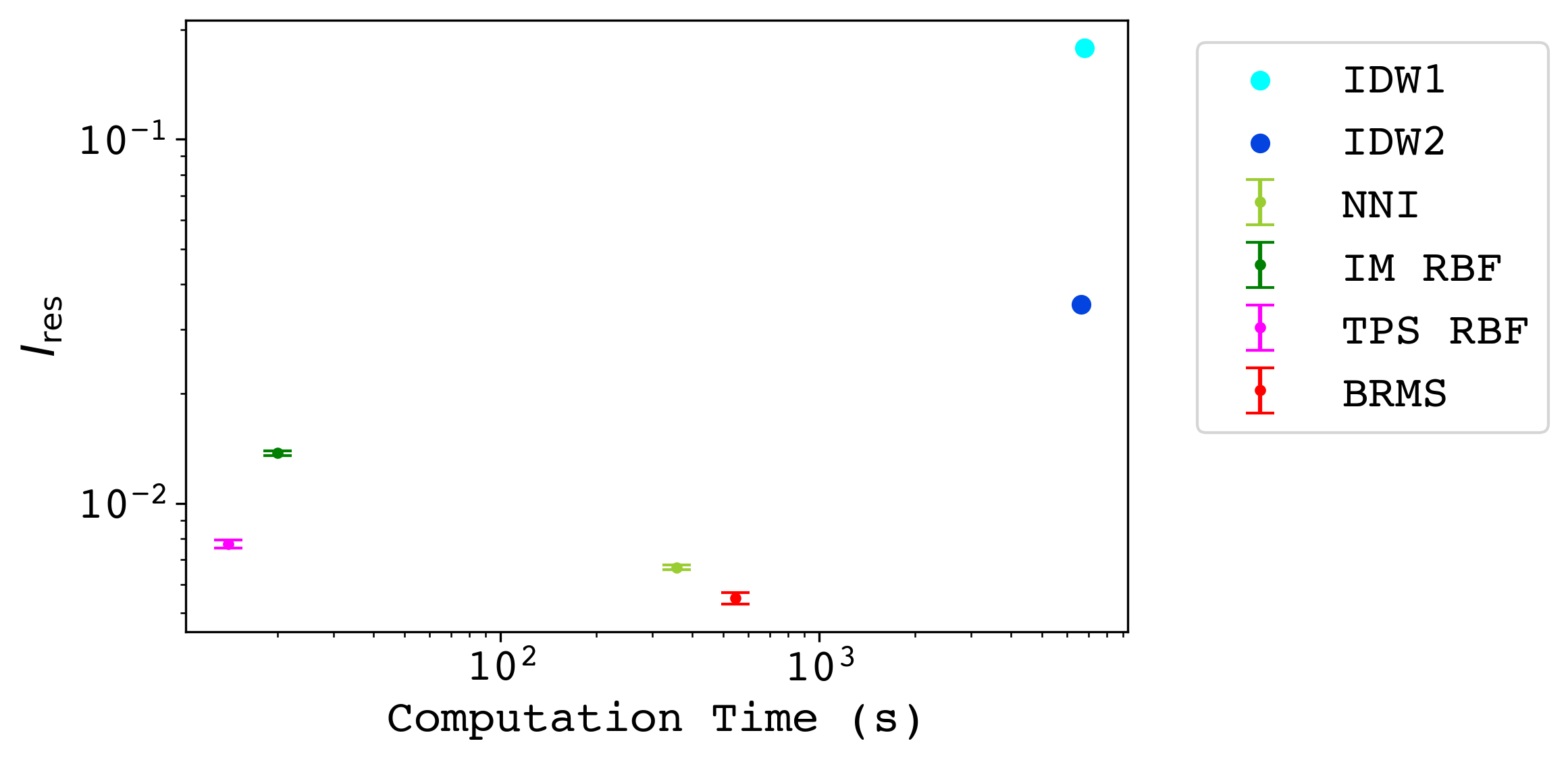}}
    \caption{The accuracy in terms of the goodness-of-fit parameter $I_\mathrm{res}$ against the computational time in seconds. The plots show the results for data sets that have extragalactic contributions for the patchy (left) and filamentary (right) foregrounds.}
    \label{fig:comp}
\end{figure}

\section{Varying Noise Standard Deviations}
\label{app:c}
\begin{figure}
    \centering
    \gridline{\includegraphics[width = 0.49\textwidth]{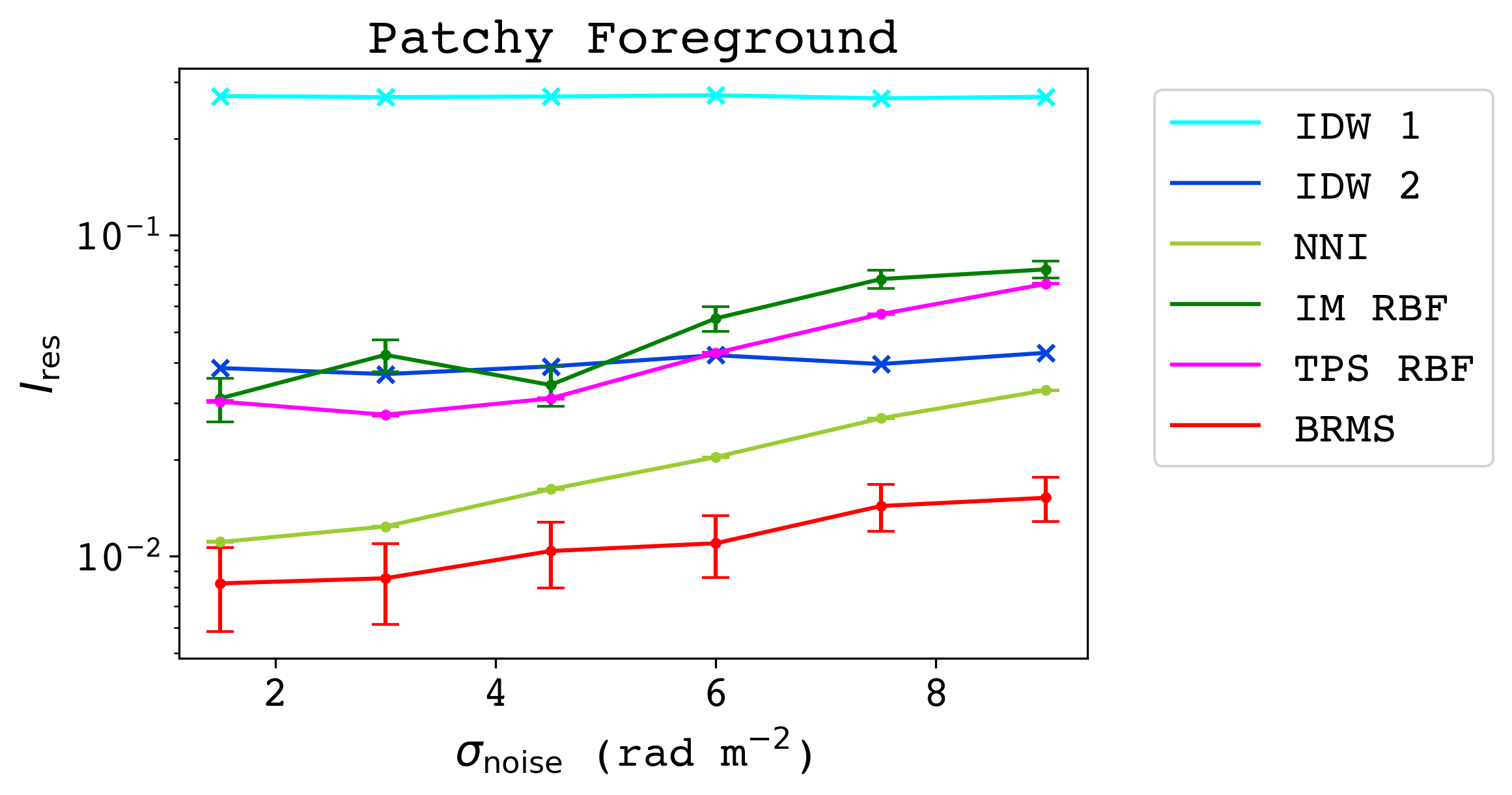} \includegraphics[width = 0.49\textwidth ]{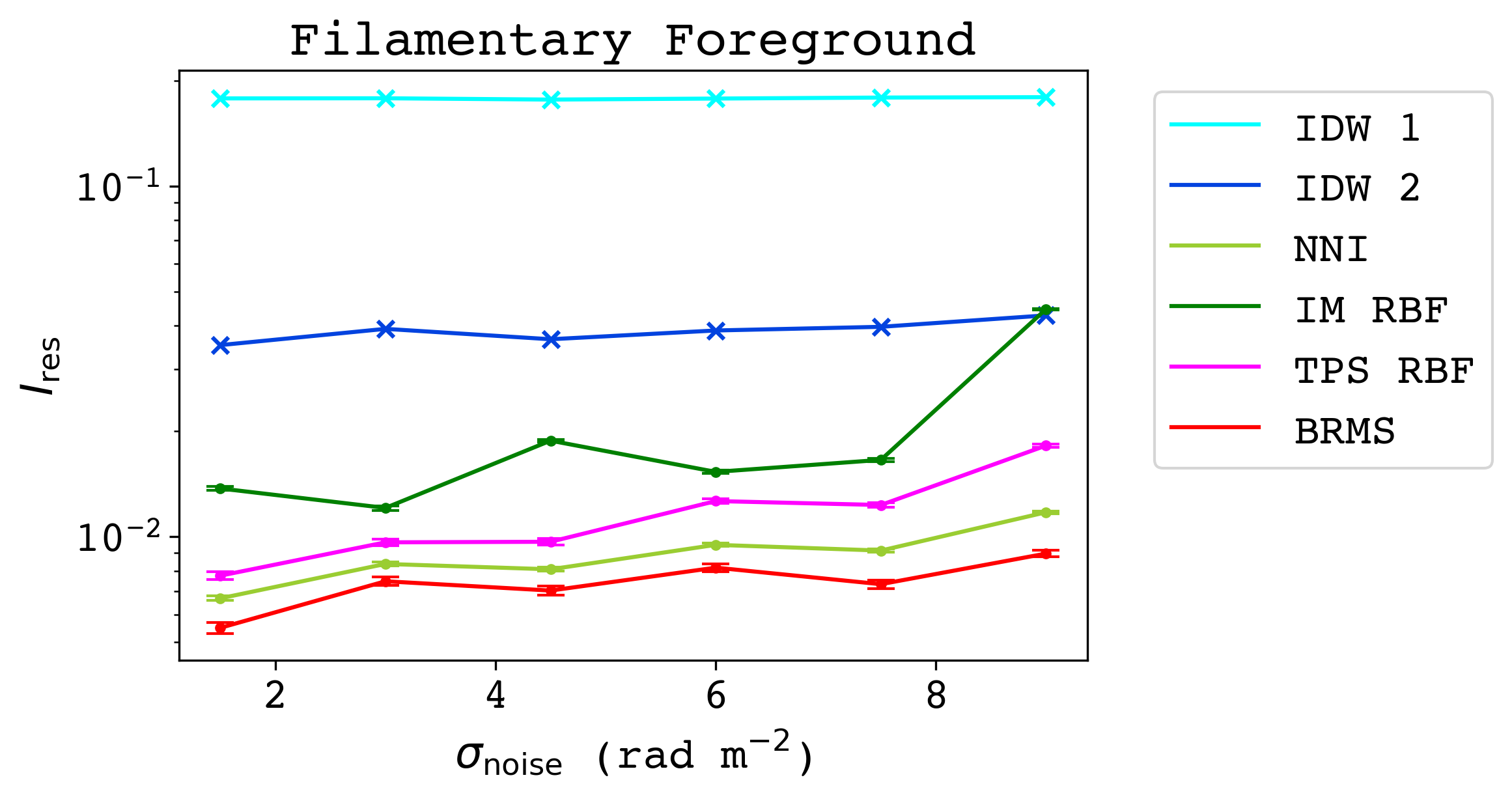}} 
    \caption{Plots of the accuracy (measured in $I_\mathrm{res}$) of the interpolation techniques against the standard deviation in the noise for the patchy (left) and filamentary (right) foreground RM structures with extragalactic RM contributions.} 
    \label{fig:noise_1}
\end{figure}

The different standard deviation in the noise that we tested were $\sigma_{\rm{noise}} = 1.5,\ 3.0,\ 4.5,\ 6.0,\ 7.5,\ 9.0$ in rad m$^{-2}$. The accuracy of the interpolation techniques for these noise sigma in the case of data samples containing extragalactic contributions is displayed in Figure \ref{fig:noise_1}. In general, the accuracy of almost all interpolation techniques reduces with an increase in $\sigma_{\mathrm{noise}}$. However, both IDW techniques do not seem to follow this trend in general as they are relatively stable against very high noise as well (with IDW2 even improving a little at higher noise for the patchy foreground). Some of the techniques do exhibit mild improvement when increasing the noise standard deviation by an increment; however, this is likely because of the randomly generated data being favourable and is likely not statistically significant.
\section{Varying data density}
\label{app:e}
We tested the accuracy of BRMS, NNI, IM and TPS against various data densities to determine an estimate of the data density that is needed to produce acceptable results. We decided to exclude the IDW techniques from these tests as they're results are far more inaccurate than the others, and their computational expense is too high. All the data samples used in this test have the same extragalactic and noise profiles as described in Section \ref{sec:data}. The initial  input densities (prior to the removal of extragalactic RM) were: $45$ pts deg$^{-2}$, $23$ pts deg$^{-2}$, $11$ pts deg$^{-2}$, and $5$ pts deg$^{-2}$. Figure \ref{fig:comp} presents the results. As expected, the accuracy of the techniques decreases drastically with a decrease in the density of sources. We will set a goodness level of $10^{-1.9}$ based on Figure \ref{fig:comp}. From this, we see that most techniques fail when the density is close to $10$ pts deg$^{-2}$. Therefore, for accurate interpolation results we recommend having a density of at least $15$ pts deg$^{-2}$. Additionally, as expected almost all of the techniques require less computation time with a decrease in the density of the number of data points; however, BRMS increases in computational time markedly for lower densities. This is likely because the inference takes longer to converge with lower source densities.

\begin{figure}
    \centering
    \gridline{\includegraphics[width = 0.49\textwidth]{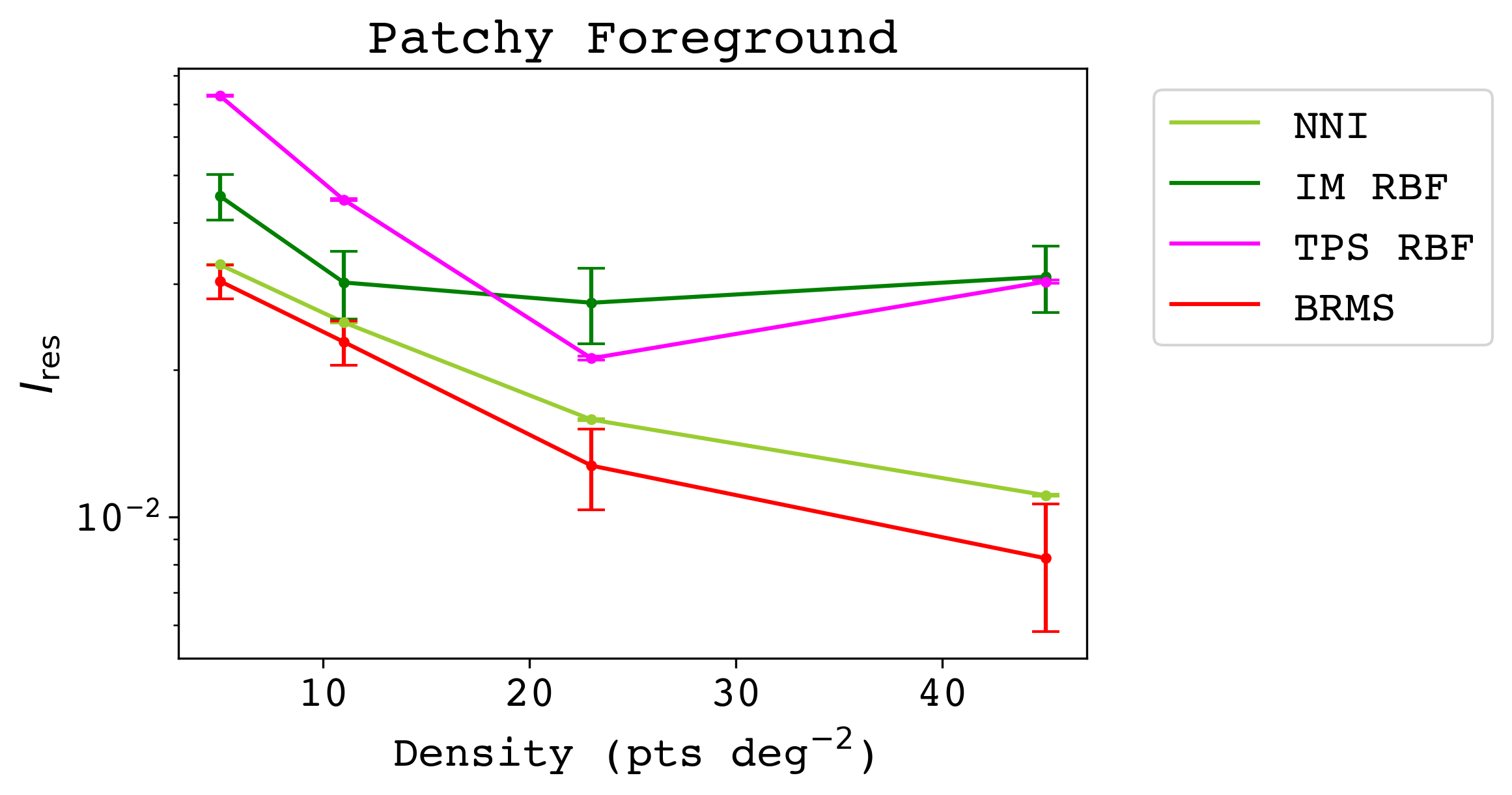} \includegraphics[width = 0.49\textwidth ]{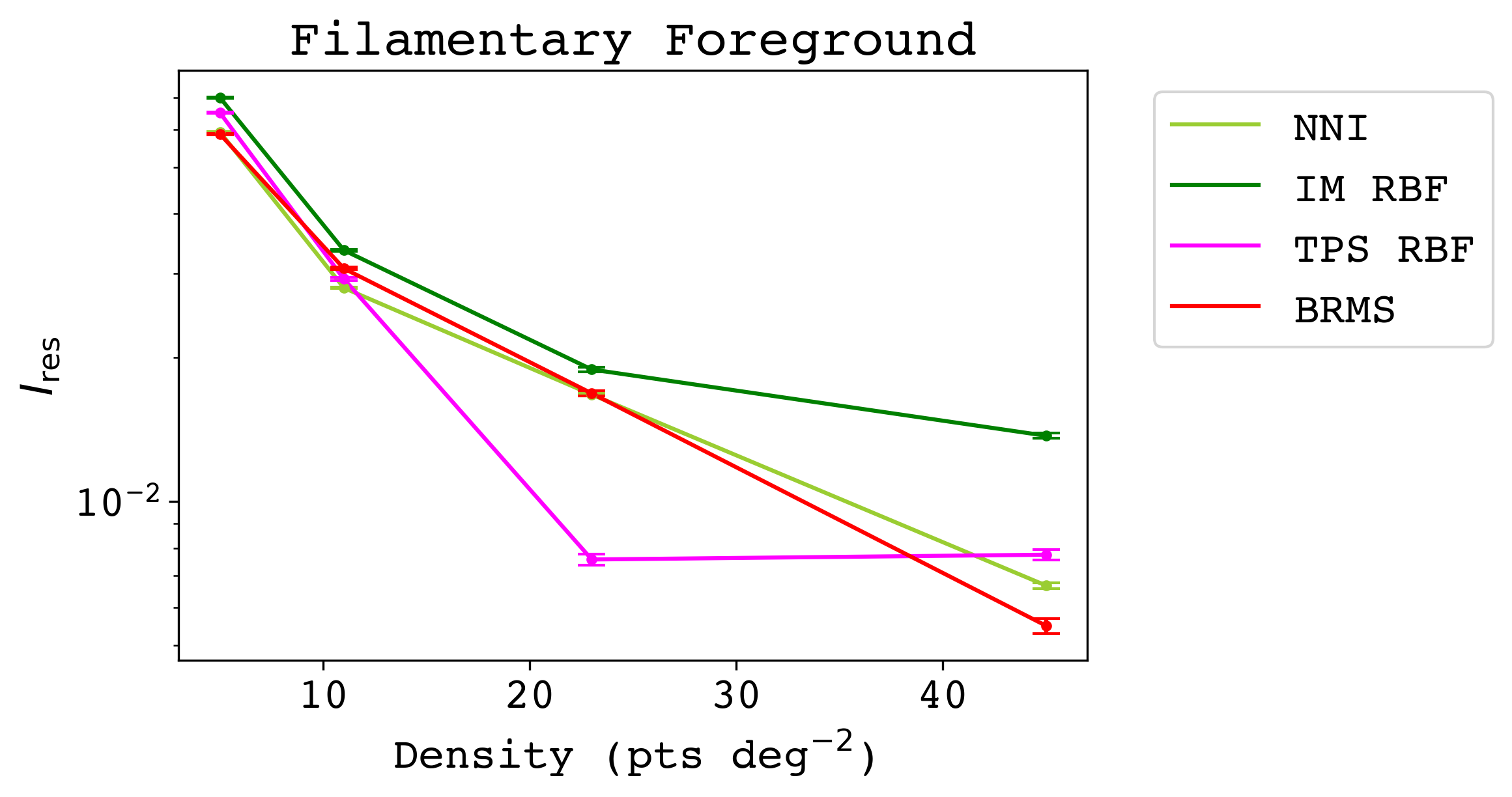}} 
    \caption{Plots of the accuracy (measured in $I_\mathrm{res}$) of the interpolation techniques against the source density for the patchy (left) and filamentary (right) foreground RM structures with extragalactic RM contributions.}
    \label{fig:comp}
\end{figure}


\FloatBarrier
\newpage

\bibliography{Interpolation_Schemes}{}
\bibliographystyle{aasjournal}



\end{document}